\documentclass[12pt,letterpaper]{JHEP3}

\usepackage{amscd,amsmath,amssymb,amsfonts,xspace,mathrsfs,amsthm}
\usepackage{color, mathtools}
\usepackage{url}
\usepackage{tikz}
\let\normalcolor\relax
\usepackage{latexsym}
\usepackage{graphicx}
\usepackage{dsfont}
\usepackage{color}
\usepackage{longtable}

\usepackage{epsf}
\usepackage[latin1]{inputenc}
\usepackage{multirow}
\usepackage{epsfig}

\numberwithin{equation}{section}

\def\bea{\begin{eqnarray}}
\def\eea{\end{eqnarray}}
\def\be{\begin{equation}}
\def\ee{\end{equation}}
\def\ba{\begin{align}}
\def\ea{\end{align}}
\def\bse{\begin{subequations}}
\def\ese{\end{subequations}}

\newcommand{\nn}{\nonumber}

\def\det{\,{\rm det}\, }

\def\sign{{\rm sgn}}

\def\Im{\,{\rm Im}\,}
\def\Re{\,{\rm Re}\,}

\DeclareMathOperator{\Res}{Res}
\newcommand{\Tr}{\mbox{Tr}}

\def\({\left(}
\def\){\right)}
\def\[{\left[}
\def\]{\right]}
\def\<{\left\langle}
\def\>{\right\rangle}
\def\hf{{1\over 2}}

\newcommand{\eps}{\epsilon}

\renewcommand{\d}{\mathrm{d}}
\newcommand{\de}{\mathrm{d}}

\newcommand{\I}{\mathrm{i}}

\newcommand{\cA}{\mathcal{A}}

\newcommand{\cD}{\mathcal{D}}

\newcommand{\cF}{\mathcal{F}}

\newcommand{\cL}{\mathcal{L}}
\newcommand{\cM}{\mathcal{M}}
\newcommand{\cN}{\mathcal{N}}
\newcommand{\cO}{\mathcal{O}}
\newcommand{\cP}{\mathcal{P}}

\newcommand{\cR}{\mathcal{R}}

\newcommand{\cV}{\mathcal{V}}

\newcommand{\IR}{\mathds{R}}

\newcommand{\IZ}{\mathds{Z}}

\newcommand{\IP}{\mathds{P}}

\newcommand{\ts}{\tilde s}

\newcommand{\tOm}{\tilde\Omega}

\def\ba{\bar a}

\def\hq{\hat q}

\newcommand\PT{\operatorname{PT}}
\newcommand\DT{\operatorname{DT}}
\newcommand\GV{\operatorname{GV}}
\newcommand\GW{\operatorname{GW}}

\def\CYm{\widehat X}

\def\Si#1{S_{#1}}

\newcommand{\q}{\mbox{q}}

\def\gmax{g_{\rm max}}
\def\gtop{g_{\rm top}}
\def\gkink{g_{\rm kink}}
\def\mkink{m_{\rm kink}}
\def\omkink{\omega_{\rm kink}}
\def\PE{{\rm PE }}
\def\y{{\rm y}}

\DeclareMathOperator{\Coh}{Coh}

\title{
Large Order Enumerative Geometry,  \\
Black Holes and Black Rings 
}

\author{Sergei Alexandrov$^1$, Albrecht Klemm$^{2,3}$, 
Boris Pioline$^4$
\\
$^1$ {\it Laboratoire Charles Coulomb (L2C), Universit\'e de Montpellier,
CNRS, \\ F-34095, Montpellier, France}\\

$^2$ {\it Department of Mathematical and Physical Sciences, University of Sheffield\\ S3 7RH Sheffield, United Kingdom}\\

$^3$ {\it Bethe Center for Theoretical Physics, Universit\"at Bonn, D-53115, Germany}\\

$^4$ {\it Laboratoire de Physique Th\'eorique et Hautes
Energies (LPTHE),  CNRS and Sorbonne Universit\'e,
Campus Pierre et Marie Curie, 
4 place Jussieu, F-75005 Paris, France} \\

\vspace*{2mm} {\tt e-mail:
\email{sergey.alexandrov@umontpellier.fr},
\email{a.klemm@sheffield.ac.uk},
\email{pioline@lpthe.jussieu.fr}
}

\vspace*{-3mm}

}

\abstract{Exploiting newly available data on Gopakumar-Vafa invariants at high genus
for one-parameter hypergeometric Calabi-Yau threefolds, we study numerically the growth
of the 5D indices, stable pair (PT) invariants and rank one Donaldson-Thomas (DT) invariants at large charges.
For the 5D index $\Omega_{5D}(d,m)$, below a critical value of the angular momentum $m$,
we find perfect agreement with the Bekenstein-Hawking-Wald entropy of rotating 5D BMPV black holes, 
including the subleading correction from 4-derivative interactions. When $m$ exceeds
the critical value, the 5D index is instead dominated by black rings with the smallest possible dipole charge.
The stable pair invariant $\PT(d,m)$, which is determined by 5D indices, has a similar black ring/hole transition at negative $m$ 
(now interpreted as the D0-brane charge) but surprisingly exhibits two other phase transitions at positive $m$: first, to a plateau and then to a polynomial growth $\sim m^{2d-1}$. 
In each phase, we derive an approximate expression for the invariant. 
Finally, the rank one DT invariant $\DT(d,m)$ is similar to $\PT(d,m)$ at negative $m$, 
and then transitions to a phase dominated by D0-branes, with entropy of order $m^{2/3}$. 
Along the way, we determine the fixed genus, large degree behavior of GV invariants (including the overall $g$-dependent constant), 
extend it to an approximate formula valid also for large $g$, 
point out the unreasonable effectiveness of a simple PT/MSW relation, and 
study the growth of topological free energies at fixed degree, 
confirming a conjecture of Mari\~no. }

\begin{document}

\section{Introduction}

Due to its relevance for understanding instanton effects
and black hole micro-states in string theory, the enumerative geometry of Calabi-Yau (CY)
manifolds has been 
a favorite arena for physicists and mathematicians. 
Historically, the first enumerative 
invariants to appear in the string theory literature were the Gromov-Witten (GW) invariants, counting holomorphic 
maps from the string worldsheet at fixed genus with image in a given curve class (see, e.g., \cite{Pandharipande:2011jz}
for a review of various approaches to curve counting). In particular, GW invariants 
determine worldsheet instanton corrections to the low energy effective action in type II strings compactified 
on a CY 
threefold~\cite{Candelas:1990rm,Bershadsky:1993cx,Antoniadis:1993ze}. 
The Gopakumar-Vafa (GV) invariants 
were introduced in \cite{Gopakumar:1998ii,Gopakumar:1998jq} in order to elucidate the integral structure 
underlying  GW  invariants,  and conjectured to count micro-states of supersymmetric spinning black holes 
in 5 dimensions~\cite{Katz:1999xq}. They are related by the GV/DT/PT correspondence~\cite{gw-dt} 
to Donaldson-Thomas (DT) \cite{thomas1998holomorphic} and Pandharipande-Thomas (PT) invariants 
\cite{Pandharipande:2007kc,Pandharipande:2011jz}, which count micro-states of certain four-dimensional 
black holes with one unit of D6-brane charge. For a more general charge vector, BPS micro-states 
are counted by generalized Donaldson-Thomas invariants~\cite{ks,Joyce:2009xv}, 
which are, at least in principle, computable from GV invariants by wall-crossing \cite{Feyzbakhsh:2020wvm,Feyzbakhsh:2021rcv}. 
The Bekenstein-Hawking-Wald (BHW) entropy of the corresponding black hole solution predicts the 
growth of these enumerative invariants at large charges,
but it is hard to test these predictions in practice, due to the difficulty of computing these invariants in the first place.

Indeed, for CY  manifolds of generic $SU(3)$ holonomy, the only known way of computing
GV invariants (and hence all other invariants) is to obtain them via mirror symmetry at genus zero~\cite{Candelas:1990rm} and genus one~\cite{Bershadsky:1993ta}, and then integrate 
the holomorphic anomaly equations \cite{Bershadsky:1993cx}
to determine them at higher genus -- a  method called \textit{direct integration} \cite{Huang:2006hq,Grimm:2007tm} 
(see also \cite{Yamaguchi:2004bt,Alim:2007qj}). Unfortunately,  the number of integration constants 
(called holomorphic ambiguities) increases linearly with the genus, and the current knowledge of the analytic
structure of the topological string partition function, and of Castelnuovo-type bounds on the maximal genus 
for given degree, only allows us  to fix these ambiguities up to a relatively low genus. 
For example, using this method, the maximal genus  attainable  in principle for the quintic threefold $X_5$ is $53$ 
(slightly
higher than the value 51 quoted in \cite{Huang:2006hq}, but much higher than the genus 
that was reachable in practice with computer resources available at the time).  
Using wall-crossing identities between GV invariants, PT invariants and 
rank 0 DT invariants\footnote{Rank 0 DT invariants 
are also called D4-D2-D0 indices, or MSW indices in reference 
to~\cite{Maldacena:1997de}.}~\cite{gw-dt,Feyzbakhsh:2021rcv,Alexandrov:2023zjb},
and enforcing modularity constraints on the latter~\cite{Gaiotto:2006wm,Alexandrov:2012au,Alexandrov:2016tnf,Alexandrov:2018lgp},
one can further fix a number of holomorphic ambiguities at higher genus and significantly extend the reach of 
the direct integration method, e.g. up to at least genus 69 for $X_5$ \cite{Alexandrov:2023zjb} 
(see \cite{Alexandrov:2023ltz,McGovern:2024kno} for related work, 
and \cite{Alexandrov:2025sig} for a review of these results and of the modular constraints). While  
direct integration is computationally intensive, 
after some algorithmic improvements the maximal genus has now become reachable, offering
a wealth of new data for precision tests of the Bekenstein-Hawking-Wald growth of BPS indices. 

Using  GV invariants available at the time for one-parameter CY threefolds of hypergeometric type 
(which includes the quintic threefold $X_5$ and 12 other models), Ref. \cite{Huang:2007sb}  
analyzed the growth of the index $\Omega_{5D}(d,m)$ counting 5D supersymmetric black holes
of electric charge $d$ and angular momentum $m/2$, as well as the growth of the DT invariants $\DT(d,m)$ 
in the regime relevant for the `entropy enigma' of \cite{Denef:2007vg} (i.e. large $d$ with $m^2/d^3$ fixed). 
Using numerical techniques  to accelerate convergence (namely Richardson transforms), the
authors of \cite{Huang:2007sb} found numerical confirmation of the BHW  prediction 
for static black holes ($m=0$), with around $2\%$ accuracy on the coefficient of 
the leading $\cO(d^{3/2})$ term, and around $10\%$ accuracy on 
the subleading $\cO(d^{1/2})$ term, which originates from
four-derivative corrections of the form $c_{a} A^a \wedge \cR^2$ to the classical supergravity action. 
In the case of spinning black holes, the evidence was weaker and limited to an angular momentum of order $\cO(1)$. 
The study of the asymptotics of rank 1 DT invariants $\DT(d,m)$ gave 
preliminary evidence supporting 
the `miraculous' cancellations suggested in \cite{Denef:2007vg}.

Our goal in this note is to update and extend the analysis of \cite{Huang:2007sb}  
in light of the new available enumerative data.
First, we revisit the asymptotics of GV invariants at fixed genus and large degree.
An asymptotic formula in this regime has been derived for the quintic
in~\cite{Candelas:1990rm} at genus 0, based on the singular behavior of the Yukawa coupling at the conifold point, and extended 
in~\cite{Bershadsky:1993cx,Klemm:1999gm} to arbitrary genus.
While our analysis confirms the general structure of the formula,
the precise value of the overall coefficient  differs from the one quoted 
for genus 0 in~\cite{Candelas:1990rm}, and is confirmed by numerical analysis using higher depth logarithmic Richardson transform.

We then compare the 5D index defined in \cite{Katz:1999xq}
with the Bekenstein-Hawking-Wald prediction for the entropy of 5D black holes. Using the new available data on GV invariants,  we obtain more precise estimates for the leading and subleading terms 
in the BH entropy for static BMPV black holes, providing a very strong confirmation of the BHW prediction.
Unfortunately, our results for the coefficient of the logarithmic correction are inconclusive, and potentially in tension with the prediction of \cite{Sen:2012cj,Sen:2012dw,Anupam:2023yns}. 
For spinning black holes, we find a remarkable agreement with
the entropy including the higher curvature correction that was computed in \cite{Cassani:2024tvk}, correcting earlier predictions \cite{Guica:2005ig,Castro:2007ci,deWit:2009de} in the 
literature.

At large spin, we confirm the existence of a kink in the curve interpolating the values of the index $\Omega_{5D}(d,m)$ at fixed degree, 
which was first observed in~\cite{Halder:2023kza} and 
interpreted as a transition into a phase governed by black rings.\footnote{It is worth noting that even though BMPV black holes carry $j_R=0$ while BPS black rings have $j_R\neq 0$, they both contribute to the index which involves summing over $j_R$ for fixed $m=2j_L^{z}$.} 
We trace back this feature to the existence of a similar kink in the distribution of GV invariants, and determine its position as $d$ becomes large (rectifying the conjecture in~\cite{Halder:2023kza}). We find strong evidence that only black rings with unit dipole charge contribute to the index, although we are not able to justify why.

Furthermore, observing that  the dependence of GV invariants at fixed degree
is well approximated by a Gaussian function, we propose an approximate
formula for these invariants, which is supposed to work at sufficiently large degree,
but does not require a fixed genus. The coefficients of this approximation
are derived by match to the asymptotic formula at fixed genus and 
the entropy of 5D static black holes tested in the previous steps.

We then analyze the growth of DT and PT invariants, counting 
4D BPS states with unit D6 or anti-D6-brane charge,  respectively. 
Their profile at fixed degree  exhibits the same phase transition at negative D0-brane charge
that was observed for the index, but also shows new unexpected features.
First, for D0-brane charge between the Castelnuovo bound and the transition point, PT invariants appear to be accurately captured by a simple linear relation \eqref{PT1} to MSW invariants.
This relation is a very special case of the PT/MSW correspondence established in~\cite{Alexandrov:2023zjb}, and is known  to hold true
only under restrictive conditions which are not satisfied in this extended range.
Second, the profile of  PT invariants turns out to 
exhibit two additional phase transitions at positive D0-brane charge 
giving rise to a `plateau' and a `ramp' (unrelated, as far as we know, to the ramp and plateau in quantum many-body physics). 
In contrast, the profile of DT invariants exhibits only one additional phase dominated by D0-branes.
The macroscopic interpretation of the new phases, however, remains unclear.
For all phases, using the GV/DT/PT correspondence, we derive simple approximate expressions of PT and DT invariants in terms of GV invariants,
which sometimes involve only a small subset of the latter.
An interesting (and mysterious) feature is that all phase transitions for the various enumerative invariants (GV, PT, DT, $\Omega_{5D}$) 
turn out to be correlated with a change in the behavior of their sign.

\medskip

The rest of the paper is organized as follows.
In \S\ref{sec-GV}, after recalling how Gromov-Witten and Gopakumar-Vafa invariants
can be computed using mirror symmetry and holomorphic anomaly equations, 
we derive their asymptotics at large degree and fixed genus, including the genus-dependent prefactor, 
and make phenomenological observations about the behavior of GV invariants at fixed degree.
In \S\ref{sec-GV5D} we study the 5D index and compare its behavior with supergravity predictions for the entropy of
black holes and black rings.
Besides, we propose a formula approximating GV invariants at large $d$ and all genera up to $\gkink$ where
there is a phase transition, and discuss its immediate implications.  
In \S\ref{sec-PT} we turn to PT invariants, observe the existence of several phase transitions in their behavior at fixed degree
and obtain an approximate expression for the PT invariants in each phase.
In \S\ref{sec-DT} we repeat the same for DT invariants, and in \S\ref{sec-disc} we summarize our results and discuss open problems.
In Appendix \ref{sec_Schwinger}, we revisit (mostly for the purpose of properly normalizing the conifold gap condition) the Schwinger one-loop computation of the topological free energy. In Appendix \ref{sec_Richardson}, 
we describe several numerical techniques for accelerating the convergence of series, generalizing the standard Richardson transform. In Appendix \ref{ap-growthFg}, we  
study the growth of topological free energies at fixed degree, confirming
a conjecture of Mari\~no. Finally, in Appendix \ref{ap-McMahon}
we determine the asymptotic behavior of the Fourier coefficients of powers of the MacMahon function, which controls the growth of rank 1 DT invariants at large D0-brane charge.

\section{Phenomenology of GW and GV invariants}
\label{sec-GV}

In this section, we recall how Gromov-Witten and Gopakumar-Vafa invariants
can be computed using mirror symmetry and holomorphic anomaly equations. 
Using the behavior of the topological amplitudes at the conifold point, 
we then determine the growth of GW and GV invariants at fixed genus, large degree. Finally, we make some numerical observations on the profile of GV invariants at fixed degree (and genus less than the Castelnuovo bound).

\subsection{Reminder on Gromov-Witten and Gopakumar-Vafa invariants}
\label{subsec-reminder}

Let $X$ be a smooth CY threefold with $h_{1,1}(X)=1$, such that $H_{\rm even}(X)\simeq \IZ^4$. 
Let $H\in H_4(X,\IZ)$ be the primitive 
effective divisor and $C\in H_2(X,\IZ)$ the primitive effective curve such that $C.H=1$. 
Let  $\kappa:=H^3$ be the cubic self-intersection, $c_2:=c_2(TX).H$ the second Chern class
and $\chi_X:=c_3(TX)$ the Euler number.
The mirror $\CYm$ is a one-parameter family of CY threefolds with 
$\chi_{\CYm}=-\chi_{X}$,  $h_{2,1}(\CYm)=1$ and $H_3(\CYm,\IZ)\simeq \IZ^4$. 
We denote by $\Omega_{3,0}$ the holomorphic 3-form on $\CYm$.
It is defined up to rescaling by $\Omega_{3,0}\mapsto e^{f(z)} \Omega_{3,0}$ where
$f$ is a holomorphic function of the modulus $z$ parametrizing the family $\CYm$, 
hence valued in a line bundle $\cL$ over  $\cM_{cx}(\CYm)$. The K\"ahler potential
for the special K\"ahler metric on $\cM_{cx}(\CYm)$, given by $e^{-K}=\I \int_{\CYm} \Omega_{3,0}\wedge {\overline \Omega _{3,0}}$, transforms as $K\rightarrow K-f-\bar f$ under changes of trivialization of $\cL$.
For simplicity, we restrict to cases, classified in \cite{Doran:2005gu},
 where $\CYm$ is parametrized by a three-punctured
sphere $z\in \cM_{cx}(\CYm)=\IP^1 \backslash \{0,\mu,\infty\}$. It is straightforward,
however, to extend the analysis to arbitrary one-parameter models, such as those
in the AESZ database \cite{cycluster}.

\begin{table}
	$$
	\begin{array}{|l|ccc|ccc|c|cccc|c|c|c|c|}
		\hline
		X & \chi &\ \kappa\ &\ c_2\ &\ g_{\rm avail} & d_C & d_{\rm mod}& \mu^{-1}&a_1&a_2&a_3&a_4 &T_\infty &\frak{w}^+ & \frak{b} & \cV \\
		\hline
		X_{5} & -200 & 5 & 50 & 64 & 22 & 26 &5^5&\frac{1}{5}&\frac{2}{5}&\frac{3}{5}&\frac{4}{5}&o_5&
		320.8713 &265.5937&1.208128\ \\
		X_{6} & -204 & 3 & 42 & 48 & 15 & 17 &2^43^6&\frac{1}{6}&\frac{1}{3}&\frac{2}{3}&\frac{5}{6}&o_6&
		372.2764&261.9897 &1.420957 \\
		X_{8} & -296 & 2 & 44 & 64 & 15 & 17  &2^{16}&\frac{1}{8}&\frac{3}{8}&\frac{5}{8}&\frac{7}{8}&o_8&
		439.9947& 259.5665&1.695113 \\
		X_{10} & -288 & 1 & 34 & 71 & 11 & 13  &2^85^5&\frac{1}{10}&\frac{3}{10}&\frac{7}{10}&\frac{9}{10}&o_{10}&
		538.2249& 256.4336&2.098885  \\
		X_{3,3} & -144 & 9 & 54 & 34 & 20 & 21&3^6&\frac{1}{3}&\frac{1}{3}&\frac{2}{3}&\frac{2}{3}& k &
		264.2581&270.9159& 0.975424 \\
		X_{4,2} & -176 & 8 & 56 & 64 & 28 & 31 &2^{10}&\frac{1}{4}&\frac{1}{2}&\frac{1}{2}&\frac{3}{4}&c&
		277.4779&269.7075 &1.028810 \\
		X_{4,3} & -156 & 6 & 48 & 26 & 14 & 15  &2^63^3&\frac{1}{4}&\frac{1}{3}&\frac{2}{3}&\frac{3}{4}&o_{12}& 
		297.7398&267.1438&1.114530\\
		X_{4,4} & -144 & 4 & 40 & 34 & 14 & 16 &2^{12}&\frac{1}{4}&\frac{1}{4}&\frac{3}{4}&\frac{3}{4}&k&
		331.3076&263.9961&1.254971 \\
		X_{6,2} & -256 & 4 & 52 & 49 & 17 & 20 &2^83^3&\frac{1}{6}&\frac{1}{2}&\frac{1}{2}&\frac{5}{6}&c&
		351.9173&263.8589&1.333732 \\
		X_{6,4} & -156 & 2 & 32 & 17 & 7 & 8  &2^{10}3^3&\frac{1}{6}&\frac{1}{4}&\frac{3}{4}&\frac{5}{6}&o_{12}&
		405.9683&259.6941 &1.563255\\
		X_{6,6} & -120 & 1 & 22 & 26 & 6 & 7 &2^83^6 &\frac{1}{6}&\frac{1}{6}&\frac{5}{6}&\frac{5}{6}&k& 
		480.8078&256.5551&1.874091 \\
		X_{3,2,2} & -144 & 12 & 60 & 14 & 13 & - &2^4 3^3&\frac{1}{3}&\frac{1}{2}&\frac{1}{2}&\frac{2}{3}&c& 
		244.0637&273.9891 &0.890778 \\
		X_{2,2,2,2} & -128 & 16 & 64 & 32 & 25 & - &2^8&\frac{1}{2}&\frac{1}{2}&\frac{1}{2}&\frac{1}{2}&m&
		223.9220& 277.4779&0.806990 \\
		\hline
	\end{array}
	$$
	\caption{Relevant data for the 13 hypergeometric models, including the genus $g_{\rm avail}$ up to 
		which direct integration has been achieved at the time of writing, the maximal degree 
		$d_C$ for which all GV invariants are known, the maximal degree 
		$d_{\rm mod}$ for which all GV invariants can be computed using predictions of modularity. This is followed
		by the data defining the hypergeometric system \eqref{eq:hypergeom}, which have a maximal unipotent monodromy at $z=0$, a 
		conifold singularity at $z=\mu$ and $T_\infty$ indicates the type (orbifold, conifold, k-point or m-point) 
		of the singularity at $z=\infty$. The parameters $\frak{w}^+$ and 
		$\frak{b}$ entering the transition matrix \eqref{Transitionmatrix} from the large volume basis to the 
		conifold basis are as in \cite[Table 2]{Bonisch:2022mgw}, after multiplying $\frak{b}$ by $\I$. 
		The ratio $\cV=\frak{w}^+/\frak{b}$ governs the exponential growth of GV invariants at fixed genus \eqref{asympGVg}. 
		The transition matrix is available to high order on the  website \cite{akstringhome}.} 
	\label{tab_cydata}
\end{table}

For any 3-cycle $\Gamma\in H_3(\CYm,\mathbb{Z})$, the period $\Pi_\Gamma=\int_\Gamma \Omega_{3,0}$   
is a multivalued function of $z\in \IP^1$  annihilated by a  linear differential operator 
$\cD$ of fourth order, called Picard-Fuchs (PF) operator.  
We choose the coordinate $z$ such that $z=0$ corresponds to the point of maximal unipotent monodromy 
(MUM point, or large volume point for $X$) 
with vanishing local exponents, 
while the singularity at $z=\mu$ has local exponents $(0,1,1,2)$, corresponding to a conifold singularity. 
Then the PF operator takes  the standard hypergeometric form
\begin{equation}
{\cal D}=\theta_z^4-\mu^{-1} z \prod_{k=1}^4 (\theta_z+a_k), 
\label{eq:hypergeom}
\end{equation} 
where the rational numbers $a_k$ are the local exponents at $z=\infty$
and satisfy $a_1+a_4=a_2+a_3=1$. 
There are 13 possibilities\footnote{We ignore the 14th case,  
which does not correspond to any smooth CY threefold.} for the numbers
$(a_1,a_2,a_3,a_4)$, tabulated along with other relevant data in Table \ref{tab_cydata}.
The local exponents at all singularities are recorded in the Riemann symbol
\be
{\cal  R}\left\{\begin{array}{ccc}
0& \mu& \infty\\ \hline
0& 0 & a_1\\
0& 1 & a_2\\
0& 1 & a_3\\
0& 2 & a_4 
\end{array}\right\}.
\ee

Near the MUM/LV point $z=0$, the Frobenius method provides a canonical basis of solutions of $\cD \Pi=0$ of the form
\begin{equation} 
\varpi^{\rm LV}_n(z)=\sum_{k=0}^n f_k(z)\,\frac{\log(z)^{n-k}}{(n-k)!}\,, 
\qquad n=0,\ldots, 3\,, 
\label{eq:Frobeniusatmum} 
\end{equation} 
where $f_k(z),\ k=0,\dots,3$, are analytic functions around $z=0$, normalized such that   $f_0(0)=1$ and $f_1(0)=f_2(0)=f_3(0)=0$.  
The $\Gamma$-class conjecture (see e.g. \cite{Pichonpharabod:2025} and references therein) provides a  transition matrix $T^{\rm LV}$ so that the entries in the vector
$\Pi:=T^{\rm LV} {\varpi}^{\rm LV}$ correspond to
integral periods, i.e. to periods 
$(\Pi_{\Gamma_0},\Pi_{\Gamma_1}, \Pi_{\Gamma^0},\Pi_{\Gamma^1})$ on a symplectic integral 
basis $(\Gamma_\Lambda, \Gamma^\Lambda)$ of $H_3(\CYm,\IZ)$~\cite{galkin2016gamma}.\footnote{The integrality 
	of this basis can be checked numerically, by calculating the monodromy matrices around the three singular points  
	and finding nearly integer entries. }
These periods are usually denoted by $(F_0,F_1,X^0,X^1)$ and are given by
\begin{equation}
\Pi= 
\begin{pmatrix*}[r]
F_0 \\ 
F_1 \\ 
X^0 \\ 
X^1 
\end{pmatrix*}
%=
%\begin{pmatrix*}[r]
%\Pi_{\Gamma_0} \\ 
%\Pi_{\Gamma_1} \\  
%\Pi_{\Gamma^0} \\ 
%\Pi_{\Gamma^1}
%\end{pmatrix*}
=(2\pi \I)^3
    \begin{pmatrix}
    -\frac{\zeta(3)\chi(X)}{(2\pi \I)^3} & -\frac{c_2}{24 \cdot 2\pi \I} & 0 & -\frac{\kappa}{(2\pi \I)^3} \\
- \frac{c_2}{24} & -\frac{ \sigma}{2 \pi \I} & \frac{\kappa}{(2\pi \I)^2} & 0 \\
 1 & 0 & 0 & 0 \\
 0 & \frac{1}{2 \pi \I} & 0 & 0
    \end{pmatrix}
\begin{pmatrix*}[r]
        \varpi_0^{\rm LV} \\
        \varpi_1^{\rm LV} \\
        \varpi_2^{\rm LV} \\
        \varpi_3^{\rm LV} 
    \end{pmatrix*} .
    \label{def:integralperiods}
\end{equation}
where $\sigma=0$ if $\kappa$ is even, and $\sigma=\frac12$ if $\kappa$ is odd.  With this choice of integral basis, 
the entries of $\Pi$ correspond to the central charge of the D6, D4, D0 and D2-branes wrapped on $X$, 
divisor $H$, a point and curve $C$, respectively. 

Griffiths' transversality implies that for $k\in\{0,1,2\}$, 
\be 
\int_{\CYm} \Omega_{3,0} \wedge \partial^k_z \Omega_{3,0} =\Pi^T \Sigma \partial_z^k \Pi  = 0\, ,
\qquad 
\Sigma=\begin{pmatrix} 0 & 1_2 \\ -1_2 & 0 \end{pmatrix}.
\label{eq:Griffithtransversality} 
\ee
The topological triple coupling  defined by the third derivative
\be
\kappa_{zzz}=  \int_{\CYm}   \partial^3_z \Omega_{3,0} \wedge\Omega_{3,0} 
\label{kzzz}
\ee
is a rational function
of $z$ with rational coefficients, up to overall normalization. 
For the hypergeometric models, using  \eqref{def:integralperiods}, one can show that
\be
\kappa_{zzz} =  \frac{(2 \pi \I)^3\kappa}{ z^3 (1 -\mu^{-1} z)}\, .
\label{eq:normalizationkappa}
\ee 
The conditions \eqref{eq:Griffithtransversality} imply that the vector $\Pi$ lives in a homogeneous Lagrangian cone
generated by a homogeneous prepotential $F(X^0,X^1)=(X^0)^2 \cF^{(0)}(X^1/X^0)$ such that 
\be
\begin{pmatrix*}[r]
F_0 \\ 
F_1 \\ 
X^0 \\ 
X^1 
\end{pmatrix*}=
\begin{pmatrix*}[c]
\partial_{X^0} F  \\ 
\partial_{X^1} F\\ 
X^0 \\ 
X^1 
\end{pmatrix*}=
X^0
\begin{pmatrix*}[c]
2 {\cal F}^{(0)}-t \partial_t {\cal F}^{(0)}\\ 
\partial_t {\cal F}_0  \\ 
1 \\ 
t 
\end{pmatrix*},
\ee
where $t=X^1(z)/X^0(z)$ is the flat coordinate near the MUM point.
The function $\cF^{(0)}(t)$ is the genus zero topological free energy and given by 
\be
\cF^{(0)}(t) = \frac{\kappa}{6}\, t^3 - \frac{\sigma}{2}\, t^2 - \frac{c_2}{24}\, t - 
\frac{\chi_X \zeta(3)}{2(2\pi\I)^3} + \frac{1}{(2\pi\I)^3} \sum_{d=1}^\infty 
\GW^{(0)}_d q^d, 
\qquad 
q:=e^{2\pi\I t}.
\label{F0}
\ee
The coefficients $\GW^{(0)}_d$ are rational numbers called genus 0, degree $d$ Gromov-Witten 
invariants. They determine the genus 0 Gopakumar-Vafa invariants via
the multi-covering formula
\be
\label{multicover0}
\GV^{(0)}_d := \sum_{k|d} \frac{\mu(k)}{k^3}\,  \GW^{(0)}_{d/k}\, , 
\ee
where  $\mu(k)$ is the M\"obius multiplicative function. The genus 0 GV invariants $\GV^{(0)}_d$ 
are integers, and  coincide with the  Donaldson-Thomas invariants counting stable coherent sheaves 
supported on degree $d$ curves; in particular, the latter are 
independent of the D0-brane charge.\footnote{We collect mathematical references for some of the claims at the end of Section \ref{sec_hae} below.} 

The third derivative of  the prepotential $\cF^{(0)}$ defines the Yukawa coupling
\be
\kappa_{ttt}:= \partial^3_t \cF^{(0)}(t) = \kappa + \sum_{d=1}^\infty \GW^{(0)}_d d^3  q^d 
= \kappa + \sum_{d=1}^\infty \GV^{(0)}_d  \frac{d^3 q^d}{1-q^d}\, .
\label{d3F0}
\ee
It is related to the topological triple coupling \eqref{kzzz} by a rescaling
\be 
\kappa_{ttt}=\left(\frac{1}{X^0}\right)^2\left(\frac{\partial z}{\partial t}\right)^3 \kappa_{zzz}(z(t)),
\label{eq:kttt2}
\ee 
reflecting the fact that $\kappa_{zzz}$ is a section of $\cL^{2} \otimes {\rm Sym}^3(T^*\cM_{cx}(\CYm))$,
where $\cL$ is the line bundle in which $\Omega_{3,0}$ is valued.

More generally, the higher genus topological free energies are defined as\footnote{The normalization factor 
$(2\pi\I)^{3g-3}$ can be justified as follows.
Contributions to  $F^{(g)}(z,\bar z)$ can be calculated from graphs, 
that represent the stable  degenerations of the  genus  $g$ Riemann surfaces $\Sigma_g$ 
 in terms of vertices associated to $\kappa_{zzz}$ and propagators $S^{zz}$ such that $\bar \partial_{\bar z} S^{zz}= C_{\bar z}^{zz}$, where the right-hand side is defined below \eqref{eq:hae}~\cite{Bershadsky:1993cx}. Since all graphs must have the same transcendentality property, 
 we focus on the graph
 
 \begin{tikzpicture}[every node/.style={circle, draw}]
    \node (A) at (0,0){\tiny $\kappa_{zzz}$};
    \node (B) at (2,0){\tiny $\kappa_{zzz}$};
    \node (C) at (4,0){\tiny $\kappa_{zzz}$};
   \node (D) at (8,0){\tiny $\kappa_{zzz}$};
   \node (E) at (10,0){\tiny $\kappa_{zzz}$};
   \node (F) at (12,0){\tiny $\kappa_{zzz}$};
      \draw (A) to[loop, out=140, in=220, looseness=6] (A);
      \draw (A) to[out=0, in=180] (B);
     \draw (B) to[out=30, in=150] (C);
     \draw (B) to[out=-30, in=-150] (C);
     \draw (D) to[out=30, in=150] (E);
     \draw (D) to[out=-30, in=-150] (E);
     \draw (E) to[out=0, in=180] (F);
     \draw (F) to[loop, out=40, in=-40, looseness=6] (F);
     \draw[dotted,line width=2pt] (5,0) -- (7,0);
    \end{tikzpicture}
    
\noindent contributing $(S^{zz})^{3g-3} (\kappa_{zzz})^{2g-2}$ times a rational symmetry factor. In our normalisation 
$\kappa_{zzz}$ is proportional to $(2 \pi \I)^3$ times a rational function and  $S^{zz}\propto (2 \pi \I)^{-3}$. Since $F^{(g)}\in \cL^{2g-2}$
and $X^0\propto (2 \pi \I)^3$, we conclude that $\cF^{(g)}$ is proportional to $(2 \pi \I)^{3g-3}$ times a $q$-series with rational coefficients, identified as GW invariants. 
}
 
\be
\label{defFg}
\cF^{(g)}(t)  = \cF^{(g)}_{\rm pol}(t)  + (2\pi\I)^{3g-3} \sum_{d=1}^\infty \GW^{(g)}_d  q^d ,
     \ee 
where $\GW^{(g)}_d$ are the genus $g$ Gromov-Witten invariants\footnote{ 
For genus $g>1$, GW invariants are  defined as  integrals over the virtual canonical class 
on the moduli stack of bi-holomorphic maps from a genus $g$ curve $\Sigma$ to  
$X$ so that the image is a degree $d$ curve in $X$.  For genus zero, the existence of conformal Killing vectors 
requires fixing three punctures on $\Sigma_{0,3}$, such that  each marked point maps to a divisor $H\subset X$. 
Similarly, for genus 1, the existence of a conformal Killing vector requires
fixing one puncture. The corresponding generating series determine $\partial^3_t \cF^{(0)}$
and $\partial_t \cF^{(1)}$, respectively. The constant for $g>1$ originates from constant map 
contributions.}, 
$\cF^{(g)}_{\rm pol}(t)$ is the cubic polynomial in \eqref{F0} for $g=0$, a linear polynomial
for $g=1$ and a constant for $g>1$:
\be
\label{Fgpol}
\cF^{(1)}_{\rm pol} = \frac{c_2}{24}\, t, 
\qquad 
\cF^{(g>1)}_{\rm pol} = \frac{(-1)^{g+1} (2\pi\I)^{3g-3} B_{2g} B_{2g-2}}{2 g (2g-2) (2g-2)!}\, \frac{\chi_X}{2}\, . 
\ee
The multicover formula~\cite{Gopakumar:1998ii,Gopakumar:1998jq}\footnote{Note that 
the formula \eqref{multicover} with $\GV^{(0)}_0=-\chi_X/2$, interpreted using Zeta function regularization, 
yields the constant term in \eqref{F0} and $\cF^{(g)}_{\rm pol}$ for $g>1$ given in \eqref{Fgpol}. 
For $g=1$, the constant is ambiguous.} 
\be
\label{multicover}
\sum_{g=0}^{\infty} \lambda^{2g-2} \sum_{d=0}^\infty \GW_d^{(g)} q^d 
=\sum\limits_{g=0}^\infty\sum\limits_{k=1}^\infty\sum\limits_{d=0}^\infty 
\frac{\GV^{(g)}_d}{k}\left(2\sin\tfrac{k\lambda}{2}\right)^{2g-2} q^{k d} 
\ee
generalizes \eqref{multicover0} and defines the genus $g$ GV invariants $\GV^{(g)}_d$.
It is motivated by a one-loop Schwinger-type computation of the free energy in a self-dual graviphoton background,
with D2-D0 branes (or equivalently, M2-branes) running in the loop.\footnote{We revisit this computation in Appendix \ref{sec_Schwinger}, in order to fix normalizations.} 
The invariants  $\GV^{(g)}_d$ are integers
counting BPS states with fixed electric charge and angular momentum (see \S \ref{sec-GV5D} 
below for more details). They vanish when the genus $g$
is large enough. 
More precisely, for smooth CY threefolds of Picard rank 1, 
$\GV^{(g)}_d$ vanishes unless the following `Castelnuovo bound' is satisfied
\be
\label{gCast}
g\leq \gmax(d) := \left\lfloor\frac{d^2}{2\kappa} + \frac{d}{2} + 1 \right\rfloor.
\ee
This bound, however,  need  not be sharp. In general, one expects that $\GV_d^{(g+1)}$ vanishes whenever $\GV_d^{(g)}=0$.
 
 \subsection{Holomorphic anomaly and direct integration}
 \label{sec_hae}
 
The genus $g$ topological free energy \eqref{defFg} can in fact be viewed as the \textit{large volume frame holomorphic limit} 
of a non-holomorphic free energy $F^{(g)}(z,\bar z)$, which is a global section of the line bundle $\cL^{2-2g}$ over 
$\cM_{cx}(\CYm)$. The non-holomorphic dependence is determined by  holomorphic anomaly equations of the form 
\be
\label{eq:hae}
\partial_{\bar z} F^{(g)} = \frac12\,C_{\bar z}^{z z}
\left[  D^2_z F^{(g-1)} + \sum_{h=1}^{g-1} D_z F^{(h)} D_z F^{(g-h)} \right] 
\ee
for genus $g\geq 2$, where $C_{\bar z}^{zz}:=e^{2K} G^{z\bar z} G^{z \bar z} \overline{\kappa_{zzz}}$,
$G_{z\bar z}=\partial_z \partial_{\bar z}K$ is the special K\"ahler metric 
induced by the prepotential $\cF^{(0)}$
and $D_z F^{(h)}=(\partial_z + (2-2h) \partial_z K)F^{(h)}$ is the K\"ahler covariant derivative. 
For $g=1$, the corresponding holomorphic anomaly equation can be integrated to get
\be
\label{eqF1}
F^{(1)}(z,\bar z)=-\frac12 \(4-\frac{\chi_X}{12} \) K -\frac12\, \log \det G_{z\bar z} + \log | f_1(z)|^2,
\ee
where $f_1(z)$ is a holomorphic ambiguity, determined by the discriminant (see \eqref{eqf1} below). 
The original genus $g$ topological free energy \eqref{defFg} arises by expressing $F^{(g)}(z,\bar z)$ in 
terms of the large volume flat coordinate $t,\bar t$ and taking the limit $\bar t\to -\I\infty$. More generally,
near any singular point $z_*$ on the boundary of  $\cM_{cx}(\CYm)$, one may introduce  
a preferred symplectic frame $(F_0^*,F_1^*,X^0_*,X^1_*)$ and a local flat coordinate $t_*=X^1_*/X^0_*$,
and define the holomorphic free energy in the corresponding frame via
\be 
 \cF_*^{(g)}(t_*) = \lim_{{\bar t}_* \rightarrow -\I \infty} F^{(g)}(z(t_*),\bar z({\bar t}_*)). 
 \ee
 For $z_*=0$, the MUM point in our conventions, we omit the subscript and set $t_*=t$.
 Given the set of holomorphic free energies $\cF_*^{(g)}$ in one frame, one can construct the 
 unique solution of \eqref{eq:hae} by summing over graphs corresponding to stable degenerations of genus $g$ curves,
 and obtain the holomorphic free energies at any other frame by taking the corresponding limit ${\bar t}'_* \rightarrow -\I \infty$. Equivalently, one can use the fact that the topological string partition function
\be 
Z(t,\lambda)=\lambda^{\frac{\chi_X}{24}-1} \exp\left(\sum_{g=0}^\infty \cF^{(g)}(t) \lambda^{2g-2}\right),
\ee
transforms as a wave function under change of 
polarization~\cite{Witten:1993ed,Aganagic:2006wq,Schwarz:2006br}. The same property also determines
the behavior of the $\cF^{(g)}$'s under monodromies in $\cM_{cx}(\CYm)$~\cite{Pioline:2025uov}. 

Unfortunately, for compact CY threefolds the full set of holomorphic free energies $\cF_*^{(g)}$ is not known in any frame.
Nonetheless, by providing a sufficient number of boundary conditions in various frames, one can determine
the full topological free energies $F^{(g)}(z,\bar z)$ to a fairly high genus, by integrating the equations \eqref{eq:hae}
and  fixing the holomorphic ambiguity arising at every step~\cite{Huang:2006hq}. 
These boundary conditions include the constant map contributions and Castelnuovo vanishing conditions mentioned above, 
as well as the so-called gap behavior near conifold loci~\cite{Huang:2006si} 
\be
\label{Fgap}
\cF_c^{(g)}(t_c ) \stackrel{t_c\to 0} \sim \frac{(-1)^{g-1} (2\pi\I)^{g-1}  B_{2g}}{2g(2g-2)
}\, \frac{m_c}{t_c^{2g-2}} + \cO(t_c^0), 
\qquad 
\forall g\geq 2,
\ee
where $t_c$ is a flat coordinate which vanishes at the conifold locus  $z=\mu$,
and $m_c$ is an integer, which is equal to one for the hypergeometric models of interest in this work.\footnote{In general, 
	the integer $m_c$, called  \textit{conifold multiplicity}, is fixed by the monodromy 
	$M_c={\rm Id}_4- m_c (\Sigma v) v^T$, where $v$ is the integral
charge vector of the object becoming massless, and $\Sigma$ is the symplectic form.
It is equal to 1 whenever the state which becomes massless is a spherical object (typically
the structure sheaf on $X$ tensored by some line bundle, corresponding to a
fluxed D6-brane, or a Lagrangian sphere $S^3\subset \CYm$ on the mirror, corresponding to a D3-brane. 
More generally, the singularity could originate from a D3-brane wrapped on a lens space $S^3/\IZ_m \subset \CYm$, 
leading to a multiplicity $m>1$~\cite{Gopakumar:1997dv}. The corresponding massless object in 
$D_b\Coh(X)$ for such `hyperconifold' singularities is not currently understood.
\label{foot-mult}}
The leading term is obtained from a Schwinger one-loop computation keeping only the BPS state becoming massless at the conifold point running in the loop. In Appendix \ref{sec_Schwinger}, we recall this computation in order to fix the precise normalization. We note that the topological free energy near the conifold point also coincides, up to the overall factor $m_c$,  with the free energy of the $c=1$ 
string, as observed in \cite{Ghoshal:1995wm}. 
The gap condition is the statement that the coefficients of potential sub-leading terms $1/t_c^{2g-1},\dots, 1/t_c$ 
actually vanish. It is crucial that the l.h.s. of \eqref{Fgap} is the topological free energy in the conifold frame.
Fortunately, the change of polarization from the conifold frame to the large volume frame does not affect the leading
singular behavior, so the singular behavior of the large volume frame free energy $\cF^{(g)}(t)$ near the conifold point 
$t(\mu)$ is still given by \eqref{Fgap} with $t_c\propto t-t(\mu)$, up to subleading logarithmic corrections. This will be key for determining
the growth of GW invariants at large degree in the next subsection. 

\medskip
Before moving on, for the benefit of a mathematically minded reader, we comment on the mathematical status of the various claims made up to here. Mirror symmetry at genus 0 was proven in~\cite{MR1408320,MR1621573} and at genus 1 in~\cite{MR2505298}. The direct integration method using modular 
generators has been proven for hypersurfaces in weighted projective space, see a summary in \cite{Leithesis}. The integrality of the GV invariants defined by \eqref{multicover} was proven in  \cite{IonelParker:2013}, and their vanishing at large genus in \cite{Doan:2021}. The coincidence between genus zero Gopakumar-Vafa invariants and DT invariants supported on curves was conjectured 
in~\cite{Katz:2006gn} and proven in~\cite{maulik2018gopakumar}.
The Castelnuovo bound \eqref{gCast} was proven in  \cite{Alexandrov:2023zjb}
subject to the validity of the Bayer-Macr\`i-Toda inequality 
(which is known to hold for the quintic threefold $X_5$ \cite{li2019stability} 
and a couple of other CY threefolds \cite{koseki2020stability,liu2021stability}, 
but widely expected to hold in general). 
The same bound \eqref{gCast}
was proven for $X_5$ in \cite{Liu:2022agh} using similar methods.
A weaker bound, unconditional on the BMT inequality,
has been established more recently in \cite{Liu:2024dol}. We are not aware of a mathematically rigorous treatment of the wave function property of $Z(t,\lambda)$.

\subsection{GW invariants at large degree, fixed genus}
\label{subsec-highd}

Based on the singular behavior \eqref{Fgap} near the conifold point, it was argued 
in \cite[(7.8)]{Bershadsky:1993cx} that for fixed genus $g$, GW
or, equivalently\footnote{Since the multicover formula \eqref{multicover} relates $\GV_d^{(g)}$ to $\GW_d^{(g')}$ with $g'\leq g$, 
	and since  $\GW_d^{(g')}$ grows slower than $\GW_d^{(g)}$ when $g'<g$, for fixed genus the ratio 
$\GW_d^{(g)} /\GV_d^{(g)}$ must go to 1.},
GV invariants  should grow as\be
\GW_d^{(g)} \sim \GV_d^{(g)} \sim a_g \, d^{2g-3} (\log d)^{2g-2} e^{2\pi \cV  d},
\label{asympGVg}
\ee
where the positive real constant $\cV$ is equal to the imaginary part of the flat coordinate
$t(z)$ at the conifold point $z=\mu$, and $a_g$ is an unspecified constant. For $g=0$,
the constant $a_0$ was computed for the quintic threefold $X_5$ in  \cite[(5.17)]{Candelas:1990rm}, however
the result turns out to be erroneous. In this section, we revisit this computation and determine
$a_g$ for any genus and any hypergeometric model.  

Recall that for ordinary differential equations with regular singularities such as \eqref{eq:hypergeom}, 
the growth of the coefficients in the power series $f_i(z)$ of solutions around one singular 
point is determined by the  behavior of these solutions at the nearest 
singular point~\cite{darboux1878}. For the models of Table \ref{tab_cydata}, the nearest singularity 
to the maximal unipotent monodromy point at $z=0$ is 
the conifold point at $z=\mu$. The fact that the nearest point to a given MUM point is a  conifold point
appears to be a general feature of the one-parameter CY operators \cite{cycluster}.  

At genus 0, the behavior of the period vector $\Pi$ in the large volume basis at the conifold point 
is encoded in the transition matrix $T^{(c)}$ with  $\Pi=T^{(c)}  \varpi^{(c)}$, which relates 
$\Pi$ to the canonical basis of periods at the conifold point, satisfying 
\be
\begin{split}
\varpi_0^{(c)}  =&\, 1+\cO(\delta^3), 
\qquad\;
\varpi_1^{(c)}=\delta+\alpha\delta^2+\cO(\delta^3), 
\\
\varpi_2^{(c)}= &\,\delta^2+\cO(\delta^3), 
\qquad 
\varpi_3^{(c)}=\varpi_1^{(c)} \log\delta+\cO(\delta^3),
\end{split}
\ee
where $\delta:=1-z/\mu$. The period $\varpi_1^{(c)}$ vanishes at the conifold point,
and is chosen to be proportional to the central charge $\nu(\delta)=\sqrt{\kappa} \varpi_1^{(c)}$
of the massless object at $z=\mu$; 
this fixes the coefficient $\alpha$ in $\varpi_1^{(c)}$. As shown in \cite[(3.6)]{Bonisch:2022mgw}\footnote{Compared to this reference, 
	we have redefined $\frak{a}^\pm,\frak{b},\frak{c},\frak{d},\frak{e}^\pm,\frak{w}^\pm$ such that they are now real positive numbers.
The numerical values of $\frak{w}^+$ and $\frak{b}$, relevant for the growth of $\GV_d^{(g)}$, are recorded in Table \ref{tab_cydata} above 
and the values of the transition matrices for these models with 500 significant digits can be obtained in \cite{akstringhome}.}, 
assuming that the massless object is the structure sheaf $\cO_X$ (i.e. a
pure D6-brane wrapped on $X$), the transition matrix has the following structure:
\be
\label{Transitionmatrix} 
\begin{pmatrix} F_0 \\ F_1 \\ X^0 \\ X^1 \end{pmatrix}
= T^{(c)} \begin{pmatrix} \varpi_0^{(c)}  \\  \varpi_1^{(c)}  \\  \varpi_2^{(c)}  \\  \varpi_3^{(c)}  \end{pmatrix} =
\begin{pmatrix} 0 & (2\pi\I)^2 \sqrt\kappa & 0 & 0  \\
\sigma \frak{w}^+ -\I  \frak{w}^- & \sigma \frak{a}^+  -\I \frak{a}^- & - \sigma \frak{e}^+  +\I  \frak{e}^- & 0 \\
-\I \frak{b} & \I \frak{c} & -\I \frak{d} & -2\pi\I \sqrt{\kappa}\\
\frak{w}^+ & \frak{a}^+ & -\frak{e}^+ & 0  \end{pmatrix}
\begin{pmatrix} \varpi_0^{(c)}  \\  \varpi_1^{(c)}  \\  \varpi_2^{(c)}  \\  \varpi_3^{(c)}  \end{pmatrix}.
\ee
Here the vanishing of the (1,1) entry is necessary for the vanishing of $F_0$ at $z=\mu$,
while the other zeros in the third and fourth column 
are a consequence of Griffiths' transcendentality. From \eqref{Transitionmatrix}, we 
can read off immediately the value of the large volume flat coordinate at the conifold point, or `quantum volume', 
\be
\label{tmu}
t(\mu) = \lim_{\delta\to 0} \frac{X^1}{X^0} = \I \frac{\frak{w}^+}{\frak{b}} := \I \cV. 
\ee

One of the key results in~\cite{Bonisch:2022mgw}, further amplified in \cite{Bonisch:2025cax}, is that the
entries in $T^{(c)}$ have deep arithmetic properties. In particular,  
the constants $\frak{w}^\pm$ and $\frak{e}^\pm$ are related to the 
Eichler-Manin-Zagier period polynomials
% for closed cycles in the fundamental  domain  ${\cal F}$
of a holomorphic weight four Hecke eigenform $f\in S_4(\Gamma_0(N))$ and its dual meromorphic Hecke eigenform 
with vanishing residues $F\in \mathbb{S}_4(\Gamma_0(N))$, respectively. 
Here $N$ is the conductor, which together with $f$ is determined by the Hasse-Weil $\zeta$-function 
of the conifold fiber $\CYm_{z=\mu}$. In particular, the constant $\frak{w}^+$ 
turns out to be proportional to the value of the Hecke L-function of $f$ at $s=1$~\cite{Bonisch:2022mgw}
\be 
\label{wplus}
\frak{w}^+ =- \kappa (2 \pi \I)^2 L(f,1)\, .
\ee
Taking $X_5$ as  an illustrative example, 
the cusp form $f\in S_4(\Gamma_0(25))$,  found earlier in~\cite{MR817640},
can be expressed in terms of the Dedekind $\eta$-function as\footnote{See  
	\href{https://www.lmfdb.org/ModularForm/GL2/Q/holomorphic/25/4/a/b/}{{\rm LMFDB  25.4.a.b}} for some more properties of $f$.}  
\be
f=\frac{\eta(5 \tau)^{10}}{\eta(\tau) \eta( 25 \tau)}+5 \eta(\tau)^2 \eta(5 \tau)^4 \eta(25 \tau)^2= q+q^2 +7q^3-7q^4+7q^6  +\cdots ,
\ee
More recently, the relation of $\frak{w}^\pm$ and $\frak{e}^\pm$ to the period polynomials 
was proven for eight of the hypergeometric cases listed above
\cite{Bonisch:2025cax}. 
Moreover, for $X_5$, a meromorphic modular form $G$ 
under $\Gamma_0(50)$ with non-vanishing 
residue was defined so that the relations of $\frak{a}^\pm$ 
to its period polynomial over closed cycles could be established, 
and the coefficients $\frak{b},\frak{c}$ and $\frak{d}$ were related to integrals of $f,F,G$ 
between points of complex multiplication 
$\tau^\pm$ in ${\cal F}$, fixing now all constants in \eqref{Transitionmatrix}. In particular, one has
\be 
\label{frakbX5}
\frak{b} =\I (2\pi\I)^3 \int_{\tau^-}^{\tau^+} \d \tau\, (5 f(\tau)-20 f(2 \tau)).
\ee   
Thus, the quantum volume \eqref{tmu} has a clear arithmetic meaning.

\subsubsection{Growth of genus 0 GW invariants}

In order to extract the growth of genus 0 GW invariants, we need to determine the asymptotic behavior 
of the Yukawa coupling \eqref{d3F0} at the conifold point. For this, we use the relation \eqref{eq:kttt2} 
to the triple topological coupling, which  for hypergeometric models is given by the simple rational
function \eqref{eq:normalizationkappa}. Expanding in $\delta=1-z/\mu$, we immediately get 
\be
\kappa_{zzz} \approx \frac{(2\pi\I)^3 \kappa}{\mu^3 \delta}\, .
\ee
Now, expanding the large volume flat coordinate $t(z)=X^1/X^0$ up to order $\delta$, we get 
\be
t(z)\approx\I \cV-\frac{2 \pi \I}{\frak{b}}\, \cV \sqrt{\kappa} \delta \log\delta\, ,
\qquad
\frac{\partial t}{\partial z}\approx  \frac{2 \pi \I}{\frak{b} \mu}\, \cV \sqrt{\kappa} \log\delta \, .
\ee
It is convenient to introduce
\be 
x=2 \pi \Im(t(z)-t(\mu))
\approx -\frac{(2\pi)^2}{\frak{b}}\, \cV \sqrt{\kappa}\, \delta \log \delta \, .
\label{eq:leadingconifold}
\ee
This relation has the form $\delta \log(\delta)=-A x$ and therefore yields, 
in the relevant branch for the inversion, 
\be 
\delta\approx -\frac{Ax}{\log(Ax)} 
\qquad {\rm with }\qquad 
A= \frac{\frak{b}}{ (2\pi)^2 \cV \sqrt {\kappa}}\, ,
\label{eq:deltax}
\ee 
which also implies that in the leading order $\log(\delta)\approx\log(x)$. 
Inserting these results into \eqref{eq:kttt2}, we get 
\be
\kappa_{ttt}\approx -\frac{\frak{b}}{\cV^3 \sqrt{\kappa}\,  \delta (\log \delta)^3}
\approx \left(\frac{2 \pi}{\cV}\right)^2\frac{1}{x (\log x)^2}\, .
\label{eq:A0t}
\ee
In fact, defining the flat coordinate in the conifold frame 
\be
\label{eq:tflatcon}
t_c=\frac{F_0}{\varpi_0^{(c)}} \approx (2\pi\I)^2 \sqrt{\kappa}\delta\, ,
\ee
we see that the Yukawa coupling  in the conifold frame has a simple pole, as predicted by 
the Schwinger computation \eqref{Fgap2},  
\be
\kappa_{t_c t_c t_c } = \left( \frac{X^0}{\varpi_0^{(c)}} \right)^2 \left(
\frac{\partial t}{\partial t_c} \right)^3 \kappa_{ttt} \approx -\frac{1}{2\pi\I t_c}\, .
\ee

Now, assuming  as in \cite{Candelas:1990rm} the ansatz
\be
\GW_d^{(0)} \approx  \GV_d^{(0)} \approx a_0 d^{\rho-3} (\log d)^\sigma e^{ 2\pi \cV d }, 
\ee
the sum over $d$ in \eqref{d3F0} has a finite radius of convergence,
and diverges on the boundary, given by the conifold point, as~\cite{MR538168} 
\begin{equation} 
\begin{array}{rl} 
\kappa_{ttt} & \approx \displaystyle{a_0 \sum_{d=1}^\infty d^\rho (\log d)^\sigma e^{- d x }}
\approx \displaystyle{ a_0 \int_0^\infty \de y\, y^\rho (\log y)^\sigma e^{- y x }}
\\[4 mm] 
&\approx \displaystyle{\frac{a_0 (-\log x)^\sigma}{x^{\rho+1}}  \int_0^\infty \de {\tilde y}\, {\tilde y}^\rho e^{-\tilde y}
= \frac{a_0 (-\log x)^\sigma}{x^{\rho+1}}\,\Gamma(\rho +1). }
\label{eq:resum}
\end{array}
\end{equation}
Comparing \eqref{eq:resum} with \eqref{eq:A0t}, 
we conclude that $\rho=0$, $\sigma=-2$ and $a_0=\left(\frac{2 \pi}{\cV}\right)^2$, i.e. 
\be
\GW^{(0)}_d \mathop{\approx}\limits_{d\gg 1} 
\nu_d^{(0)}:=\left(\frac{2 \pi}{\cV}\right)^2 \frac{e^{2 \pi \cV d}}{d^3 (\log d)^2}\, .
\label{eq:asymg0}
\ee
This agrees with \eqref{asympGVg}, but the value of the coefficients $a_0$ differs from the one 
found in \cite{Candelas:1990rm}, which depends on the additional parameter $\frak{b}$ and reads $\(\frac{\frak{b}}{4\pi^2 \cV}\)^2$. In the next subsection, we present numerical evidence for the correctness of our result.

\subsubsection{Genus one}

At genus one we apply a similar reasoning to the 
first derivative $\partial_t \cF^{(1)}(t)$. For this, we must specify the 
holomorphic ambiguity $f_1(z)$ appearing in the non-holomorphic genus one free energy \eqref{eqF1}, 
\be
f_1(z)= z^{-\frac{c_2+12}{24}}\prod_{i} (1-z/\mu_i)^{-m_i/12},
\label{eqf1}
\ee
where $\mu_i$ are the distinct conifold points and 
and $m_i$ their respective multiplicities (see footnote \ref{foot-mult}). 
For hypergeometric models, $\Delta_i=1-z/\mu$ and $m_i=1$. 
By explicit computation, one may check that the 
non-holomorphic Weil-Petersson metric  $G_{z\bar z}$ as well as the K\"ahler potential $K$ 
are regular at the conifold point, such that $\cF_1$ is dominated near $z=\mu$ by
the holomorphic ambiguity $f_1(z)$.
Using \eqref{eq:leadingconifold}, we get 
\be 
\frac{1}{2 \pi \I}\, \partial_t {\cal F}_1=\frac{1}{2 \pi \I} \,\partial_t\log\(\delta^{-\frac{1}{12}}\) \approx \frac{1}{12 x} \, .
\ee
Assuming the ansatz $\GW^{(1)}_d \approx a_1 d^{\rho-1} (\log d)^\sigma e^{ 2\pi \cV d}$ in 
\eqref{defFg} and using the same manipulations as in \eqref{eq:resum} yields $\rho=\sigma=0$ and $a_1=\frac{1}{12}$,
which implies the result in agreement with \eqref{asympGVg}
\be
\GW^{(1)}_d \mathop{\approx}\limits_{d\gg 1} 
\nu^{(1)}_d: =\frac{1}{12} \frac{e^{2 \pi \cV d}}{d}\, .
\label{eq:asymg1}
\ee

\subsubsection{Higher genus}

For general genus $g>1$, following the discussion below \eqref{Fgap}, we have that
the 
holomorphic free energy $\cF^{(g)}(t)$ in the large volume frame is related to 
the holomorphic free energy $\cF_c^{(g)}(t_c)$ in the conifold frame 
via 
\be
\label{FLVfromFc}
 \cF^{(g)}(t) \approx \left(\frac{X^0}{ \varpi_0^{(c)}}\right)^{2g-2}
\cF_c^{(g)}(t_c),
\ee
up to terms  involving $\cF_c^{(g')}(t_c)$ with $g'<g$ on the r.h.s., which originate from the change of polarization 
and are subleading near the conifold point. 
Combining \eqref{FLVfromFc} with the
conifold gap condition \eqref{Fgap}, we get 
\be 
\cF^{(g)}(t) 
\approx \left(\frac{X^0}{\varpi_0^{(c)}}\right)^{2g-2} \frac{(-1)^{g-1} (2 \pi \I)^{g-1}  B_{2 g}}{2 g (2g-2) t_c^{2g-2}} \, . 
\ee
Using the relation \eqref{eq:tflatcon} between the conifold flat coordinate $t_c$ and $\delta$, the relations  \eqref{eq:leadingconifold}, \eqref{eq:deltax}
between the large volume flat coordinate $t$ and $\delta$,  
as well as the behavior of $X^0$ and $\omega_0^{(c)}$ in  \eqref{Transitionmatrix}, 
we see that the leading behavior of the large volume frame free energy 
${\cal F}^{(g)}(t)$ near $t=t(\mu)$ is given by
\be 
{\cal F}^{(g)}(t)\approx \frac{(2 \pi \I)^{g-1}  B_{2 g}}{2 g (2g-2)}
\left(\frac{\cV\log(x)}{x}\right)^{2g-2}.   
\ee
Then substituting the ansatz $\GW_d^{(g)}\approx a_g d^\rho(\log d)^\sigma e^{ 2\pi \cV d}$ for $g>1$ into \eqref{defFg}
and repeating the calculations in \eqref{eq:resum} fixes $\rho=2g -3$, $\sigma=2g -2 $ and 
\be
a_g= \frac{(2 g - 1) |B_{2 g}|}{(2 g)!} \left(\frac{\cV}{2 \pi}\right)^{2g-2}.
\label{aggrowth}
\ee
For $g=0$ and $g=1$, this result agrees with the coefficients found in 
\eqref{eq:asymg0} and \eqref{eq:asymg1}. Thus, we conclude that 
for any fixed genus $g$, in agreement with \eqref{asympGVg}, 
the GV and GW invariants have the following
universal leading asymptotic growth at high degree 
\begin{equation} 
\GW_d^{(g)} \approx \GV_d^{(g)} \mathop{\approx}\limits_{d\gg 1} 
\nu_d^{(g)} :=  \frac{(2 g - 1) |B_{2 g}|}{(2 g)!} \left(\frac{\cV}{2 \pi}\right)^{2g-2}
d^{2 g - 3} (\log d)^{2 g - 2} e^{2 \pi \cV d} .
\label{allgenusleading}
\end{equation}
For later reference, we note that for large genus, the coefficient $a_g$ 
grows as 
\be 
a_g\approx \frac{g}{\pi^2}  \( \frac{\cV}{4\pi^2} \)^{2g-2}  ,
\label{ag-largeg}
\ee 
where we used 
\be  
|B_{2g}| \approx 4\sqrt{\pi g}\(\frac{g}{\pi e}\)^{2g},
\qquad 
(2g)!\approx 2\sqrt{\pi g}\(\frac{2g}{e}\)^{2g}.
\ee 

\subsubsection{Numerical evidence} 

\begin{table}
$$
\begin{array}{|l|c|c|c|c|c|c|}
\hline
N  & g=0 & g=1& g=2& g=3& g=6 & g=10\\
\hline
0 &  .4090&.4661973&.52972& .58789&.63434 & .39094\\
2& .8134&.9685760&1.1743&1.4531&2.9149 & 5.2448\\  
4& .8983&.9887690&1.0740& 1.1396&1.1079 &3.0926\\ 
8& .9519& .9954161&1.0329&1.0618& 1.1028&2.3845\\ 
16&.9778&.9983877&1.0164& 1.0316&1.0475 &.82138\\
32&.9893&1.000129&1.0101&1.0194&1.0455 &1.1024\\ 
64&.9948&1.000003&1.0048&1.0093& 1.0200&1.0301\\
128&.9975&.9998904&1.0021&1.0043&1.0088 & 1.0049\\
256&.9988& .9999583&1.0011&1.0022&1.0053 &1.0069 \\
 \hline
\end{array}
$$ 
\caption{Values of the depth $N$ logarithmic Richardson transform  $L_N [r^{(g)}_d]$
of the ratio $r^{(g)}_d=\GW_d^{(g)}/\nu_d^{(g)}$, 
for $X_5$, with $g\leq 10$ and $d=d_{\rm max}=650-N$. 
For fixed $g$, the values tend to 1 as $N$ increases,
as long as the ratio $N/d_{\rm max}$ stays small enough.  
 \label{Richardson}}
\end{table}

\begin{figure}[h]
	\begin{center}
		\includegraphics[height=5cm]{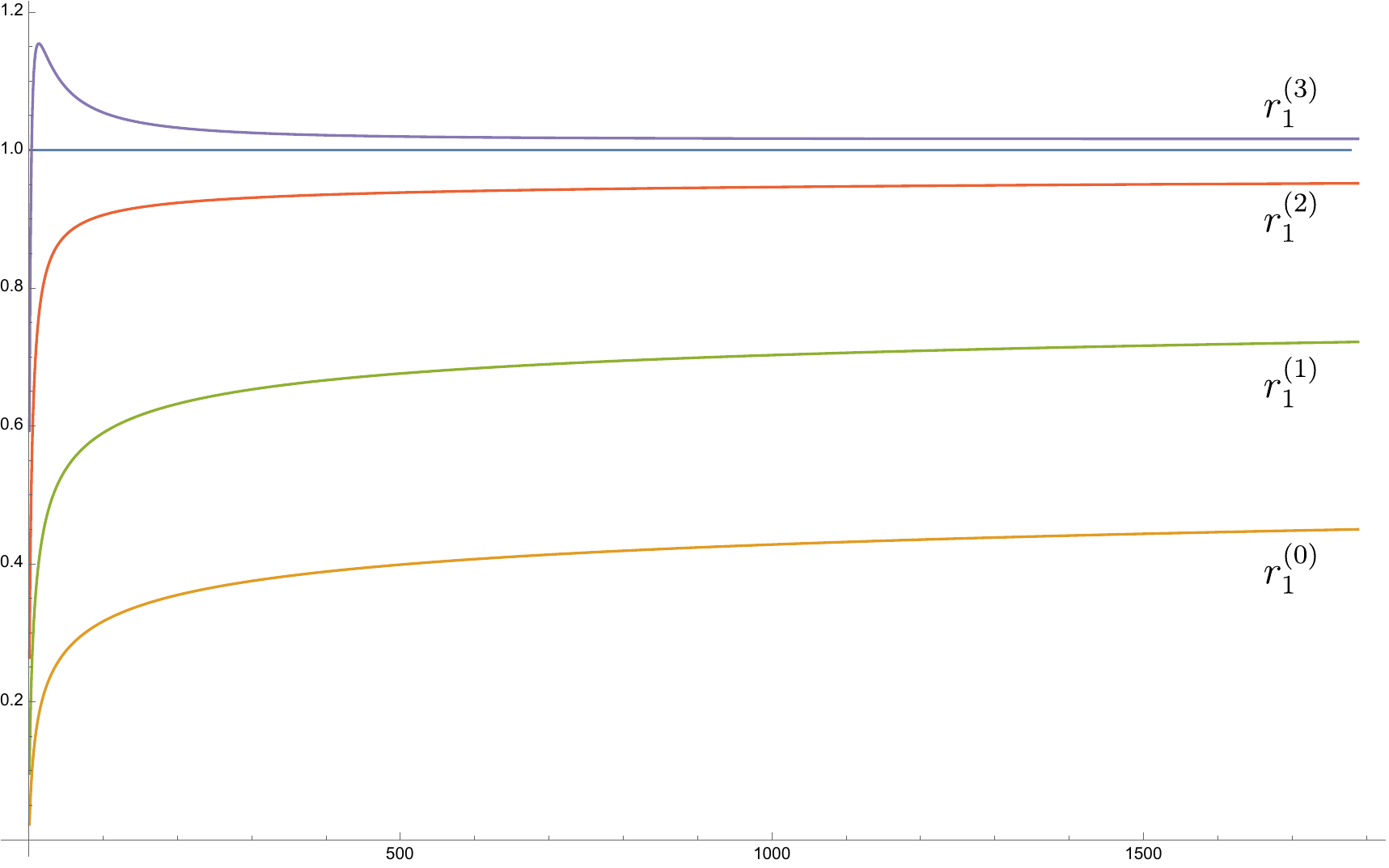} \includegraphics[height=5cm]{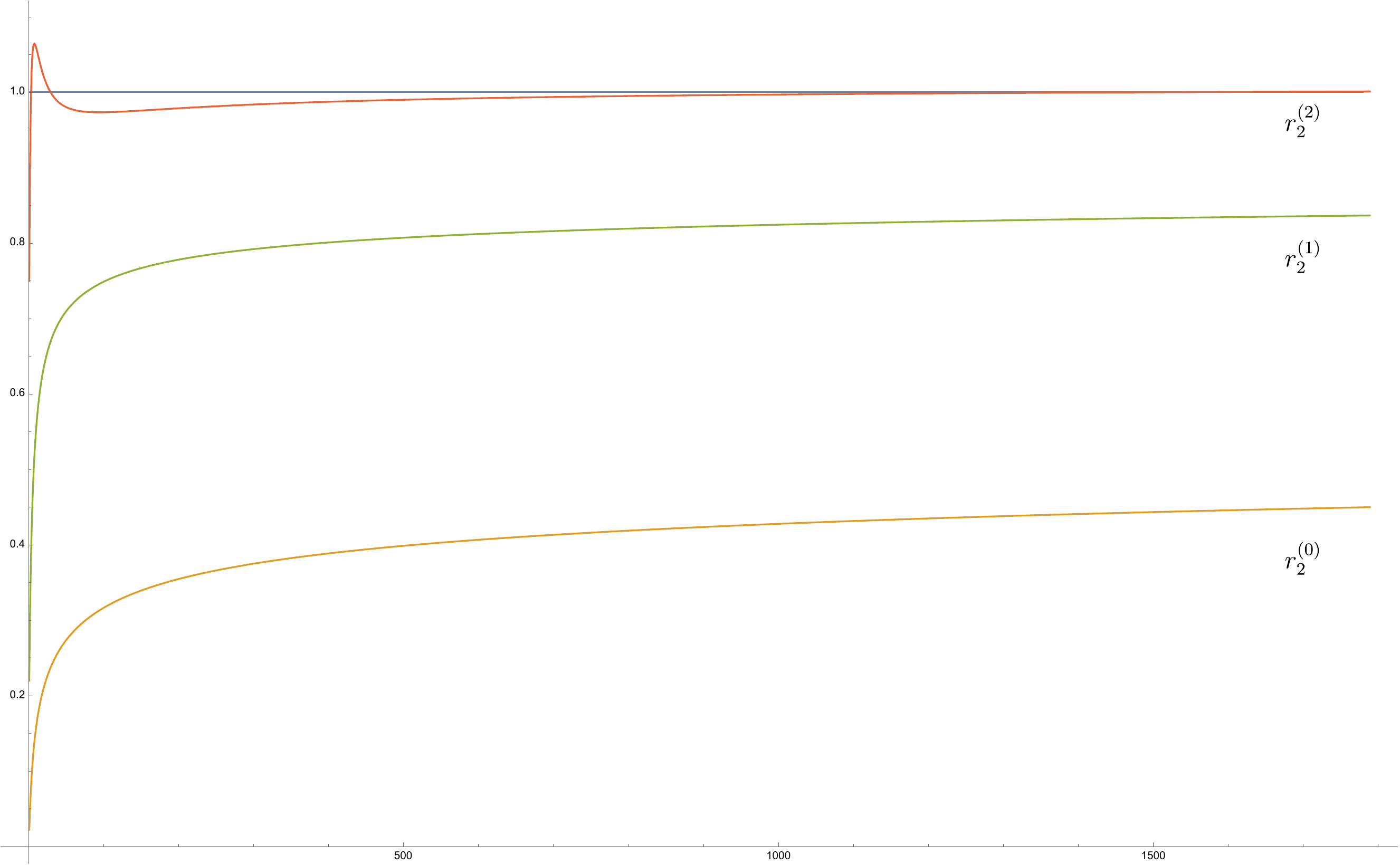} 
	\end{center}
	\vspace{-0.7cm}
	\caption{Asymptotics of genus zero GV invariants for the quintic $X_5$. On the left, we plot the 
	$N$-th iteration of the depth 1 logarithmic Richardson transform $r_1^{(N)}:= (L_1)^N [r_{d}^{(0)}]$, 
	for $N=0,1,2,3$. This is  identical to the plot in \cite{Candelas:1990rm}, but extended up to degree $d=1780$. 
	On the right, we plot the $N$-th iteration of the depth $2$ logarithmic Richardson transform 
	$r_2^{(N)}:=(L_2)^N [r_{d}^{(0)}]$,  for $N=0,1,2$. This leads to much  
	better convergence as the ratio approaches $1$ with an error of $0.002$, 
	confirming \eqref{eq:asymg0} to high accuracy.    
 \label{Quinticgenuszero}}
\end{figure}

We have checked the asymptotic formula \eqref{allgenusleading} for the quintic $X_5$  
up to genus $g=10$, by computing the ratio $r^{(g)}_d = \GW_d^{(g)}/{\nu_d^{(g)}}$ 
up to degree $d=650$.
After applying a logarithmic Richardson transform\footnote{See Appendix \ref{sec_Richardson} 
	for a brief introduction to Richardson transforms and more general 
acceleration algorithms.} 
$L_N$  of sufficiently high depth $N$, we find that $L_N [r^{(g)}_d] $ rapidly converges
to 1 (see Table \ref{Richardson}). For genus 0, 
a similar check was made in \cite[Fig. 5.1]{Candelas:1990rm} up to much lower degree $d\leq 25$
and using instead \text{iterated} logarithmic Richardson transforms of depth 1, which appear 
to be much less efficient than a single higher depth Richardson transform. 
This check, however, was not sufficient to probe the overall normalization, 
and to detect the error in $a_0$, which becomes manifest at higher degree 
(see  Fig. \ref{Quinticgenuszero}). 

\begin{figure}[h]
	\begin{center}
		\includegraphics[height=4.2cm]{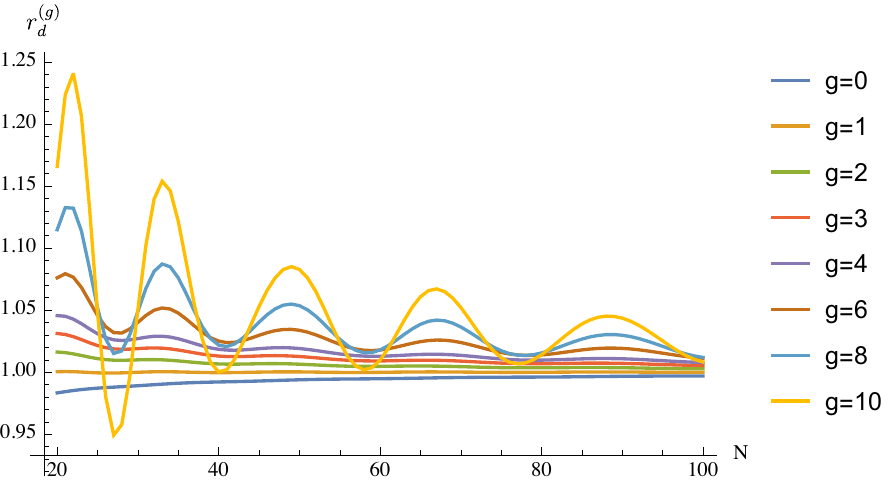} 
	%	\hfill
		\includegraphics[height=4.2cm]{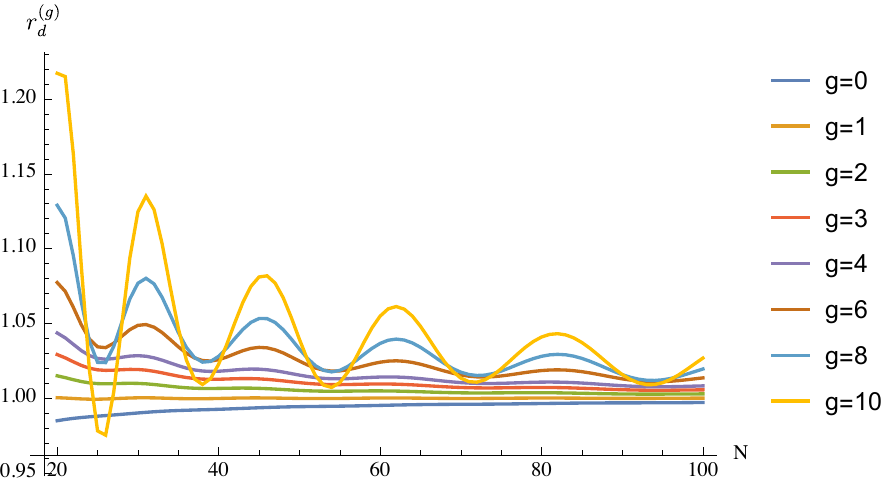} 
	\end{center}
	\vspace{-0.7cm}
	\caption{Ratio of the exact GW invariant and asymptotic estimate 
	for $X_5$ (left) and $X_{4,2}$ (right), for genus 0 up to 10, as a function of the order $N$ of the 
	logarithmic Richardson transform, using degree $d\leq 1600$.
 \label{Quinticgenuszeroten}}
\end{figure}

It is worth noting that as the depth $N$ of the Richardson transform is increased,
there are  oscillations of increasing amplitude and extent at small degree, 
but $L_N [r^{(g)}_d] $
eventually settles to 1 with high precision (see Fig. \ref{Quinticgenuszeroten}). 
This convergence allows us to probe the coefficient $\cV$ inside the exponential
with a precision of $\pm 10^{-18}$, the exponent $\rho$ of the polynomial correction
with a precision of $\pm 10^{-6}$ and  the exponent $\sigma$ 
of the logarithmic correction with a precision of $\pm 10^{-3}$  to  degree $d=650$,
leaving no doubt that the formula \eqref{allgenusleading} 
correctly captures the leading asymptotic behavior as $d\to \infty$ for any genus. 
We have performed a similar analysis for the degree $(4,2)$ 
complete intersection $X_{4,2}$ up to $d=1600$ with similar
results, and for the other hypergeometric models, but for much smaller degrees as 
the computation of high degree invariants, while straightforward in principle, is time-consuming in practice.
In all cases we find agreement with the prediction \eqref{allgenusleading}.

\subsection{Phenomenology of GV invariants at fixed degree}
\label{subsubsec-observ}

\begin{figure}[h]
	\begin{center}
		\includegraphics[height=5.3cm]{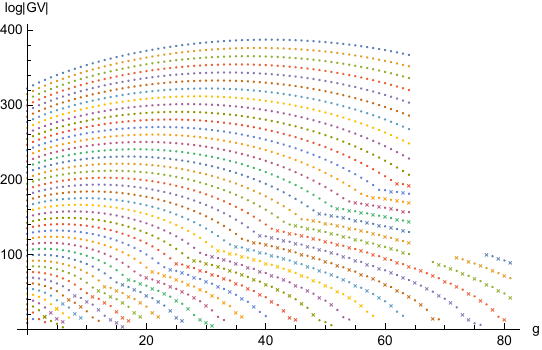} 
		\includegraphics[height=5.3cm]{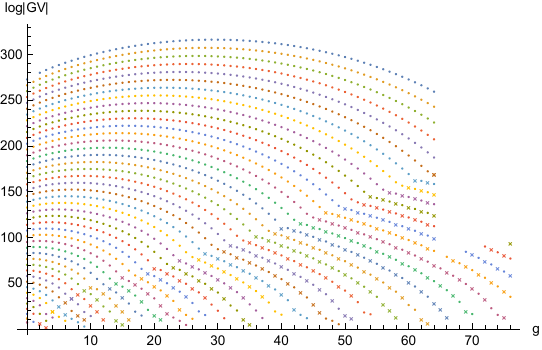} 
			\end{center}
\vspace{-0.7cm}
\caption{$\log|\GV_d^{(g)}|$ as a function of $g$ for $X_5$ (left) and $X_{4,2}$ (right). 
	Different degrees correspond to different colors. In both cases the maximal shown degree is 44. 
	Positive GV invariants are shown by dots, while negative ones by crosses. Gaps correspond to unknown invariants.
 \label{Fig-GVgenus}}
\end{figure}

Let us now fix the degree $d$ and consider the dependence of GV invariants $\GV_d^{(g)}$ on the genus $g$, bounded by $\gmax(d)$. In Fig. \ref{Fig-GVgenus} we plotted $\log|\GV_d^{(g)}|$ 
(keeping track of the sign of $\GV_d^{(g)}$) for two CY threefolds, $X_5$ and $X_{4,2}$.
These threefolds are distinguished by the fact that the degree for which all GV invariants 
(from $g=0$ up to $\gmax(d)$) are known is larger than for other manifolds in our list 
($d_{\rm mod}=26$ and 31, respectively; see Table \ref{tab_cydata}).
We will use these manifolds to exemplify our results throughout this work.
However, qualitatively all other manifolds exhibit a similar behavior. 
From the plots, one can make the following (phenomenological) observations:
\begin{itemize}
\item
The most striking feature is the existence of a kink: while GV invariants at each fixed 
degree nicely lie along a piecewise smooth curve, the slope has a discontinuity at a particular genus, 
which we denote $\gkink(d)$. 
It is worth noting that this kink does not seem to appear for local CY threefolds \cite{akp-to-appear}. 

\item
Remarkably, the point where the slope jumps coincides (up to one or two units) with the point 
where the sign of GV invariants changes its behavior: it stays positive for all $g<\gkink(d)$ 
but starts alternating after the kink. 
We thus define $\gkink(d)$ as the first genus where $\GV_d^{(g)}$ becomes negative. 
%(For local $\IP^2$, the sign alternates for all genera, so one may assign $\gkink(d)=0$.) 
Close to the Castelnuovo bound (hence near the horizontal axis), the sign of GV invariants becomes erratic.

\item
Another important feature is that up to the kink, i.e. for $0\leq g\leq \gkink(d)$, 
the smooth curves along which GV invariants lie are very well approximated by inverted parabolas. 
This feature was noticed already in \cite[App. A]{Huang:2007sb}
and will be used in \S\ref{subsec-apformula} as a basis for our approximate formula for GV invariants.

\item 
Let $\gtop(d)$ be the position of the maximum of $\log|\GV^{(g)}_d|$ at fixed degree. 
Its dependence on $d$, as well as the asymptotic form of the value $\log|\GV^{(\gtop(d))}_d|$ 
at the maximum, can be guessed from the numerics. In the next Section, we shall derive that both grow like $d^{3/2}$, 
but $\gtop(d)$ is suppressed in addition by a factor of $\log(d \log(d))$ (see \eqref{gtop} and \eqref{GVtop}),
contradicting the guess put forward in \cite[(A.2)]{Huang:2007sb}.

\item 
At large degree, we find numerically that $\gkink(d)$ grows like $d^{3/2}$ (see Fig. \ref{Fig-kink}). 
In fact, in the next section we shall find that a similar kink arises in the 5D BPS index and 
the positions of the two kinks are very close.
For the second kink we find an analytic formula \eqref{omkink}.
Its large $d$ expansion, given in \eqref{omkink-app} in terms of 
$\omega=\frac{m}{2(w d)^{3/2}}$ with $w=\left( \frac{2}{9\kappa} \right)^{1/3}$,
confirms the scaling.

\begin{figure}[h]
	\begin{center}
		\includegraphics[height=4.2cm]{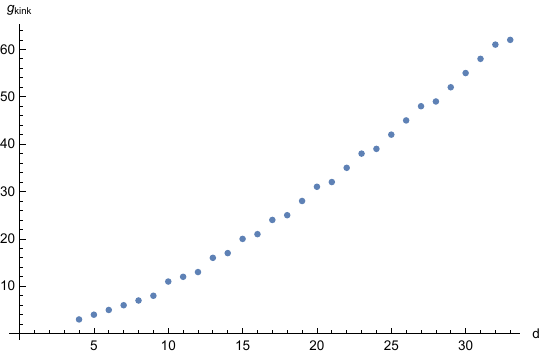} 
		\includegraphics[height=4.2cm]{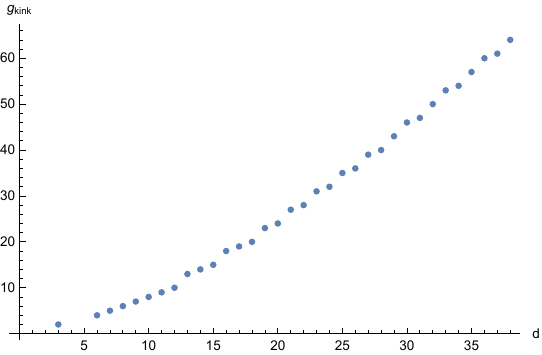} 
		\\
		\includegraphics[height=4.5cm]{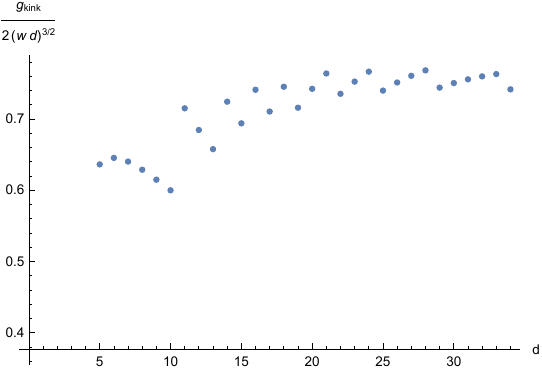} 
		\includegraphics[height=4.5cm]{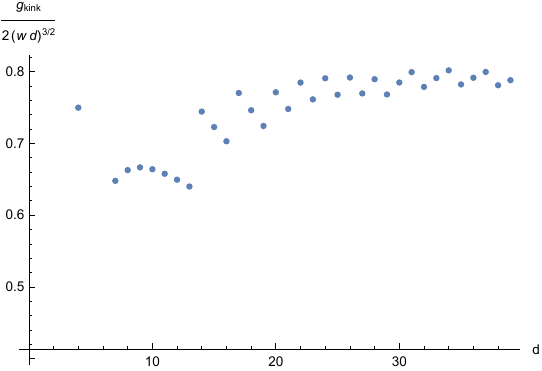} 
			\end{center}
\vspace{-0.7cm}
\caption{The values of $\gkink(d)$ for $X_5$ (left) and $X_{4,2}$ (right). At the bottom, $\gkink(d)$ is normalized by $2(wd)^{3/2}$. 
\label{Fig-kink}}
\end{figure}

\item
Finally, unfortunately, we do not have enough data to guess the form of GV invariants after the kink,
i.e. for $\gkink(d)< g\leq \gmax(d)$. 
It is tempting to claim that they are again captured by a Gaussian, 
but we do not have a sufficient set of constraints to determine its parameters.

\end{itemize}
These observations, especially the existence of the kink at $g=\gkink(d)$, will become relevant in the analysis of the 5D BPS index, 
to which we now turn.

\section{5D index, black holes and black rings}
\label{sec-GV5D}

In this section, we turn to the study of the index $\Omega_{5D}(d,m)$  
counting BPS states in 4+1 dimensions with fixed quantized electric charge $d$ and angular momentum $J_3=m/2$. 
After reviewing its relation to GV invariants, we study its asymptotics at large $d$, 
first in the static case $(m=0)$ and then allowing for angular momentum 
(with either $m$ or $m/d^{3/2}$ held fixed in the large $d$ limit).
Finally, we use the index to derive an approximate expression for GV invariants supposed
to hold at large $d$ and $g<\gkink(d)$.

\subsection{5D index from GV invariants}
\label{subsec-5dbh}

The GV invariants are related to counting of five-dimensional BPS states in M-theory 
compactified on $X$ as follows~\cite{Katz:1999xq}. Since the little group for a massive particle 
in 4+1 dimensions is $SO(4)= (SU(2)\times SU(2))/\IZ_2$, any massive state is 
characterized by its (quantized) electric charge $d$ 
and by a pair of angular momenta $(j_L,j_R)$, such that $J_\psi=j_L^z+j_R^z$ 
and $J_\phi=j_L^z-j_R^z$ generate rotations in the spatial planes $(x^1,x^2)$ and $(x^3,x^4)$, respectively. 
Following \cite{Gopakumar:1998jq}, let $N^{j_L,j_R}_d$
be the number of BPS states whose highest weight in the multiplet
carries momentum $(j_L,j_R)$. The Gopakumar-Vafa invariants
are defined by tracing over $j_R$ with an insertion of $(-1)^{2j_R}$, and expanding the character of the $SU(2)_L$
action in powers of the character of the representation $[1/2]\oplus [0] \oplus [0]$ of $SU(2)_L$,
\be
\label{eqNjLjRGV}
\sum_{j_L,j_R} (-1)^{2j_L+2j_R} (2j_R+1) 
N^{j_L,j_R}_d\chi_{j_L}(y) = 
\sum_{g=0}^{\gmax(d)} \GV_d^{(g)} \, (2-y-1/y)^{g},
\ee
where $\chi_j(y)=y^{-2j}+y^{-2j+2}+\dots + y^{2j}$ is the character of a spin $j$ representation of $SU(2)$. 
Adding in the descendants of the highest weight state in the BPS multiplet, which fill in the representation
$[1/2]\oplus [0] \oplus [0]$ of $SU(2)_L$, 
one defines the 5D index $\Omega_{5D}(d,m)$ through the 
relation~\cite{Katz:1999xq,Huang:2007sb}
\bea
\Tr (-1)^{2(j_L+j_R)} y^{2j_L^z}  &=& \sum_{g=0}^{\gmax(d)} \GV_d^{(g)} \, (2-y-1/y)^{g+1}
\nn\\
&=&  \sum_{g=0}^{\gmax(d)} \GV_d^{(g)} \,(-1)^{g+1}  \left(\sqrt{y}-\frac{1}{\sqrt{y}}\right)^{2g+2}
\label{Om5Dindex}\\
&:=&  
%\sum_{g=0}^{\gmax(d)} \GV_d^{(g)} %\sum_{m=-(g+1)}^{g+1}\binom{2g+2}{g-m+1} (-y)^m.
\sum_{m=-(\gmax(d)+1)}^{\gmax(d)+1}\Omega_{5D}(d,m) (-y)^m. 
\nn
\eea
The index counts 5D BPS states with electric charge $d$ 
and angular momentum $j_L^z=m/2$, weighted by  $(-1)^{2j_R}$, 
and is given by a  convolution of GV invariants and binomial coefficients,  
\be
\label{Om5D}
\Omega_{5D}(d,m) =  \sum_{g=|m|-1}^{g_{\rm max}(d)} \GV_d^{(g)} \binom{2g+2}{g-m+1}.  
\ee
Note that $\Omega_{5D}(d,m)$ is manifestly symmetric under $m\to -m$ as expected, and vanishes if $|m|>\gmax(d)+1$.
Moreover, it gets contributions from all angular momentum multiplets $(j_L,j_R)$ with $j_L\geq |m|/2$. 
As explained in~\cite{Sen:2012cj}, the number of states with 
fixed total angular momentum $j_L=m/2\geq 0$ (and fixed projection $j_L^z=m/2$), weighted by  $(-1)^{2j_R}$,  
can be obtained by taking the difference
\be
\begin{split}
\tOm_{5D}(d,m) =&\,  \Omega_{5D}(d,m) - \Omega_{5D}(d,m+2) 
\\
=&\,2(m+1)\sum_{g=|m|-1}^{g_{\rm max}(d)} \GV_d^{(g)} \binom{2g+2}{g-m+1}\, \frac{2g+3}{(g+m+2)(g+m+3)}\, .
\end{split}
\label{newindex-def}
\ee
In practice, we shall find that $|\Omega_{5D}(d,m)|$ sharply drops as $|m|$ increases, 
so the difference between the two ensembles is negligible, except when discussing certain subleading corrections.

\subsection{Supergravity predictions}

At large charge $d\gg 1$, we expect that the BPS states are captured by BPS solutions 
in the effective $\cN=1$ supergravity theory obtained by compactifying M-theory on $X$.
This includes spherically symmetric black holes, but also potentially other black objects such as black
rings, multi-centered black holes, black Saturns, 
etc.\footnote{See e.g. the recent discussion in \cite{Cassani:2024kjn,Cassani:2025iix,Boruch:2025sie}, in the context of the gravitational path integral.}

\subsubsection{BMPV black holes}

At low angular momentum, one may expect the dominant contribution to come from
the spherically symmetric BMPV black hole~\cite{Breckenridge:1996is}. Its
macroscopic entropy, computed in classical, two-derivative supergravity,  is given by 
\be
S_0^{\rm bh} = 2\pi \sqrt{Q^3-J^2} ,
\label{Sbh}
\ee
where the `graviphoton' electric charge $Q$ and  angular momentum $J=j_L$  are related to $d$ and $m$ through 
\be
Q= w d,
\qquad  
J= \frac{m}{2}\, ,
\qquad 
w:= \left( \frac{2}{9\kappa} \right)^{1/3} .
\label{QJw}
\ee
Importantly, BMPV black holes have vanishing angular momentum\footnote{The general 
	rotating black hole solutions of \cite{Cvetic:1995kv} carry both $j_L$ and $j_R$, 
	but they are BPS only when $j_R=0$.}
$j_R$, so they give rise to positive contributions to the 5D index.

Higher derivative interactions in the 5D supergravity action of the form 
$c_{a} A^a \wedge \cR \wedge \cR$ generate corrections suppressed by a factor of $1/Q$ compared to the leading 
$\cO(Q^{3/2})$ classical entropy \eqref{Sbh}.
Unfortunately, there are conflicting predictions in the literature~\cite{Guica:2005ig,Castro:2007ci,deWit:2009de,Cassani:2024tvk},
and we do not know {\it a priori} which one is correct.
They are all can be written as\footnote{For one-parameter models, we use the notation $c_a=c_2$, 
	since it is determined by the second Chern class of the CY threefold, $c_2=\int_{\cD} c_2(TX)$.
	Note that to put the result of \cite{Cassani:2024tvk} into the form \eqref{Sexp}, we expanded it to first order in the coefficient of the higher curvature correction, and fixed that coefficient in terms of $c_2$ so as to match the correction to the entropy in the static case.
	\label{foot-expand}}
\be 
S^{\rm bh}  = 2\pi Q^{3/2}\sqrt{1-\omega^2} \( 1+  \frac{3 w c_2}{16 Q} \,g(\omega) \)  + \cO(c_2^2),
\label{Sexp}
\ee 
where $\omega=\frac{J}{Q^{3/2}} = \frac{m}{2(wd)^{3/2}}$ and the function $g(\omega)$ is given by
\begin{subequations}
\bea
\cite[(15)]{Guica:2005ig} 
%further discussed in \cite[(2.9)]{Huang:2007sb}
&\qquad&
g(\omega)  = 1 + \frac13\, \omega^2,
\label{Sexp1}
\\ 
\cite[(3.27)]{Castro:2007ci}
&&
g(\omega)  =\frac{1}{1-\omega^2}\, ,
\label{Sexp2}
\\
\cite[(7.17)]{deWit:2009de}
&&
g(\omega)  = 1\, ,
\label{Sexp3}
\\
\cite[(7.12)]{Cassani:2024tvk}
&&
g(\omega)  = \frac{1-\tfrac{4}{3} \omega^2}{1-\omega^2} .
\label{Sexp4}
\eea
\label{Sexpq}
\end{subequations}
For $J=0$, all these predictions agree and amount to shifting the 
electric charge $d\mapsto d+ \frac{c_2}{8}$ in the leading order result~\cite{Castro:2007hc,Alishahiha:2007nn}. 
The discrepancy in the $J$-dependence of the subleading contribution seems to arise due to subtleties 
in the definition of physical charges in the presence of Chern-Simons and higher derivative terms. As we shall see below, our data strongly support the latest prediction \eqref{Sexp4}.

\subsubsection{Black rings}
\label{subsubsec-br0}

In addition, $\cN=1$ supergravity in 5D admits BPS black ring solutions, which have horizon topology $S^1\times S^2$ and in general carry both $j_L$ and $j_R$
angular momentum~\cite{Elvang:2004rt,Elvang:2004ds,Bena:2004de,Gauntlett:2004qy}. 
These solutions are characterized by electric charges $q_a$, dipole charges $p^a$ and a parameter $q_0$ 
related to angular momentum. 
Indeed, they can be obtained from D4-D2-D0 BPS black hole solutions of $\cN=2$ supergravity 
in 4D by applying the 4D/5D lift~\cite{Gaiotto:2005xt}. As a result, their entropy takes the same form as for
D4-D2-D0 black holes\footnote{In \eqref{S0br} and especially \eqref{Sbr} 
	we correct numerical factors and signs which affect some of the formulae in~\cite{Cyrier:2004hj,Halder:2023kza}.}
\be
\label{S0br}
S_0^{\rm br}=2\pi\sqrt{-\frac{\kappa_{abc} p^a p^b p^c }{6}\, \hq_0}, 
\qquad 
\hq_0:=q_0-\frac12\, \kappa^{ab} q_a q_b,
\ee
where $(p^a,q_a,q_0)$ are the D4-D2-D0 brane charges, $\hat q_0$ is a combination invariant under "spectral flow",
and $\kappa^{ab}$ is the inverse of the quadratic form $\kappa_{ab}=\kappa_{abc} p^c$. 
For one-parameter models the brane charges can be written as 
\be 
(p^a,q_a,q_0)=\(r\, ,\, d+\frac{\kappa r^2}{2}\,,\,\frac{\kappa r^3}{6}+\frac{c_2 r}{24}-n\),
\label{D420charge}
\ee 
where $r,d,n\in\IZ$ and the rational shifts are required
by charge quantization \cite{Alexandrov:2010ca}\footnote{The D0-charge quantization is ensured by the well-known fact 
that $\chi(\cO_{rD})=\frac{\kappa r^3}{6}+\frac{c_2 r}{12}\in\IZ$, 
where $\cD$ is the effective divisor on which the D4-brane is wrapped.}
\be 
p^a\in\IZ, 
\qquad
q_a\in \IZ+\frac12\, \kappa_{abc}p^b p^c,  
\qquad
q_0\in \IZ-\frac{c_{a} p^a}{24}\, .
\ee 
Substituting the charges into \eqref{S0br}, the entropy reduces to
\be
\label{Sbr}
S_0^{\rm br}=2\pi\sqrt{-\frac{\kappa r^3 }{6}\, \hq_0},
\qquad 
\hq_0 = -n -\frac{d^2}{2\kappa r} - \frac{r d}{2} + \frac{\kappa r^3+c_2r}{24} \, .
\ee
In terms of the 5D black ring solution, the D2-brane charge $d$
is interpreted as the (quantized) electric charge, while $n$ coincides with $2j_L^z=J_\phi+J_\psi$, up to an integer 
shift which will be fixed in \S\ref{subsubsec-ring}. 
The angular momentum $J_\phi=j_L^z- j_R^z$ is instead equal to 
$\hf p^a (q_a -\frac12 \kappa_{abc} p^b p^c)=\hf rd$, giving access to the dipole charge $r$.
Note that the charge $n$ is bounded from below by the condition that the entropy \eqref{Sbr} is real 
and hence $\hq_0$ must be negative, but it is also bounded from above by the condition $j_L^z\leq 0$. 
Although solutions with positive $j_L^z$ exist, they are either not supersymmetric 
or do not bend to form a ring \cite{Bena:2007kg,Bena:2009ev,Compere:2010fm}. 
In either case they do not contribute to the index.\footnote{Of course, one can get a black ring solution
with the opposite sign of $j_L^z$ by flipping the sign of all charges, in agreement with 
the symmetry of $\Omega_{5D}(d,m)$.}
In the presence of the higher derivative term $c_{2} A \wedge \cR \wedge \cR$, 
the entropy formula \eqref{Sbr} for the black ring receives the same corrections as for D4-D2-D0 black holes~\cite{Maldacena:1997de}, 
namely~\cite{Guica:2005ig,Bena:2005ae} 
\be
\label{Sbrc2}
S^{\rm br}=2\pi\sqrt{-\frac{\kappa r^3 + c_2 r}{6}\, \hq_0}\, ,
\ee
where $\hq_0$ is given by the same expression as in \eqref{Sbr}.

\subsubsection{Logarithmic corrections}
\label{subsubsec-log}

In addition to higher derivative corrections, the entropy of both black holes and black rings 
receives logarithmic corrections of the form 
$S \to S + \alpha \log Q$
due to massless modes. The coefficient $\alpha$ depends on the ensemble  considered
and on the scaling of charges \cite{Sen:2012cj,Sen:2012dw,Anupam:2023yns}.
In particular, it was found that in the ensemble with both the angular momentum $J=j_L$
and third component $j_L^z$ fixed,  scaling as $J\sim Q^{3/2-\gamma}$ as $Q\to\infty$,
the coefficient $\alpha$ is given by~\cite[(1.4)]{Sen:2012cj}
\be 
\alpha_{j^z,j}=-\frac14\, (n_V-3+4\gamma),
\label{alph-pred-2j}
\ee 
where $n_V$ is the number of $U(1)$ gauge fields and in the subscript 
we indicated the fixed quantities.
Furthermore, in \cite[(6.2)]{Sen:2012cj} it is shown that $\alpha_{j^z,j}$
is related to the coefficient $\alpha_{j^z}$ in the ensemble where  only $j_L^z$ is fixed by
\be 
\alpha_{j^z,j}=\alpha_{j_3}-\gamma.
\label{alph-pred-diff}
\ee 
Combining the two predictions and restricting to one-parameter models with $n_V=1$, one obtains 
\be 
\alpha_{j^z}=\hf\, .
\label{alph-pred-j}
\ee

\subsection{Quantitative checks}

\begin{figure}[h]
	\begin{center}
		\includegraphics[height=5.3cm]{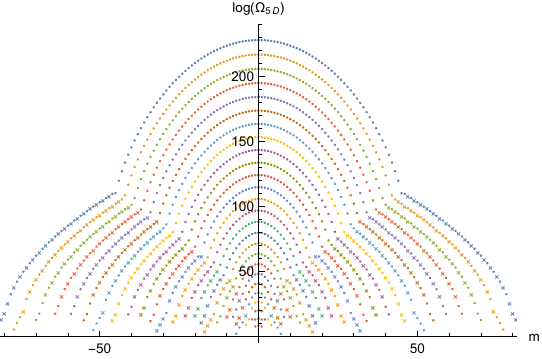} 
		\includegraphics[height=5.3cm]{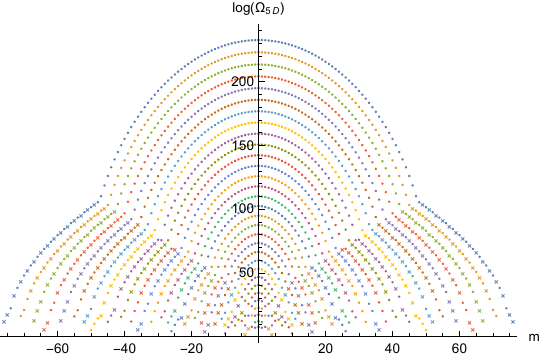} 
	\end{center}
	\vspace{-0.7cm}
	\caption{$\log|\Omega_{5D}(d,m)|$ as a function of $m$ for $X_5$ (left) and $X_{4,2}$ (right). 
		Different degrees correspond to different colors. The maximal shown degrees are 26 and 31, respectively. 
		Positive BPS indices are shown by dots, while negative ones are shown  by crosses. 
		\label{Fig-Om5D}}
\end{figure}

In Fig. \ref{Fig-Om5D} we plot the values of    
$\log|\Omega_{5D}(d,m)|$ as a function of angular momentum $m$ for our two favorite models $X_5$ and $X_{4,2}$,
up to the maximal degree where GV invariants are known for all genera. 
One notices a striking similarity with Fig. \ref{Fig-GVgenus}: 
the discrete numerical values lie on a smooth curve, up to some critical
value $\mkink(d)$ where the slope changes abruptly and the sign of 
$\Omega_{5D}(d,m)$
starts to alternate. This change of slope was in fact
already observed in \cite{Halder:2023kza} where it was interpreted as
a phase transition between BMPV black holes, which dominate at low angular momentum, to black rings,  which appear to
dominate at large angular momentum.
%\footnote{Unfortunately, the authors of \cite{Halder:2023kza} could not provide convincing quantitative evidence 
%for this interpretation, partly due to the ambiguity in fixing the dipole charge $r$.}

\begin{figure}[h]
	\begin{center}
		\includegraphics[height=5.3cm]{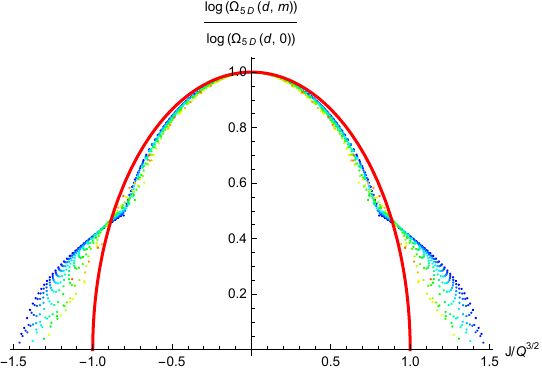} 
		\includegraphics[height=5.3cm]{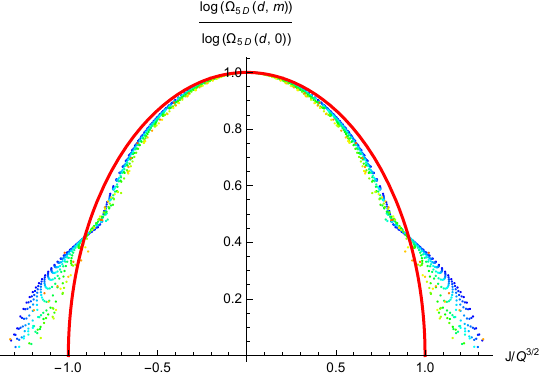} 
			\end{center}
\vspace{-0.7cm}
\caption{The ratio $\frac{\log|\Omega_{5D}(d,m)|}{\log|\Omega_{5D}(d,0)|}$ 
as a function of $J/Q^{3/2}$, for $X_5$ (left) and $X_{4,2}$ (right) 
as $d$ ranges from $6$ (orange) to 26 and 31 (blue), respectively. 
The red curve is the ratio of the classical black hole entropies given by $\sqrt{1-J^2/Q^3}$.
\label{Fig-Om5DN}}
\end{figure}

Thus, one expects that at large enough $d$, keeping the ratio $m/d^{3/2}$ fixed,
the logarithm of the 5D index \eqref{Om5D} should reproduce the black hole entropy \eqref{Sexp} for $|m|<\mkink$, 
while it should be related somehow to the black ring entropy \eqref{Sbrc2} 
for $|m|>\mkink$. The precise relation, however, is not clear as the black ring entropy depends on an additional dipole
charge.\footnote{In \cite{Halder:2023kza} it was proposed to extremize $S_0^{\rm br}$ 
	with respect to $r$ keeping $d$ and $m$ fixed.
Below we propose a different prescription which appears to produce much better agreement.}
We will discuss the precise match between the actual data and the analytic expectations
in the subsections below. For now, it is instructive to present the data 
for the 5D index (more precisely, the logarithm of its absolute value)
normalized with respect to its value at $m=0$, in terms of the ratio  $\omega=J/Q^{3/2}$,
see Fig. \ref{Fig-Om5DN}.
Remarkably, in the region before the kink the data for all degrees follow essentially the same curve, 
which is very close to $\frac{S_0^{\rm bh}(Q,J)}{S_0^{\rm bh}(Q,0)}=\sqrt{1-\omega^2}$.
The deviation from the classical entropy can be attributed to quantum corrections.
Note that the spread of the data after the kink is much larger, which indicates 
that the corresponding entropy is not a function of the single parameter $\omega$.

\subsubsection{Static black holes}
\label{subsubsec-static}

For static black holes, setting $m=0$ and taking into account that $g(0)=1$ for all choices in \eqref{Sexpq}, 
the entropy \eqref{Sexp}, supplemented by the logarithmic correction discussed in \S\ref{subsubsec-log}
and a constant term,
predicts at large $d$
\be
\log \Omega_{5D}(d,0) = b_0 d^{3/2} + b_1 d^{1/2} + \alpha \log d + \beta+ \frac{b_2}{\sqrt{d}}+\cdots
\label{expOm5}
\ee
with
\be
b_0 = \frac{4\pi}{3\sqrt{2\kappa}}\, ,
\qquad
b_1 = \frac{\pi c_2}{4\sqrt{2\kappa}}\, .
\label{coeff-b}
\ee
%while Eq. \eqref{SexpHL} predicts
%\be
%\log \Omega_{5D}(d,0) = b_0 d^{3/2} + \frac43 b_1 d^{1/2} + \alpha \log d  + \beta +  \frac{b_2}{\sqrt{d}}+\dots
%\ee
%with $b_2=-\frac{\pi c_2^2}{24\sqrt{2\kappa}}$. 
%For the quintic, the coefficients evaluate to
%\be
%b_0 \sim 1.32, \quad b_1\sim 12.4, \quad b_2 \sim -103
%\ee
As noted in \cite[(3.3)]{Huang:2007sb} and mentioned below \eqref{Sexp}, 
the subleading coefficient $b_1$ (originating from 
$\cA\wedge \cR^2$-type corrections) 
could be reproduced by shifting $d\mapsto d+\frac{c_2}{8}$ in the first term.\footnote{In \cite{Halder:2023kza}, 
	the authors entertain two possibilities for the coefficient $b_1$, the one presented in \eqref{coeff-b} 
	(called SUGRA prescription) and another value differing by a factor $3/4$ (called OSV prescription), 
	which would come by shifting $d\mapsto d+\frac{c_2}{6}$. 
	They argue that numerics supports the OSV prescription (Figure 5, {\rm loc. cit.}), 
	but we find overwhelming evidence for the SUGRA value, after a suitable Richardson transform. 
	We are also unconvinced by the 'black ring' curves in the same Figure 5, which arise from $\cO(c_2)$ 
	corrections to the black ring entropy with some \textit{ad hoc} choice of the dipole charge $r$.
	}
The same shift would then produce 
the $\cO(1/\sqrt{d})$ term in \eqref{expOm5} with coefficient $b_2=\frac{3c_2^2 b_0}{512}$. 
While there is no reason to believe that this is the correct value of $b_2$, 
it provides an estimate which suggests that the $1/\sqrt{d}$ corrections could 
overshadow the $\log d$ term up to quite large values of $d$. It  
is therefore advisable to choose a model with the smallest possible value of 
$\frac{c_2}{8 d_{\rm mod}}$, such that the sub-subleading correction proportional to $b_2$ is more likely to be negligible. 
The most promising models in this respect are $X_{4,2}, X_5, X_{2,2,2,2}, X_{3,3}, X_6$ in this order.

\begin{table}
$$
\begin{array}{|l|c|c|c|c|}
\hline
X & \ N_0\ &\;\% \mbox{error in } b_0\; &\ N_1\ &\; \%\mbox{error in } b_1\; \\
\hline
 X_{5} & 12 & 0.036 & 15 & -2.39 \\
 X_{6} & 7  & 0.024 & 11 & -0.02 \\
 X_{8} & 10 & -0.11 & 9 & -5.89  \\
 X_{10} & 8 & 0.19 & 11 & -6.00  \\
 X_{3,3} & 4 & 0.41 & 10 & -2.33  \\
 X_{4,2} & 14 & -0.015 & 18 & -0.52  \\
 X_{4,3} & 4 & 0.76 & 8 & -2.95  \\
 X_{4,4} & 4 & 0.41 & 7 & -3.34 \\
 X_{6,2} & 9 & 0.29 & 15 & -0.92  \\
 X_{6,4} & 2 & 0.51 & 6 & -4.07 \\
 X_{6,6} & 2 & 0.69 & 5 & -3.55 \\
 X_{3,2,2} & 2 & -0.49 & 6 & 0.32 \\
 X_{2,2,2,2} & 4 & 0.30 & 6 & -3.59  \\
 \hline
\end{array}
$$ 
\caption{The minimal deviations from the theoretical predictions \eqref{coeff-b} 
of the coefficients $b_0$ and $b_1$ of the asymptotics  \eqref{expOm5} of the 5D index 
obtained by the Richardson transform of the series \eqref{defss} evaluated at the maximal available degree. 
$N_0$ and $N_1$ are depths of the Richardson transform at which these values are achieved.
 \label{tab_b0b1}}
\end{table}

\medskip

First, we verify the predictions \eqref{coeff-b}.
In order to suppress transient effects at finite $d$ and accelerate convergence, following
\cite{Huang:2007sb} we apply the depth $N$ Richardson transform $R_N$  \eqref{defRichardN} to
the two series 
\be
s_0(d)=\frac{\log\Omega_{5D}(d,0)}{b_0 d^{3/2}}-1,
\qquad
s_1(d) = \frac{\log\Omega_{5D}(d,0) - b_0 d^{3/2}}{b_1 d^{1/2}}-1.
\label{defss}
\ee 
The values of the Richardson transforms $R_N[s_i](d)$
at the largest value of $d$ and $N\ll d$, where we can compute them reliably, 
give an estimate of the deviations of the coefficients $b_0$ and $b_1$, respectively, from their expected values. 
One finds that upon increasing $N$, convergence improves at first, 
but beyond a certain optimal $N_{\rm opt}$,  a further increase in $N$ leads to erratic results. 
This fact can be easily understood by noticing that each increase of the depth shortens the data set by one. 
Thus, increasing depth, one eventually restricts oneself
to small degrees where the asymptotics \eqref{expOm5} is not a valid approximation.

In Table \ref{tab_b0b1} we present the depth $N_{\rm opt}$ of the Richardson transform
for which one achieves the best convergence of the two series \eqref{defss},
and the corresponding deviations from 0.
The agreement with the supergravity prediction for the leading coefficient $b_0$ 
is within less than $0.5\%$ over all models except $X_{4,3}$ and $X_{6,6}$,
and for $X_5$, $X_6$ and $X_{4,2}$ it is even better by another order.
These results significantly improve upon the errors reported in 
\cite{Huang:2007sb} by a factor of 2 to 100, depending on the model. 
The agreement for the subleading coefficient $b_1$ is less impressive, 
though it is improved by a factor of 3 or so compared to the previous analysis. 

\medskip

\begin{figure}[h]
\begin{center}
\includegraphics[height=4.7cm]{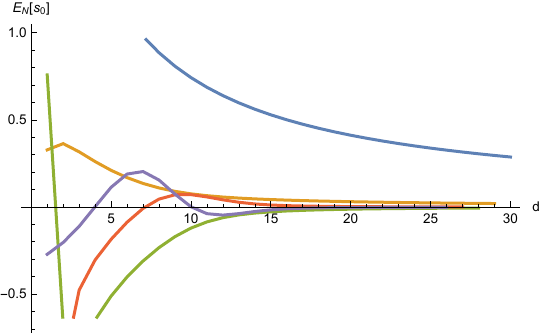} 
\hspace*{2mm} 
\includegraphics[height=4.7cm]{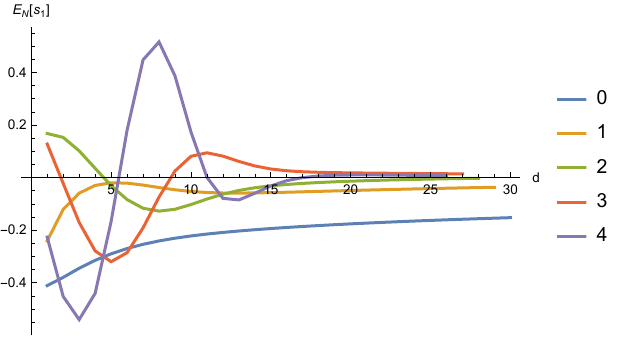}
\end{center}
\vspace{-0.7cm}
\caption{The $E_N$ transforms of the series \eqref{defss} for $X_{4,2}$ with $N=0,\dots,4$. 
	The sets of functions used in the transform are
$\{d^{-1},d^{-3/2}\log(d), d^{-3/2},d^{-2}\}$ (left)
and $\{d^{-1/2},d^{-1},d^{-3/2},d^{-2}\}$ (right).
\label{fig-bb}}
\end{figure}

In fact, one can apply the more general $E$-algorithm described in Appendix \ref{sec_Richardson} to further strengthen our results. 
For example, if one applies it to eliminate the terms 
$\{f_k(d)\}=\{d^{-1},d^{-3/2}\log(d), d^{-3/2}\}$ in the expansion of $s_0(d)$ 
(terms that follow from the expected subleading terms in \eqref{expOm5}),
a very good match ($0.038\%$ for $X_5$ and $0.067\%$ for $X_{4,2}$) is achieved already at depth 3.
Similarly, applying the $E$-algorithm with 
$\{f_k(d)\}=\{d^{-1/2},d^{-1}\}$\footnote{It is worth noting that the log-term is not necessary to include 
to achieve this remarkable convergence. \label{foot-nolog}} 
to $s_1(d)$, one obtains depth 2 results which are better than the ones presented in Table \ref{tab_b0b1} by a few orders
($0.025\%$ for $X_5$ and $0.23\%$ for $X_{4,2}$); upon 
subtracting further terms, the agreement can be  improved even further.
In Fig. \ref{fig-bb} we show examples of applying the $E$-algorithm 
to $s_0(d)$ and $s_1(d)$ for $X_{4,2}$. They leave no doubt that 
the numerical evaluation of the 5D index is perfectly consistent with the predictions
\eqref{coeff-b}.

\medskip

\begin{figure}[h]
\begin{center}
\includegraphics[height=4.7cm]{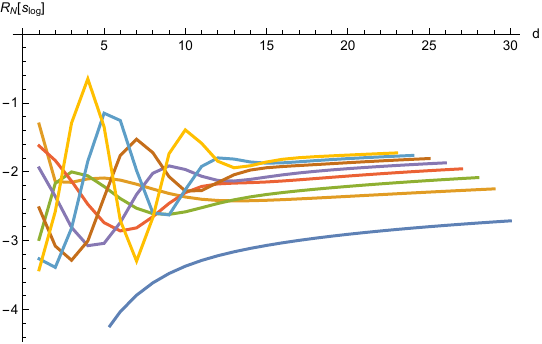} 
\hspace*{2mm} 
\includegraphics[height=4.7cm]{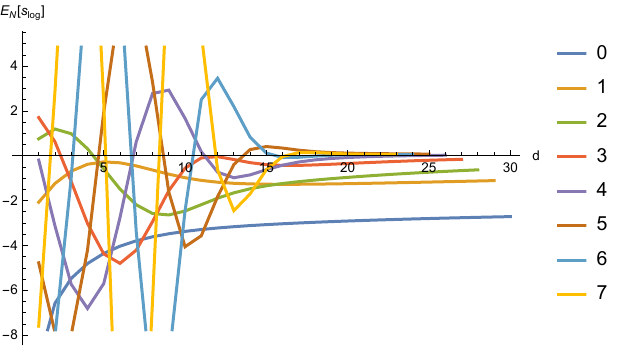}
\end{center}
\vspace{-0.7cm}
\caption{The Richardson and $E_N$ transforms of $s_{\log}(d)$ for $X_{4,2}$ with $N=0,\dots,7$. 
	The set of functions used in the $E$-algorithm is
$\{d^{1-k}/\log(d)\}_{k\geq 1}$.
\label{fig-log}}
\end{figure}

Next, we turn to the logarithmic correction in \eqref{expOm5}.
Similarly to \eqref{defss}, we define the series
\be 
s_{\log}(d)=\frac{\log\Omega_{5D}(d,0) - b_0 d^{3/2}-b_1 d^{1/2}}{\log(d)}
\label{def-slog}
\ee 
and analyze its various transforms.
Unfortunately, this analysis does not lead to a definite limiting value of the series.
On the one hand, the standard Richardson transform $R_N$ suggests a slow convergence to the value around $-3/2$, 
which is quite far from the expected value $\alpha=1/2$ in \eqref{alph-pred-j}. 
On the other hand, it is not expected to work anyway because it eliminates 
the first terms in the Taylor expansion in $1/d$,
whereas the expansion \eqref{expOm5} indicates that all subleading terms in $s_{\log}(d)$
should involve the factor of $1/\log(d)$.
Therefore, it is natural to try and apply the $E$-algorithm with functions of this form, e.g. $f_k(d)=d^{1-k}/\log(d)$.
Indeed, with this choice one finds an apparent convergence towards 0 
(see Fig. \ref{fig-log} for the relevant plots in the case of $X_{4,2}$).
However, this value is still different from the supergravity prediction \eqref{alph-pred-j} 
and other subtraction schemes, e.g. the one based on $f_k(d)=d^{(1-k)/2}/\log(d)$, 
fail to produce convergent results with the currently available data. 
Hence, it is not clear whether one should take
seriously the fact that the $E$-algorithm with the above choice of 
$\{f_k(d)\}$ apparently leads to vanishing $\alpha$. 
(Note however that this value is also supported by the remark in 
footnote \ref{foot-nolog}.)

\begin{figure}[h]
\begin{center}
\includegraphics[height=4.7cm]{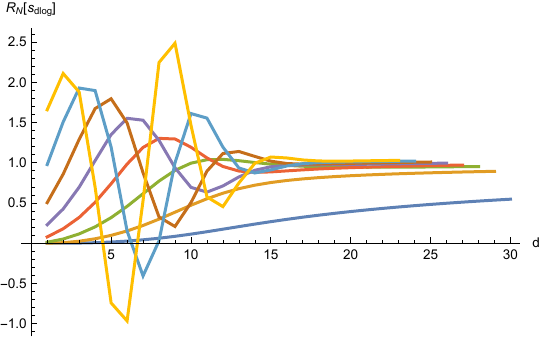} 
\hspace*{2mm} 
\includegraphics[height=4.7cm]{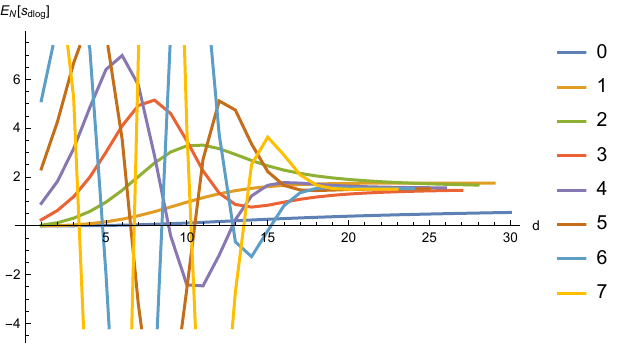}
\end{center}
\vspace{-0.7cm}
\caption{The Richardson and $E_N$ transforms of $s_{\de\log}(d)$ for $X_{4,2}$ with $N=0,\dots,7$. 
	The set of functions used in the $E$-algorithm is
$\{d^{1-k}/\log(d)\}_{k\geq 1}$.
\label{fig-dlog}}
\end{figure}

Given the uncertainty on the value of $\alpha$, it seems pointless to try and investigate further subleading terms.
However, one can also try to check the prediction \eqref{alph-pred-diff} 
for the difference of the logarithmic corrections in two ensembles.
In the static case, the angular momentum does not scale with degree and therefore one should set $\gamma=3/2$. 
Since the change of ensemble does not affect the two leading terms 
in the large degree expansion of the index, the series to be analyzed is  
\be 
s_{\de\log}(d)=\frac{\log\Omega_{5D}(d,0) -\log\tOm_{5D}(d,0)}{\log(d)}\, .
\label{def-sdlog}
\ee 
We performed the same analysis for this series as for \eqref{def-slog}
and found somewhat similar results: the Richardson transform converges around 1, 
the $E$-algorithm with $f_k(d)=d^{1-k}/\log(d)$ converges towards 3/2, 
and other subtraction schemes are not convergent.
(See Fig. \ref{fig-dlog} for the relevant plots in the case of $X_{4,2}$.)
Thus, the Richardson transform again disagrees with the supergravity
prediction, but this time the simplest version of the $E$-algorithm does agree! 
This fact might be considered as an indication that it should also produce correct results for $s_{\log}(d)$, 
which is yet another argument supporting $\alpha=0$.

\subsubsection{Slow-spinning black holes}
\label{subsubsec-slowspin}

Let us now consider 5D index $\Omega_{5D}(d,m)$ at large $d$ keeping $m$ non-vanishing but fixed.
Following \cite{Huang:2007sb}, we define
\be
\rho(d,m) = \frac{d^{3/2}}{m^2} \log\left| \frac{\Omega_{5D}(d,0)}{\Omega_{5D}(d,m)}\right|. 
\ee
Assuming that the entropy in this regime is dominated by BMPV black holes, 
from \eqref{Sexp} we find that this quantity is expected to have the following expansion
\be 
\rho(d,m)=p_0+\frac{p_1}{d}+\frac{p_{3/2}}{d^{3/2}}+ \cO(1/d^2),
\label{exprho}
\ee 
where\footnote{This corrects a factor 1/4 in $p_0$ \cite[(3.20)]{Huang:2007sb}}
\bea
p_0 &=& \frac{\pi}{4 w^{3/2}}=\frac{3\pi}{4}\, \sqrt{\frac{\kappa}{2}}\, ,
\nn \\
p_1&=&\frac{3\pi c_2}{64 w^{3/2}}\,(1-2g_1)=\frac{9\pi c_2}{64}\,\sqrt{\frac{\kappa}{2}}\,(1-2g_1),
\label{allp}\\
p_{3/2}&=&\frac{\alpha}{12 w^3}=\frac{3\alpha\kappa}{8}\, ,
\nn 
\eea
and $g_1$ is the first Taylor coefficient in the expansion $g(\omega)=1+\sum_{k>0}g_k\omega^{2k}$.
Note that the dependence on $m$ is absent in the leading terms shown in \eqref{exprho}, 
as it would start to appear at order $m^2/d^3$.
Similarly to the static case, we define the series
\be 
\ts_0(d)=\frac{\rho(d,1)}{p_0}-1,
\qquad 
\ts_1(d)=\frac{64 w^{3/2}}{\pi c_2}\,d\, (\rho(d,1)-p_0).
\ee 
The first series measures the deviation of the first expansion coefficient from its supergravity prediction \eqref{allp}, 
while the second extracts the value of the second coefficient. 
We normalized the latter in such a way that the limiting values of $\ts_1$
implied by the four functions \eqref{Sexpq} are all integer and given by $1,-3, 3$ and $5$, respectively.

\begin{figure}[h]
\begin{center}
\includegraphics[height=4.7cm]{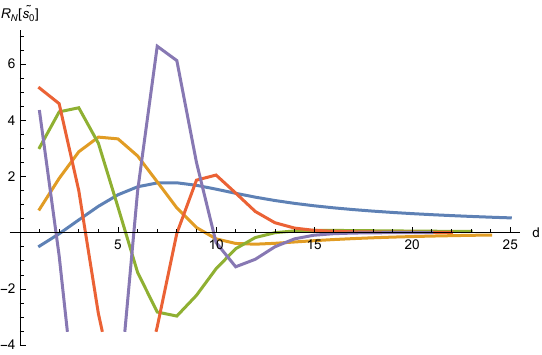} 
\hspace*{2mm} 
\includegraphics[height=4.7cm]{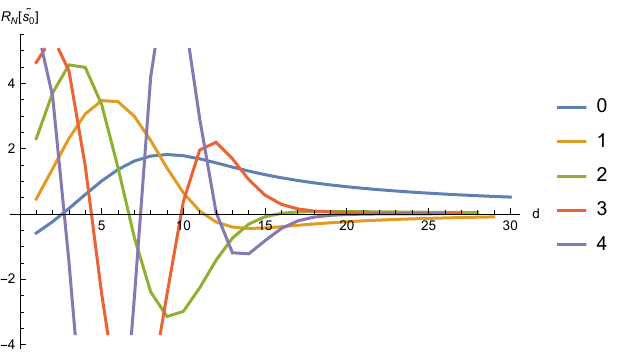}
\end{center}
\vspace{-0.7cm}
\caption{The Richardson transforms of $\ts_0(d)$ for $X_5$ and $X_{4,2}$ with $N=0,\dots,4$. 
\label{fig-RNp0}}
\end{figure}

\begin{figure}[h]
\begin{center}
\includegraphics[height=4.7cm]{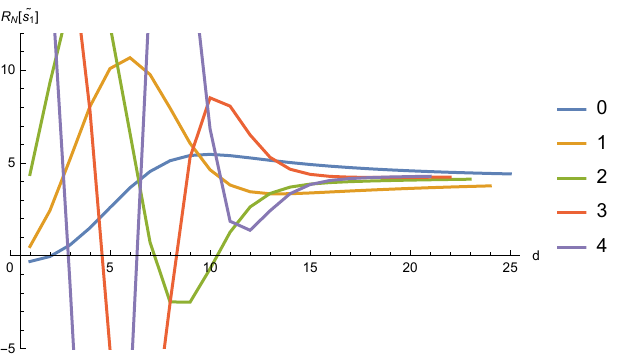} 
\hspace*{2mm} 
\includegraphics[height=4.7cm]{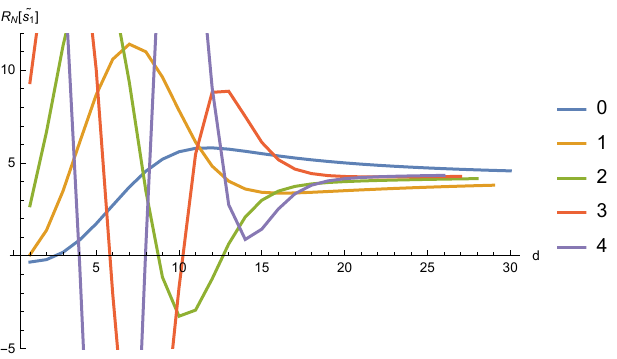}
\\
\includegraphics[height=4.7cm]{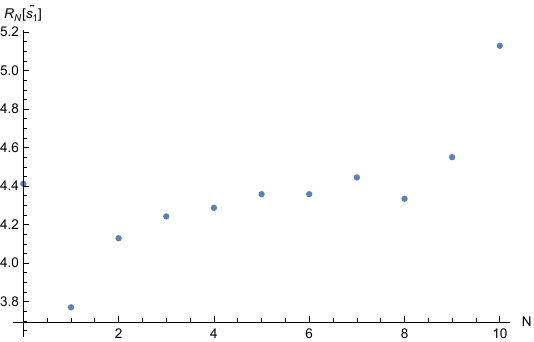} 
\hspace*{1cm} 
\includegraphics[height=4.7cm]{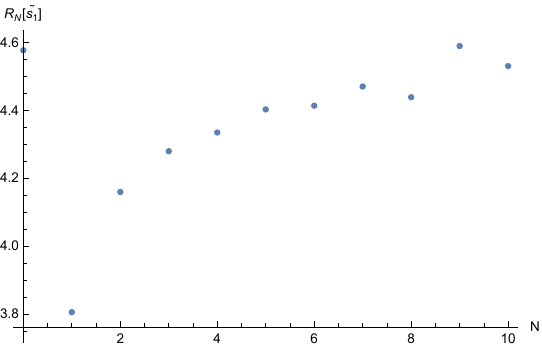}
\end{center}
\vspace{-0.7cm}
\caption{The Richardson transforms of $\ts_1(d)$ for $X_5$ and $X_{4,2}$ with $N=0,\dots,4$ (top)
and the values of $\(R_N[\ts_1]\)_d$ for maximally available $d$ for $N$ up to 10 and the same manifolds (bottom). 
\label{fig-RNp1}}
\end{figure}

The results of several Richardson transforms of $\ts_0(d)$ are shown in Fig. \ref{fig-RNp0}. 
They clearly demonstrate that this series converges to 0 with an error of less than $0.7\%$, 
a fact that the analysis in \cite{Huang:2007sb} was not able to demonstrate.
The same Richardson transforms applied to $\ts_1(d)$,
shown in Fig. \ref{fig-RNp1}, indicate a slow convergence towards the value $5$, which supports the last prediction \eqref{Sexp4}. 
Unfortunately, our attempts to accelerate convergence
by using the $E$-algorithm to eliminate half-integer inverse powers of $d$ 
were unsuccessful due to the limited available data. 
In the next subsection we shall find further support for this result
at finite angular momentum.

\subsubsection{Fast-spinning black holes}
\label{subsubsec-fastspin}

\begin{figure}[h]
	\begin{center}
		\def\leng{3.4cm}
		\def\lengg{3.1cm}
	\begin{tabular}{|c|c|c|} \hline
	\includegraphics[height=\lengg]{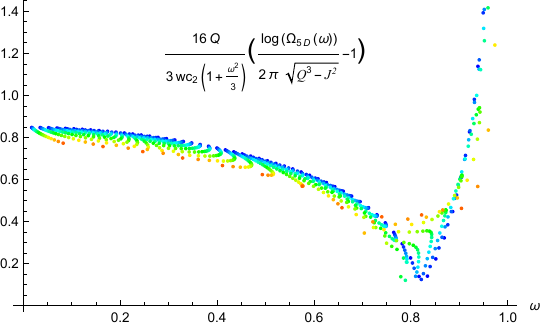} &
	\includegraphics[height=\lengg]{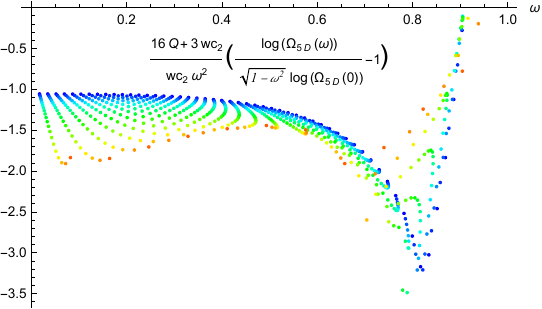} &
	\includegraphics[height=\lengg]{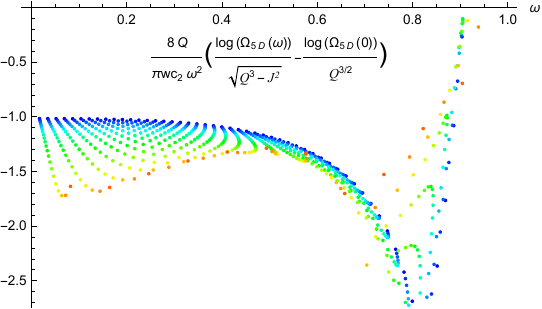} 
	\\ \hline 
	\includegraphics[height=\lengg]{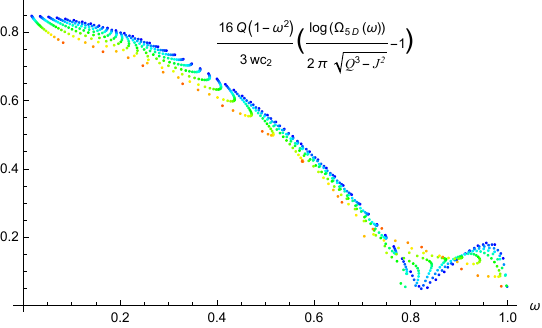} &
	\includegraphics[height=\lengg]{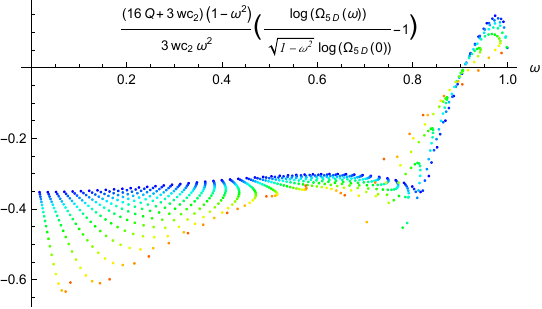} &
	\includegraphics[height=\lengg]{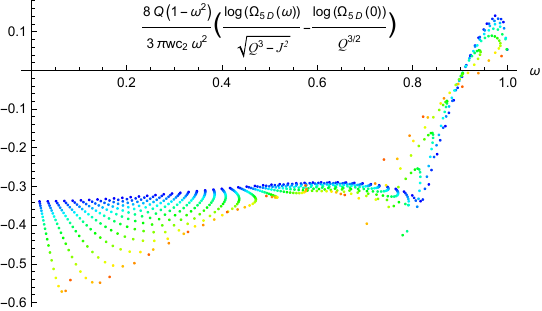} 
	\\ \hline
	\includegraphics[height=\lengg]{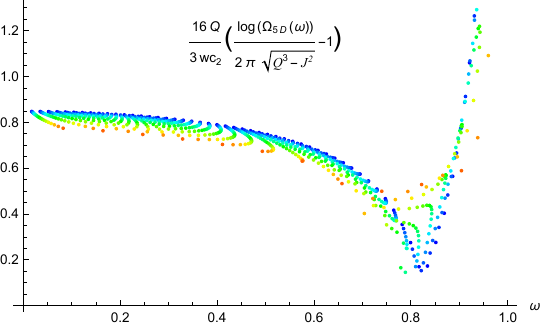} &
	\includegraphics[height=\lengg]{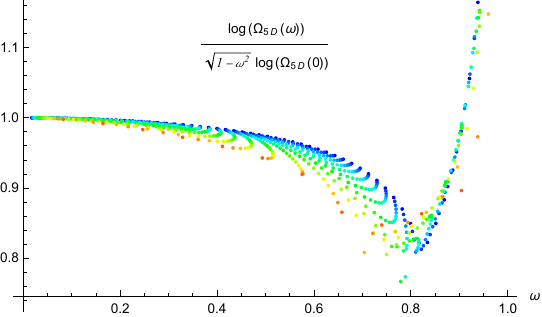} &
	\includegraphics[height=\lengg]{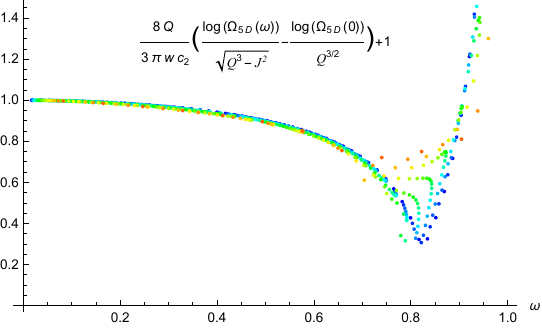} 
	\\  \hline
	\includegraphics[height=\lengg]{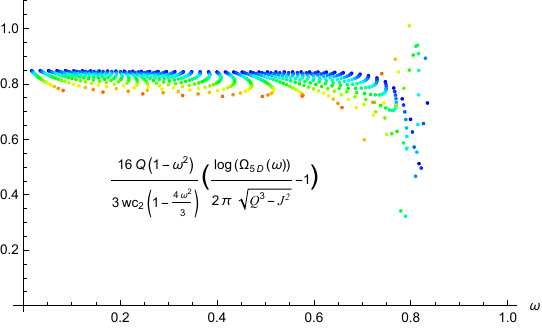} &
	\includegraphics[height=\lengg]{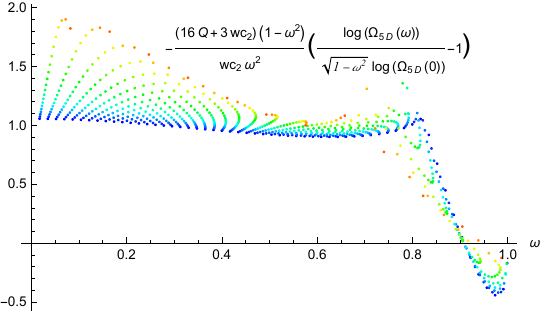} &
	\includegraphics[height=\lengg]{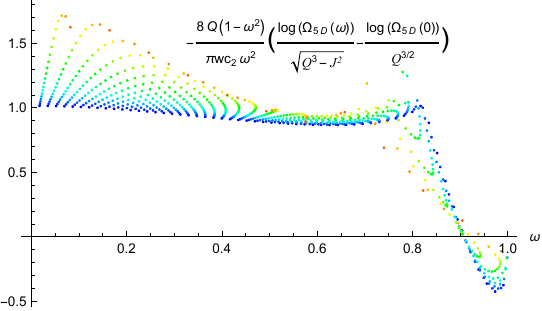} \\
	\hline
	\end{tabular}
	\end{center}
	\vspace{-0.3cm}
	\caption{Tests of the quantum correction to the entropy of rotating black holes. 
		Different columns correspond to three different ways to check whether the tested function represents the correction.
		Different rows correspond to four tested functions: 
		$g(\omega)=1+\frac13\omega^2$, $\frac{1}{1-\omega^2}$, 1 and $\frac{1-\frac43\omega^2}{1-\omega^2}$.	
		All functions are plotted for $X_{4,2}$ and $d$ ranging from 11 (orange) to 31 (blue). 
		Due to symmetry, we restrict to positive $\omega$. 
		The plateau in the last row with values close to 1 
		gives strong evidence for the validity of the prediction \eqref{Sexp4}. 
		\label{Fig-test}}
\end{figure}

Now we turn to spinning black holes in the regime where the ratio 
$\omega=J/Q^{3/2}$ is kept finite while $Q$ is scaled to infinity,
but $\omega$ remains smaller than the critical value given by the position of the kink. 
In this regime the logarithm of the 5D index is expected to reproduce the entropy \eqref{Sexp}.
Figure~\ref{Fig-Om5DN} leaves little doubt that it correctly reproduces the classical contribution.
The real challenge is in  the quantum correction.
To test the predictions existing in the literature encoded in the function $g(\omega)$ \eqref{Sexpq},
we computed three different combinations which, in the absence of further corrections to \eqref{Sexp}, for $|m|<\mkink$
should equal 1 provided $g(\omega)$ is the correct function (here 
$\Omega(\omega):=\Omega_{5D}(d,2(wd)^{3/2}\omega)$ for some fixed $d$):
\bea
1)&&\tfrac{16 Q}{3w c_2 g(\omega)}\(\tfrac{\log|\Omega(\omega)|}{2\pi\sqrt{Q^3-J^2}}-1\),
\nn\\
2) &&\tfrac{16Q+3wc_2}{3w c_2 (g(\omega)-1)}\(\tfrac{\log|\Omega(\omega)|}{\sqrt{1-\omega^2}\log|\Omega(0)|}-1\) 
\quad \mbox{ or }\
\tfrac{\log|\Omega(\omega)|}{\sqrt{1-\omega^2}\log|\Omega(0)|}
\mbox{ for } g=1, 
\\
3) &&\tfrac{8Q}{3\pi w c_2 (g(\omega)-1)}\(\tfrac{\log|\Omega(\omega)|}{\sqrt{Q^3-J^2}}-\tfrac{\log|\Omega(0)|}{Q^{3/2}}\)
\mbox{ or }\ 
\tfrac{8Q}{3\pi w c_2}\(\tfrac{\log|\Omega(\omega)|}{\sqrt{Q^3-J^2}}-\tfrac{\log|\Omega(0)|}{Q^{3/2}}\)+1
\mbox{ for } g=1. 
\nn
\eea 
The results are presented in Fig. \ref{Fig-test}. While the plots shown in the first three rows do not exhibit the expected behavior, the fourth row does manifest a plateau close to 1.
Together with the results of the previous subsection,
this provides a strong indication
that the last prediction \eqref{Sexp4} is indeed correct.\footnote{After this work appeared on arXiv, we learnt from Sameer Murthy that in the context of type IIB string theory compactified on $K3\times T^2$, the same correction  \eqref{Sexp4} to the macroscopic entropy was derived in \cite{Dabholkar:2010rm}, and matched with the microscopic counting of $\cN=4$ BPS dyons in the so-called type IIB Cardy limit. The result from \cite{Dabholkar:2010rm} however differs from \cite[(7.12)]{Cassani:2024tvk} at second order in $c_2$.
}

\begin{figure}[h]
\begin{center}
\includegraphics[height=5.2cm]{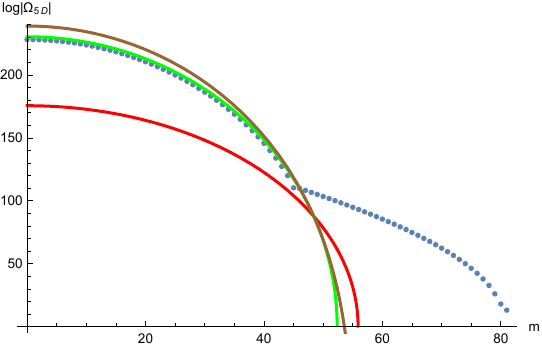} 
\hspace*{2mm} 
\includegraphics[height=5.2cm]{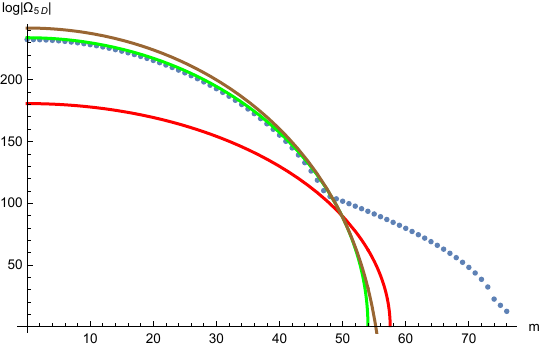}
\end{center}
\vspace{-0.4cm}
\caption{5D index (blue) vs. Bekenstein-Hawking-Wald entropy of spinning BMPV black holes, for $X_5$ (left) and $X_{4,2}$ (right). The red curve shows the classical Bekenstein-Hawking entropy \eqref{Sbh}, the brown curve includes the 4-derivative correction at linear order (\eqref{Sexp} with $g(\omega)$ given in \eqref{Sexp4}), and the green curve includes
it at non-linear order as in \eqref{Sbhfull}. 
\label{fig-SBHW}}
\end{figure}

Furthermore, let us recall (see footnote \ref{foot-expand})
that the prediction \eqref{Sexp4} was obtained by expanding a non-linear entropy formula \cite[(7.12)]{Cassani:2024tvk}, which in our notations takes the form
\be
S^{\rm bh} = 2\pi Q^{3/2} \sqrt{1+\frac{3w c_2}{8 Q} - 
\left(1+\frac{w c_2}{2 Q}\right) \omega^2} \, .
\label{Sbhfull}
\ee
Therefore, it is natural to verify whether \eqref{Sbhfull} gives a better approximation than its linearization \eqref{Sexp}.
To this end, in Fig. \ref{fig-SBHW}, we compare the logarithm of the (unnormalized) 5D index with three approximations to the entropy of rotating BMPV black holes: classical, quantum corrected at linear order in the parameter $c_2$, and \eqref{Sbhfull}.
The agreement of the last curve with the logarithm of the 5D index throughout the region before the kink is quite remarkable.
In particular, it significantly improves upon the linearized entropy formula, 
%with the higher curvature correction taken into account only
%at linear order, 
especially for small angular momenta.
This suggests that the result \eqref{Sbhfull} may hold beyond the leading order, 
even though  this was not expected {\it a priori} from its derivation.

Given the last observation, it is natural to reconsider the issue of the logarithmic correction discussed
in \S\ref{subsubsec-static} by replacing the expansion \eqref{expOm5} at vanishing angular momentum by 
\be
\log \Omega_{5D}(d,0) = b_0 d^{3/2} \sqrt{1+\frac{3 c_2}{8 d}}
+ \alpha \log d + \beta+ \cdots.
\label{expOm5new}
\ee
However, the analysis shows similar erratic results 
with the Richardson transform converging to $\alpha=-3/2$, the $E$-algorithm based on
$f_k(d)=d^{(1-k)/2}/\log(d)$ pointing towards $\alpha=0$, and other subtraction schemes not converging
to any definite value. 

\subsubsection{Black rings}
\label{subsubsec-ring}

\begin{figure}[h]
	\begin{center}
		\includegraphics[height=5.0cm]{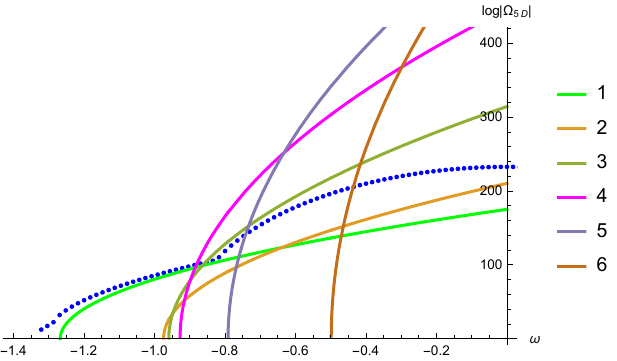} 
		\hspace*{2mm} 
		\includegraphics[height=5.0cm]{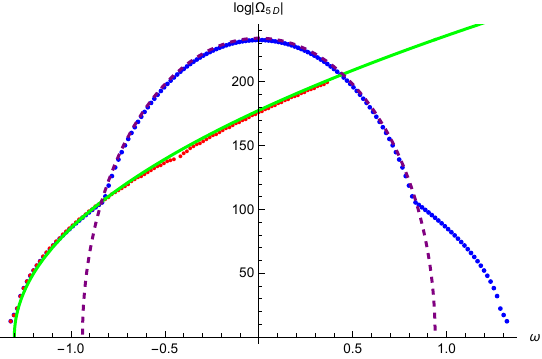}
	\end{center}
	\vspace{-0.7cm}
	\caption{Left: $\log|\Omega_{5D}|$ (blue) and the black ring entropy $S^{\rm br}$ for $r=1,\dots, 6$ with $n=m$. 
		Right: $\log|\Omega_{5D}|$ (blue), $\PT_1(d,n)$ (red) and $S^{\rm br}$ for $r=1$ (green) with $n=m+2$.
		The purple dashed curve is the quantum corrected black hole entropy \eqref{Sbhfull}.
		All graphs are made for $X_{4,2}$ and the maximal available degree $d=31$.
		\label{fig-BRentropy}}
\end{figure}

The last range of parameters to be considered is $|m|>\mkink(d)$
where the 5D index is expected to be dominated by black rings \cite{Halder:2023kza}.
Their entropy, including the leading correction from higher derivative terms, is given by \eqref{Sbrc2}. 
The problem however is that it depends on additional charge $r$, which should be either fixed or summed over.
To understand the dependence on this charge, in the left graph of Fig. \ref{fig-BRentropy} 
we plotted $S^{\rm br}$ for $r=1,\dots, 6$ and compared them against $\Omega_{5D}(d,m)$,
where $d$ is fixed to be the maximal available degree and the D0-charge $n$ is identified with the argument $m$
of the 5D index. (Due to symmetry of the index, we restricted the plot to negative $m$.)
Remarkably, the black ring entropy for $r=1$ is very close to the 5D index for $m<-\mkink$ and much smaller than it after the kink.
The curve for $r=2$ emerges only for $m$ slightly less than $-\mkink$ and remains smaller than the 5D index up to positive $m$.
In contrast, the curves for higher charges have a very steep slope and quickly overpass the index.
This suggests that black rings with $r>1$ should not contribute to the index, while its behavior for $|m|>\mkink$ 
is well captured by black rings with $r=1$.\footnote{Note that for such small
dipole charge, the horizon $S^1\times S^2$ is very anisotropic, with 
a Planck size two-sphere, which questions the validity of the classical black ring picture. Nonetheless, the fact that the macroscopic black ring entropy appears to accurately match the logarithm of the index
indicates that the interpretation of the dominant contribution to the index in terms of black rings can still be trusted.}

In fact, the match can be further improved if one takes into account that the D0-brane charge can differ 
by an integer shift from the angular momentum. To see how this shift comes about, we have to refer to some results 
presented in the next section. First, the relation \eqref{relOmPT} shows that at negative $m$ the 5D index is approximately equal
to the Pandharipande-Thomas invariant $\PT(d,m)$ with the argument $m$ shifted by $+2$.
Then in \S\ref{subsec-effPT} we demonstrate that for $m<-\mkink(d)$, PT invariants are very well approximated by 
MSW invariants (or D4-D2-D0 BPS indices) multiplied by some trivial factor resulting from the primitive wall-crossing formula.
But as discussed in \S\ref{subsubsec-br0}, the black ring entropy \eqref{Sbrc2} is precisely the entropy of four-dimensional 
D4-D2-D0 black holes encoded by MSW invariants.
These arguments imply that\footnote{This relation (with $m$ replaced by $-m$) has already been proposed in \cite[(5.17)]{Halder:2023kza}.
However, it was based on unjustified assumptions, and actually the shift by 2 does \textit{not} seem to follow
from the equations in {\rm loc. cit.}}
\be 
\Omega_{5D}(d,m)\approx \PT_1(d,m+2),
\qquad
m<-\mkink(d),
\label{OmPT1}
\ee 
where $\PT_1(d,m)$ is defined in \eqref{PT1}, and hence they suggest the following identification 
between the D0-brane charge and the angular momentum
\be 
n=m+2.
\ee 
In the right plot of Fig. \ref{fig-BRentropy} we took into account this shift 
and the resulting curves for 5D index, $\PT_1$ and $S^{\rm br}$ with $r=1$ indeed perfectly 
match up to the kink.\footnote{An apparent deviation of the red curve from the green one in the region 
	between the kink and $m=0$ is due to the vanishing of the prefactor in the definition of $\PT_1(d,m)$, 
	which does not have any physical significance since it is an artifact of the approximation.}

Finally, we analyze the position of the kink.
From Fig. \ref{Fig-Om5DN} it is obvious that $\mkink(d)\sim d^{3/2}$ 
because it always remains at finite values of $\omega=J/Q^{3/2}$.
A more precise dependence on $d$ can be determined by comparing the black hole and black ring entropies.
Indeed, the kink is nothing but the intersection point of these two curves (see the right plot in Fig. \ref{fig-BRentropy}).
Thus, we should equate $S^{\rm bh}$ given in \eqref{Sbhfull} and 
$S^{\rm br}$ given by \eqref{Sbrc2}, set $r=1$ and $n=m+2$ in the latter, 
and solve with respect to $m$, i.e.
\be 
S^{\rm bh}(d,-\mkink)=S^{\rm br}(1,d,-\mkink+2).
\ee 
Remarkably, this equation can be solved explicitly, which gives
\be 
\mkink=\frac{ 
\sqrt{ \frac{8 d^3}{\kappa} -  \(3 - \frac{4c_2}{\kappa}\)d^2 - 
3\(\kappa + \frac32\, c_2\)d 
+(\kappa+ c_2) \(\frac{5\kappa-c_2}{4}- 12 + 
\(\frac{\kappa+ c_2}{8} -6\)\frac{c_2}{d}\)}+\kappa+c_2}
{3 \(1 + \frac{c_2}{2 d}\)}\, .
\label{omkink}
\ee 
The result however looks somewhat complicated, so 
it might be advantageous to use instead 
its expansion at large $d$. Written in terms of the $\omega$ parameter, it is given by 
\be 
\omkink(d)= \frac{\mkink}{2(wd)^{3/2}} = 1-\frac{3\kappa+4c_2}{8d}
+\sqrt{\frac{\kappa}{8}}\,\frac{\kappa+c_2}{d^{3/2}}+\cO(d^{-2})\,.
\label{omkink-app} 
\ee 

\begin{figure}[h]
	\begin{center}
		\includegraphics[height=5.0cm]{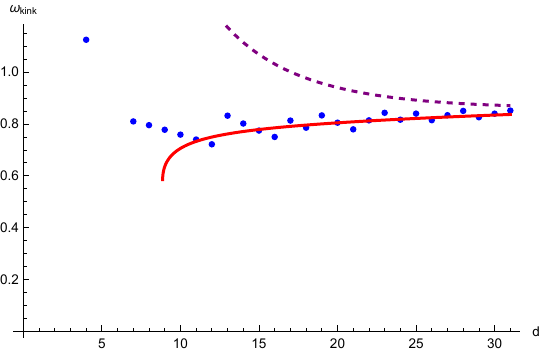} 
	\end{center}
	\vspace{-0.7cm}
	\caption{Positions of the kink $\omkink(d)$ for the 5D index for $X_{4,2}$, 
		its analytic expression \eqref{omkink} (red) and the approximation \eqref{omkink-app} (purple dashed).
		\label{fig-kink}}
\end{figure}

In Fig. \ref{fig-kink} we compare both the exact expression \eqref{omkink} and its approximation \eqref{omkink-app} against the actual positions 
of the kink for available degrees for $X_{4,2}$.
The former indeed fits the data very well, while the latter 
clearly approaches them. 
Note that \eqref{omkink-app} disagrees with the conjecture in \cite[\S 3.3]{Halder:2023kza}, 
according to which $\omkink(d)$ would 
asymptote to a universal value close to $0.9$.

\subsection{Approximate formula for GV invariants}
\label{subsec-apformula}

As was noticed in \S\ref{subsubsec-observ}, at fixed degree $d$ the GV invariants $\GV_d^{(g)}$ are well
approximated by a Gaussian function of the genus $g$. In this section, we build on this observation to propose
a formula that approximates GV invariants for all (sufficiently large) 
degrees and genera smaller than $\gkink(d)$. To fix the parameters of the Gaussian, 
we shall use the previous results, namely, the asymptotic formula at fixed genus 
\eqref{asympGVg} and the 5D index for static black holes \eqref{expOm5new}.

As a starting point, let us consider the following ansatz
\be
\GV_d^{(g)} \approx \cP(d,g) \(\xi d \log d\)^{2g-2} e^{C-A(g-B)^2},
\label{GV-gauss-new}
\ee
where the parameters $A$, $B$ and $C$ are functions of $d$,
$\xi$ is a constant and $\cP(d,g)$ is a polynomially bounded function of $(d,g)$.
Note that, up to the factor $\cP(d,g)$, the dependence on $g$ is Gaussian, 
while the power of $d\log d$ already appears in the fixed genus asymptotic formula \eqref{asympGVg}.
We want to fix the unknown parameters by imposing the two conditions, \eqref{asympGVg} and \eqref{expOm5new}.

To impose the first condition, we assume for the moment that $A$ vanishes at large $d$.
Then, to match the asymptotics \eqref{asympGVg}, 
one has to set
\be  
B=0, 
\qquad
C=2\pi \cV d,\qquad
\cP_0(d,g)= \frac{a_g}{d}\, \xi^{2-2g}, 
\label{valBC}
\ee 
where $\cP_0(d,g)$ denotes the asymptotics of $\cP(d,g)$ at large $d$ and fixed $g$.
Due to \eqref{aggrowth} and \eqref{ag-largeg}, in order to cancel the exponential dependence on the genus
we should take
\be 
\xi=\frac{\cV}{4\pi^2},
\qquad
\cP_0(d,g)=\frac{1}{d}\,\frac{|(2g-1)B_{2g}|}{ \Gamma(2g+1)} (2\pi)^{2g-2}
\approx \frac{g}{\pi^2 d}\, ,
\label{xi-P}
\ee 
where the last estimate is already accurate at small $g$.

To impose the second condition, we analyze the 5D index at zero angular momentum,
\be
\label{Om5D-m0}
\Omega_{5D}(d,0) =  \sum_{g=0}^{g_{\rm max}(d)} \frac{(2g+2)!}{((g+1)!)^2}\,\GV_d^{(g)}. 
\ee
First, using Stirling's formula $\Gamma(x+1)\approx \sqrt{2\pi x}(x/e)^x$, 
one finds that the binomial coefficient behaves at large $g$ as 
\be 
\frac{(2g+2)!}{((g+1)!)^2} \approx \frac{2^{2g+2}}{\sqrt{\pi g}}.
\label{appBinCoef}
\ee
Next, we approximate the sum in \eqref{Om5D-m0} by an integral.
As a result, after substitution of
\eqref{GV-gauss-new} with \eqref{valBC} and the approximation \eqref{appBinCoef}, the formula for the index takes the form 
\be 
\Omega_{5D}(d,0) \approx 16 \int_{g=0}^{\gmax(d)}\frac{\de g}{\sqrt{\pi g}}\,\cP(d,g)\,  
e^{2\pi \cV  d-Ag^2+(2g-2)\log (2\xi d\log d)} .
\label{Om5D-intap-new}
\ee
The approximations are justified because the integral is expected to be dominated by 
a saddle point $g_{\rm sp}(d)$ which scales as a positive power of $d$ and thus is supposed
 to be large.\footnote{In fact, the approximation \eqref{GV-gauss-new} is supposed 
 	to be true only for $g<\gkink$. Thus, writing \eqref{Om5D-intap-new}, 
 	we also assumed that the saddle point is before the kink and the integration 
 	in the interval $(\gkink,\gmax)$ does not contribute to the asymptotics.
\label{foot-sp-kink}}\;\footnote{One 
could worry that the boundary contribution
coming from the Castelnuovo bound grows faster than the saddle point contribution because 
at $g=g_{\rm max}(d)$ the binomial coefficient is of order $2^{d^2/\kappa}$,
whereas we argued previously that $\log |\GV_d^{(g)}|$ was bounded by $d^{3/2}$.
However, between the kink and the Castelnuovo bound the sign of GV invariants fluctuates 
and it appears that it leads to cancellations suppressing the growth. 
\label{foot-Cast-cancel}}
Indeed, the saddle point is found to be at
\be 
g_{\rm sp}=\frac{\cA^{-1}}{\log (2\xi d\log d)}\, ,
\label{gsp}
\ee 
and the integral evaluates to
\be 
\Omega_{5D}(d,0) \approx 
\frac{4\, \cP(d,g_{\rm sp})\, e^{2\pi \cV  d+\cA^{-1}}}{(\xi d\log d)^2(\log (2\xi d\log d))^{1/2}}\,  ,
\label{Om5D-intapp-new3}
\ee
where we introduced 
\be 
\cA=\frac{A}{(\log (2\xi d\log d))^{2}}\,.
\label{rescaleA}
\ee 
This condition allows us to fix the coefficient $\cA$ in terms of the 5D BPS index, 
namely,\footnote{Since the last term depends on $g_{\rm sp}$, which in turn depends on $\cA$, 
it looks like \eqref{invAOm} is an equation rather than an expression.
However, in the given approximation where $d$ is sufficiently large
one can replace $\cA^{-1}$ in \eqref{gsp} by \eqref{invAOm} with the last term omitted.}
\be 
\cA^{-1}
\approx \log \Omega_{5D}(d,0)-2\pi \cV d +2\log(\xi d\log d)
+\hf\, \log(\log (2\xi d\log d))-\log (4\cP(d,g_{\rm sp})).
\label{invAOm}
\ee 
Substituting \eqref{expOm5new}, we arrive at the following expression,
which is supposed to hold at sufficiently large $d$, 
\be
\begin{split} 
\cA^{-1}
\approx&\, b_0 d^{3/2} \sqrt{1+\frac{3 c_2}{8 d}}-2\pi \cV d 
 + (\alpha+2) \log d+2\log\log d
\\
&\,
+\hf\log\log\(\frac{\cV}{2\pi^2} \, d\log d\)
+\beta -\log \(\frac{64\pi^4}{\cV^2}\,\cP(d,g_{\rm sp})\).
\end{split}
\label{invA}
\ee 
This implies that at large $d$ one has 
\be 
A\sim \frac{(\log (d\log d))^{2}}{b_0 d^{3/2}}\, ,
\qquad 
g_{\rm sp}\sim\frac{b_0 d^{3/2}}{\log (d\log d)}\, ,
\label{asymp-Agsp}
\ee 
which confirms our approximations used above.

Combining the above results and substituting the value of $\xi$ from \eqref{xi-P}, 
we conclude that GV invariants can be approximated 
for $d$ sufficiently large by\footnote{This result certainly does not hold for very small $d$. 
	For the formula \eqref{GV-gauss-new-res} to make sense, 
one should at least ensure that $d$ is sufficiently large to have 
$\frac{\cV}{4\pi^2}\, d\log d>1$ and $\cA>0$.}
\be
\GV_d^{(g)} \approx \cP(d,g) \(\frac{\cV}{4\pi^2}\, d\log d\)^{2g-2} 
e^{2\pi \cV  d-\(\log\(\frac{\cV}{2\pi^2} \,d\log d\)\)^2\cA g^2},
\label{GV-gauss-new-res}
\ee
where $\cA$ is given by \eqref{invA} and the function $\cP(d,g)$ is restricted to have asymptotics \eqref{xi-P}.
Note that, upon replacing $\cP(d,g)$ by its asymptotics $\cP_0(d,g)$, 
the approximate formula differs from the asymptotics at fixed genus \eqref{asympGVg}
only by the Gaussian factor $e^{-(\log(2\xi d\log d))^2\cA g^2}$.
This shows that for that asymptotics to be valid, one should have $g^2\ll b_0 d^{3/2}$.

\begin{figure}[h]
\begin{center}
\includegraphics[height=5cm]{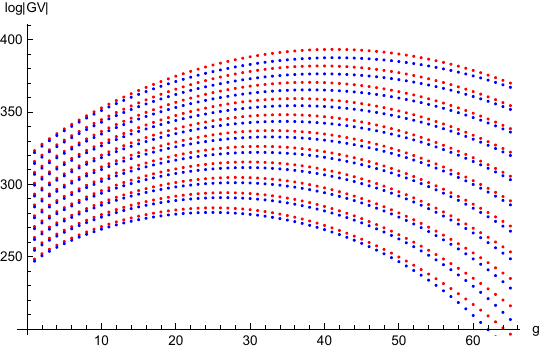} 
\hspace*{2mm} 
\includegraphics[height=5cm]{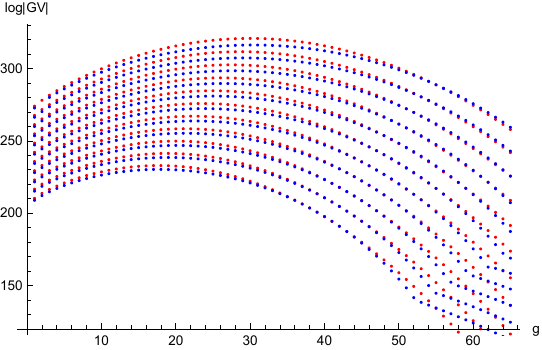}
\\
\includegraphics[height=5cm]{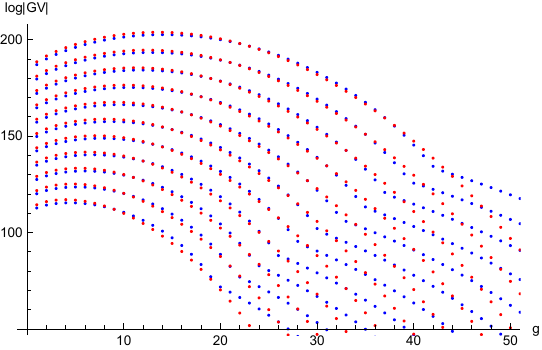} 
\hspace*{2mm} 
\includegraphics[height=5cm]{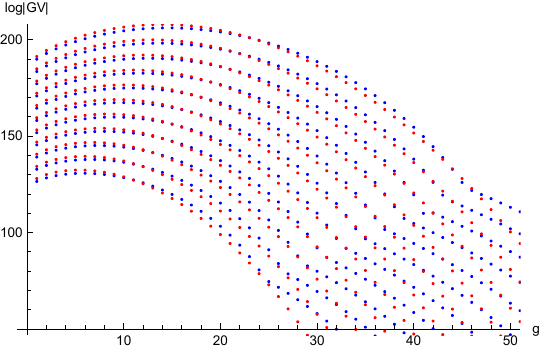}
\end{center}
\vspace{-0.5cm}
\caption{At the top: $\log|\GV_d^{(g)}|$ (blue) and its approximation 
	\eqref{GV-gauss-new-res} (red) for $X_5$ (left) and $X_{4,2}$ (right)
	 for degrees from 34 to 44. The approximation is computed using \eqref{invA} with $\alpha=\beta=0$. 
At the bottom: the same but now for degrees from 16 to 26 for $X_5$ (left) and 
from 21 to 31 for $X_{4,2}$ (right), and the approximation is computed using \eqref{invAOm}
with the actual values of the 5D index.
\label{fig-approxGV}}
\end{figure}

Unfortunately, our approach does not allow to fix the form of the function $\cP(d,g)$.
Replacing it by its large $d$ asymptotics $\cP_0(d,g)$ \eqref{xi-P} already appears to approximate  
very well the actual values of GV invariants as shown in Fig. \ref{fig-approxGV} 
where the plots in the top row are obtained using \eqref{invA} with $\alpha=\beta=0$. 
One might try to improve this agreement by tuning the factor $\cP(d,g)$,
but in fact 
it turns out that the deviation originates rather from 
the approximation of $\Omega_{5D}(d,0)$ used in \eqref{invA}.
This is demonstrated by the plots in the bottom row of Fig. \ref{fig-approxGV}
where the coefficient $\cA$ of the Gaussian approximation is evaluated instead using \eqref{invAOm}
with the actual values of the 5D index.
The plots show a rather impressive agreement, even for lower values of $d$ than 
those plotted at the top line. (We had to restrict to lower degrees because 
we need to know all GV invariants at given $d$ to be able to compute $\Omega_{5D}(d,0)$.)
Thus, to improve the validity of the 
approximation \eqref{GV-gauss-new-res}, 
one should either go to larger values of $d$ where the asymptotic formula 
\eqref{expOm5new} for the 5D index holds to greater accuracy, or improve this formula
by finding the values of $\alpha$ and $\beta$ and/or including additional corrections.

\subsubsection{Implications}

The result \eqref{GV-gauss-new-res} has several interesting implications 
which we now discuss.

\medskip 

\noindent
\textit{Maxima at fixed degree}
\smallskip 

\noindent 
The first, obvious implication of 
the approximate formula \eqref{GV-gauss-new-res} is that 
the position of the maximum at fixed degree $d$ and the value 
at the maximum are given by
\bea
\gtop(d)&=& \frac{\log(\xi d\log d)}{(\log(2\xi d\log d))^2}\,\cA^{-1} 
\sim \frac{b_0 d^{3/2}}{\log (d\log d)}\, ,
\label{gtop}
\\
\GV_d^{(\gtop(d))}&\approx & \frac{\cP(d,\gtop(d))}{ (\xi d\log d)^2}\,  
e^{2\pi \cV  d+\frac{(\log(\xi d\log d))^2}{(\log(2\xi d\log d))^2}\, \cA^{-1}}
\sim e^{b_0 d^{3/2}}.
\label{GVtop}
\eea
In particular, at large degree the leading asymptotics of $\gtop(d) $ 
is exactly the same as that of the saddle point $g_{\rm sp}$ \eqref{asymp-Agsp}.

\bigskip 
\noindent
\textit{Relation between two ensembles}
\smallskip 

\noindent 
Let us consider the index \eqref{newindex-def} counting states with 
both the total angular momentum $j_L$ and its projection $j_L^z$ fixed in the regime where $m\sim d^{3/2-\gamma}$.
At large $d$, upon approximating the sum by an integral and evaluating it by saddle point as in \eqref{Om5D-intap-new}, one easily gets 
\be 
\log|\tOm_{5D}(d,m)|\approx \log|\Omega_{5D}(d,m)|+\(\frac32-\gamma\)\log(d)-\log g_{\rm sp}.
\ee 
Since according to \eqref{asymp-Agsp} $g_{\rm sp}\sim d^{3/2}$,
one obtains 
\be 
\log|\tOm_{5D}(d,m)|\approx \log|\Omega_{5D}(d,m)|-\gamma \log d,
\ee 
which reproduces \eqref{alph-pred-diff} 
and thus provides a microscopic derivation of this relation.

\bigskip 
\noindent
\textit{Growth of topological free energies}
\smallskip 

\noindent 
In Appendix \ref{ap-growthFg}, we use the approximate formula \eqref{GV-gauss-new-res}
to estimate the sum over genera $F_d(\lambda)$ \eqref{defFd} appearing 
in the Gopakumar-Vafa representation of 
the topological free energy \eqref{multicover}, in the limit where the 
degree is large.
In particular, we confirm and make precise the expectation of \cite{Marino:2024tbx}
that at sufficiently large degree, $F_d(\lambda)$ grows quadratically in $d$ (assuming that the topological
string coupling $\lambda$ has a non-zero imaginary part).
We also point out the existence of a transient regime where the sum is 
dominated by a saddle point similar to \eqref{gsp} and behaves differently.

\section{Growth of PT invariants}
\label{sec-PT}

In this section, we study the asymptotics of the Pandharipande-Thomas invariants $\PT(d,m)$, 
which physically count bound states of a single anti D6-brane with $d$ D2-branes and $m$ D0-branes 
in a suitable large volume, large $B$-field limit (see e.g. \cite{Pandharipande:2011jz} 
for a review of their mathematical definition in terms of stable pairs). 
Since PT invariants vanish when $m$ is sufficiently large and negative,
the generating series 
\be
Z_{PT}(\y,\q)= 1 + \sum_{d>0, m} \PT(d,m)\, \y^d\, \q^m
\ee
is a well-defined Laurent series in $\q$, and Taylor series in $\y$ (a fugacity conjugate to the degree, 
not to be confused with the fugacity $y$ conjugate to angular momentum). 
Moreover, at each order in $\y$, the coefficient 
$\sum_m  \PT(d,m) \q^m$ is known to be a rational function of $\q$, invariant under $\q\mapsto 1/\q$.

PT invariants are related to GV invariants by the 
GV/PT correspondence~\cite{gw-dt}
\be
Z_{PT}(\y,\q) = Z_{PT}^{(0)} (\y,\q) \,
Z_{PT}^{(1)} (\y,\q),
\ee
where
\be
\begin{split}
\label{PTGV}
Z^{(0)}_{PT}(\y,\q)=&\,
\prod_{d=1}^\infty\prod_{k=1}^\infty \left(1-(-\q)^k \y^d\right)^{k \GV^{(0)}_d},
\\
Z^{(1)}_{PT}(\y,\q)=&
\prod_{d=1}^\infty\prod_{g=1}^{\gmax(d)}
\prod_{\ell=0}^{2g-2}
\left(1- (-\q)^{g-\ell-1} \y^d
\right)^{(-1)^{g+\ell} {\scriptsize \begin{pmatrix} 2g-2 \\ \ell \end{pmatrix}}
\GV^{(g)}_d} \, .
\end{split}
\ee
In particular, the Castelnuovo bound for GV invariants \eqref{gCast} is equivalent to the following lower bound for PT invariants
\be
\label{mCast}
m \geq m_{\rm min}(d) = -\left\lfloor\frac{d^2}{2\kappa} + \frac{d}{2}\right\rfloor .
\ee
Following \cite{Denef:2007vg}, we shall refer to the factors $Z_{PT}^{(0)}$ and $Z_{PT}^{(1)}$, 
coming from genus 0 and higher genus contributions, respectively, as "halo" and "core" states. 
For various purposes it is advantageous to rewrite the GV/PT relation as~\cite[\S $4 \tfrac12$]{Pandharipande:2011jz} 
\be
\label{PTGVpleth}
Z_{PT}(\y,\q)= \PE\[
\sum_{d=1}^\infty \sum_{g=0}^{\gmax(d)} (-1)^{g+1} \GV^{(g)}_d 
 \(1-\q\)^{2g-2} \q^{(1-g)} \y^d\]_{\q \to -\q},
\ee
where 
\be
\PE[f](\y,\q)=\exp\(\sum_{k=1}^\infty\frac{1}{k}\, f(\y^k,\q^k)\)
\label{defPE}
\ee
is the plethystic exponential. 

In particular, the representation \eqref{PTGVpleth} allows us to establish a relation to the 5D index $\Omega_{5D}(d,m)$. 
Indeed, using \eqref{Om5D}, one finds 
\be
\label{ZPTOm5}
Z_{PT}(\y,\q) = \PE\[ \frac{1}{(\q + \q^{-1} -2)^2} 
\sum_{d=1}^\infty \sum_{|m|\leq |m_{\rm min}(d)|} \Omega_{5D}(d,m) \, (-\q)^m \y^d \]_{\q \to -\q},
\ee
which can be interpreted as the partition function of a gas of 5D BPS black holes 
(and black rings) in a Taub-NUT geometry \cite{Dijkgraaf:2006um}. 
We note that a similar factor 
$(\q + \q^{-1} -2)^{-2}=(\sqrt{\q}-1/\sqrt{\q})^{-4}$ 
appears in the ratio 
\be
\frac{Z_{4D}^{\rm hair}}{Z_{5D}^{\rm hair}}=\frac{e^{2\pi\I\rho}}{\eta^{24}(\rho) 
(e^{\I \pi v}-e^{-\I \pi v})^4}
\ee
of the indices of the hair degrees of freedom living outside the horizon of a 4D BPS black hole 
at the center of Taub-NUT and a 5D BMPV black hole, respectively, computed in \cite{Banerjee:2009uk}
in a setup with twice as much supersymmetry (namely, type IIB compactified on $K3\times T^2$). 
In the present case, the BPS black holes cannot be lifted to black strings in 6 dimensions, 
so the $\rho$ dependent factor does not arise,
but they preserve the same number of supercharges, presumably leading
to the same term $(e^{\I \pi v}-e^{-\I \pi v})^4$ in the denominator. 
The plethystic exponential accounts for all possible ways of splitting the electric charge 
and angular momentum in an arbitrary number of 5D BPS states placed in the Taub-NUT geometry.

\begin{figure}[h]
	\begin{center}
		\includegraphics[height=5.3cm]{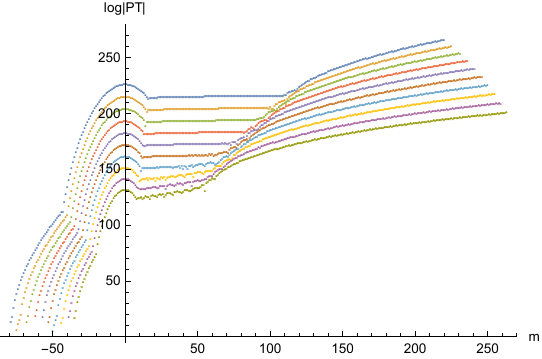} 
		\includegraphics[height=5.3cm]{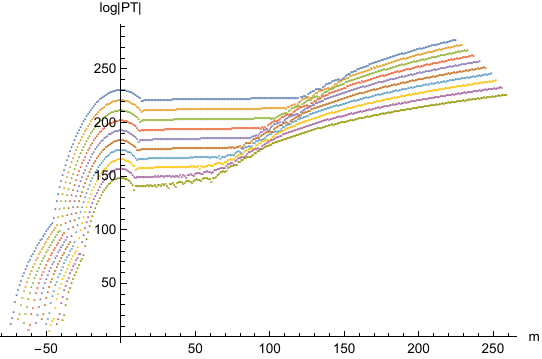} 
			\end{center}
\vspace{-0.7cm}
\caption{$\log|\PT(d,m)|$ as a function of $m$ for 10 maximal available degrees for $X_5$ (left) 
	and $X_{4,2}$ (right). Different degrees correspond to different colors. 
	Positive invariants are shown by dots, while negative ones by crosses. 
\label{Fig-PT}}
\end{figure}

Using the GV/PT relation, one can compute PT invariants up to the same degree 
for which GV invariants are known at all genera (see the 
column $d_{\rm mod}$ in Table \ref{tab_cydata} of 
hypergeometric models). In Fig. \ref{Fig-PT}, we plotted
PT invariants for fixed degree in the range $d_{\rm mod}-10\leq d \leq d_{\rm mod}$ 
for the models $X_5$ and $X_{4,2}$ with the highest 
value of $d_{\rm mod}$. 
There are several immediate observations which one can make from these plots:
\begin{itemize}
\item 
Similarly to GV invariants and the 5D index, PT invariants at fixed degree align along piecewise smooth curves.

\item
The left part of these curves, from the Castelnuovo bound
$m_{\rm min}(d)$ to small positive values of $m$, resembles similar curves 
formed by the 5D index, cf. Fig. \ref{Fig-Om5D}.
In particular, they exhibit the same kink at a negative value $m^{\PT}_{\rm kink}(d)<0$. 
As we show below, this similarity is not an accident, 
since PT invariants are indeed directly related to $\Omega_{5D}$ in this regime.

\item 
Surprisingly, for larger positive values of $m$, one finds two other phase transitions. 
At the first transition, the curve turns into a `plateau', and at the second transition, 
it starts growing logarithmically.
Note that it was necessary to reach at least $d=20$ to notice 
the plateau regime, which is hardly recognizable for small degrees.

\item 
All phase transitions are accompanied by the change in the behavior of the sign of PT invariants: 
after a few steps away from the Castelnuovo bound
where the sign is chaotic, the sign starts alternating with the parity of $m$; 
after the first kink it becomes positive,  
then turns negative at the plateau, and again starts alternating after the last transition.

\end{itemize}

\begin{figure}[h]
	\begin{center}
		\includegraphics[height=4.6cm]{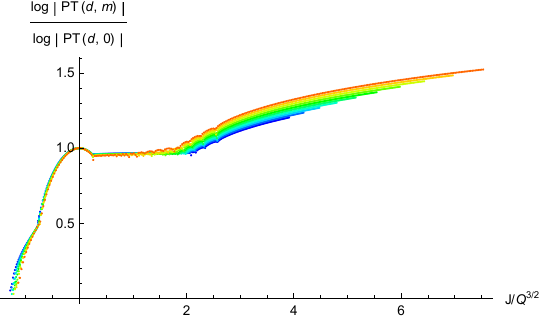} 
		\includegraphics[height=4.6cm]{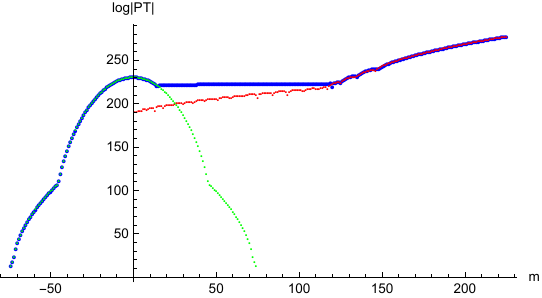} 
			\end{center}
\vspace{-0.7cm}
\caption{Left: the ratio $\frac{\log|\PT(d,m)|}{\log|PT(d,0)|}$ 
as a function of $J/Q^{3/2}$, for 10 maximal available degrees 
for $X_{4,2}$. The colors change from orange ($d=22$) to blue ($d=31$). 
Right: $\log|\PT(d,m)|$ for $X_{4,2}$ and $d=31$ calculated in three ways: 
using \eqref{PTGVpleth} (blue), restricting to $g=0$ in this formula and to $k=1$ in \eqref{defPE} (red), 
and restricting to $g\geq 1$ and $k=1$, respectively (green).
\label{Fig-PTN}}
\end{figure}

Below we analyze each of the phases of PT invariants in turn, 
and in each case derive an approximate expression for $\PT(d,m)$.
However, it is useful to comment on the two plots in Fig. \ref{Fig-PTN}.
(Since the plots for all one-parameter models look similar, we show them only for $X_{4,2}$.)
On the left, we plotted PT invariants normalized by their values at $m=0$ 
and parametrized by $J/Q^{3/2}$, as we did before for the 5D index.
One can note that the spread of the data for different degrees is not so big at the plateau, 
while it increases afterwards.
Then, on the right of this Figure, we plotted PT invariants for the maximal available degree 
together with the two approximations obtained by replacing in the MNOP formula \eqref{PTGVpleth} 
the plethystic exponential by the usual one and keeping only contributions either 
of halo states. i.e. $g=0$, or of 
core states, i.e. $g\geq 1$.\footnote{Note 
that the red and green curves do not span over all allowed values of $m$. 
This is because genus 0 GV invariants do not generate states with negative $m$ 
via the MNOP formula, while GV invariants of positive genus
generate only states with $|m|< \gmax(d)$.}
Remarkably, the red and green curves corresponding to these approximations
perfectly fit the blue one 
after and before the plateau, respectively. This implies that these approximations
generate the dominant contributions in these two regimes.

\subsection{Before the plateau}

\begin{figure}[h]
	\begin{center}
		\includegraphics[height=5.4cm]{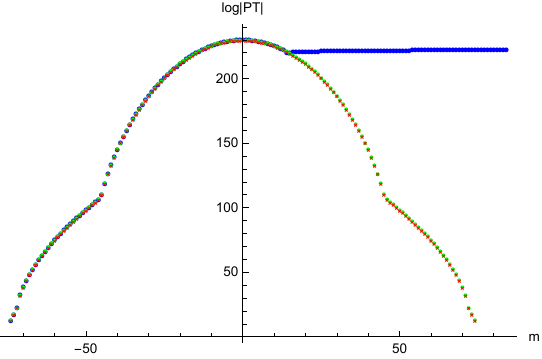} 
		\includegraphics[height=5.4cm]{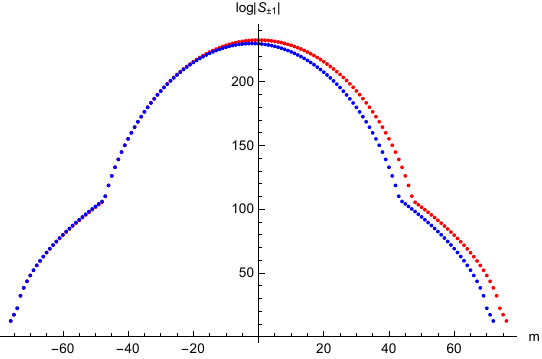} 
	\end{center}
\vspace{-0.7cm}
\caption{Left: $\log|\PT(d,m)|$ for $X_{4,2}$ and $d=31$ (blue), 
its approximation obtained by restricting to $k=1$ in \eqref{defPE} 
and to $g\geq 1$ in \eqref{PTGVpleth} (green), and one more approximation
obtained by additional restriction to $d=31$ in \eqref{PTGVpleth} (red crosses).
Right: $\log|\Si{1}(d,m+2)|$ (blue) and $\log|\Si{-1}(d,m)|$ (red) for $X_{4,2}$ and $d=31$. 
\label{Fig-PTg2}}
\end{figure}

As was just discussed, PT invariants before the plateau regime are very well approximated 
by replacing in the MNOP formula the plethystic exponential by the usual one and restricting to core states. 
In fact, the left plot in Fig. \ref{Fig-PTg2} shows that amazingly enough, in order to reproduce 
the invariants $\PT(d,m)$ in this regime to good accuracy, it suffices to keep 
only the contributions of GV invariants of the same degree $d$ inside the exponential, 
which is equivalent to dropping the exponential! 
Incorporating these restrictions in \eqref{PTGVpleth}, one immediately obtains 
the generating function for different D0-charges
\be 
Z^{(d)}_{\rm \geq 1}(\q)=\sum_{g\geq 1} \frac{(1+\q)^{2g-2}}{\q^{g-1}}\, \GV^{(g)}_d
\label{genleft}
\ee 
from which one finds
\be 
\PT(d,m)\approx \Si{1}(d,m),
\qquad
\Si{\ell}(d,m):=\sum_{g=|m|+\ell}^{g_{\rm max}(d)}
\binom{2(g-\ell)}{g-\ell-m}\GV^{(g)}_d.
\label{PTbefore}
\ee 
This approximate formula agrees with the \textit{exact} expression for PT invariants
found in \cite[(3.10)]{Alexandrov:2023ltz} near the Castelnuovo bound \eqref{mCast}, 
where the GV/PT relation becomes linear.
The remarkable fact is that it remains an extremely good approximation 
all the way up to the transition to the plateau happening at positive values of $m$.

Furthermore, the function $\Si{1}$ in the r.h.s. of \eqref{PTbefore} is closely
related to the 5D index \eqref{Om5D}, which in this notation can be written as 
\be 
\Omega_{5D}(d,m)=\Si{-1}(d,m).
\label{OmS}
\ee 
Thus, PT invariants differ from $\Omega_{5D}$ only by shifting $g\mapsto g-2$ in the binomial coefficient. 
It turns out that this shift can be traded for a similar shift in the argument $m$. 
Namely, with a very good accuracy, one has the following relations
\be  
S_{\ell_1}(d,m)\approx S_{\ell_2}(d\pm (\ell_2-\ell_1)), 
\qquad 
\pm m <0.
\label{relSS}
\ee
The point is that the two functions differ only by a ratio of two polynomials in $g$ of the same degree, which
is close to 1 for the values of $g$ providing the dominant contributions.
We demonstrated the relation for $\ell_1=-\ell_2=1$ in the right plot in Fig. \ref{Fig-PTg2}.
Combining \eqref{PTbefore}, \eqref{OmS} and \eqref{relSS}, we conclude that 
\be
\Omega_{5D}(d,m)\approx \PT(d,m+2),
\qquad 
m <0.
\label{relOmPT}
\ee 
This explains the identical form of the curves representing the 5D index and PT invariants before the plateau. 
This relation can also be easily understood from the identity \eqref{ZPTOm5}.
Indeed, it is enough, as above, to drop the plethystic exponential and in addition to pick up only the first term 
in the expansion of the prefactor, i.e. $\q^{-2}$. It is this factor that produces the shift in \eqref{relOmPT}. 

It is clear that the dominant contribution to $\PT(d,m)$ for $m$ after the kink but before the plateau 
comes from single centered black holes with D6-D4-D2-D0 charge given by $\(-1,0,-d+\frac{c_2}{24},m\)$ 
where the shift in the D2-charge is due to its non-trivial quantization.
On the other hand, before the kink the dominant contribution is generated by multi-centered black holes 
\cite{Denef:2000nb,Denef:2001xn,Bates:2003vx}.
Indeed, the relation \eqref{relOmPT} identifies this regime with the phase of the 5D index with large angular momentum
which is dominated by black rings and the latter descend by the 4D/5D relation to multi-centered black holes \cite{Gaiotto:2005xt}.
Moreover, due to observations in \S\ref{subsubsec-ring}, one can expect that the relevant black holes are bound states 
of an anti-D6-brane with unit D4-flux and D4-D2-D0 black hole with charge \eqref{D420charge} specialized to $r=1$.

\subsection{The unreasonable effectiveness of the PT/MSW relation}
\label{subsec-effPT}

In \cite[\S 4.2]{Alexandrov:2023zjb}, a relation between PT invariants and 
MSW invariants counting D4-D2-D0 bound states was established, 
by studying the chamber structure of the DT-invariants for an anti-D6-brane. 
Under a restrictive set of conditions on charges,
%the ratios $x=d/\kappa$ and $\alpha=-3m/(2d)$, 
which typically hold only close to the Castelnuovo bound \eqref{mCast},
it was shown that 
\be
\label{PT0}
\PT(d,m)=
\sum_{d',m'} (-1)^{\chi(d',m')} \chi(d',m') \PT(d',m') \, \Omega_{1,d-d'}(\hat q'_0),
\ee
where $\Omega_{r,d}(\hat q_0)$ is the rank $r$ MSW invariant with charges 
$d$ and $\hq_0$ defined as in \S\ref{subsubsec-br0},  
\be
\begin{split}
\chi(d',m') = m-m' + d+ d'-\chi_D, 
\qquad & \chi_{r\cD}=\frac{\kappa r^3}{6}+\frac{c_2 r}{12}\, ,
\\
\hat q'_0 = m'-m - \frac{(d'-d)^2}{2\kappa} - \frac{d+d'}{2} + \frac{\chi(\cD)}{24}\, ,
\qquad & \chi(r\cD)=\kappa r^3 + c_2 r,
\end{split}
\ee
and the sum over $(d',m')$ runs over a finite set satisfying certain inequalities, 
which ensure in particular that $0\leq d'<d$.
This provides a recursive way to compute PT invariants, assuming the 
rank 1 MSW invariants are known. Under even more restrictive conditions
(the so called optimal case, \cite[(4.26)]{Alexandrov:2023zjb}), 
one can show that the only term contributing on the r.h.s. of \eqref{PT0} is $(d',m')=(0,0)$.
Since $\PT(0,0)=1$, the expression for $\PT(d,m)$ reduces to 
\be
\label{PT1}
\PT_1(d,m) := (-1)^{m+d-\chi_\cD} (m+d-\chi_\cD) \, \Omega_{1,d}(\hat q_0).
\ee

\begin{figure}[h]
\begin{center}
\includegraphics[height=5cm]{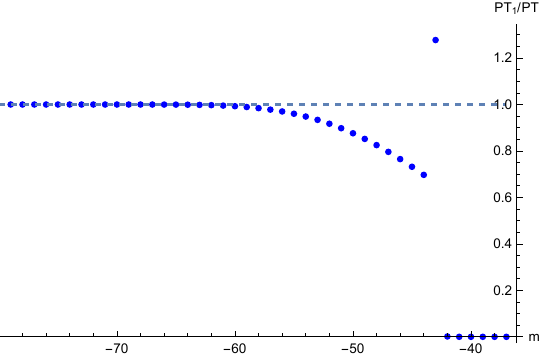}
\hspace{5mm}
\includegraphics[height=5cm]{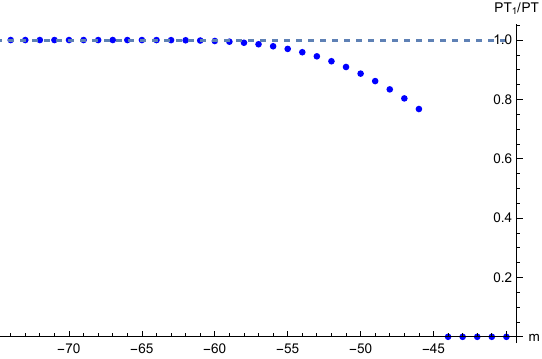}
\end{center}
\vspace{-0.7cm}
\caption{The ratio between the approximation $\PT_1(d,m)$ \eqref{PT1}
and the exact PT invariant $\PT(d,m)$ for $X_5$, $d=26$ (left) and $X_{4,2}$, $d=31$ (right).
\label{fig-PTNaive}}
\end{figure}

While this expression is exact very close to the Castelnuovo bound,
it becomes only an approximation when the conditions for its validity are violated 
and more terms in \eqref{PT0} start contributing.
Moreover, if one further relaxes the conditions for \eqref{PT0}, PT invariants get contributions from higher rank MSW invariants. 
In particular, an exact formula including rank 2 contributions and 
the corresponding conditions on charges have been derived in \cite{Alexandrov:2023ltz}.  
Nonetheless, it turns out that the approximation \eqref{PT1} works amazingly well all the way
from the Castelnuovo bound up to the kink! 
In Fig. \ref{fig-PTNaive}, we plot the ratio $\frac{\PT_1(d,m)}{\PT(d,m)}$ 
(without taking the logarithm!) as a function of $m$. 
Remarkably, the ratio is indistinguishable from unity over the full range 
before the kink despite the fact that the stringent conditions ensuring that 
$\PT(d,m)=\PT_1(d,m)$ are never obeyed in this range. 
If this property can be quantified, it might be useful in fixing holomorphic ambiguities 
in the direct integration method, at least at a heuristic level.

\subsection{Plateau}

The plateau phase of PT invariants is the only regime where 
it is important to take into account the plethystic exponential in the MNOP formula.
Indeed, acting on the genus zero contribution in \eqref{PTGVpleth},
it produces factors $1/(1-(-\q)^k)^2$. 
Let us take $k=2$. After expansion in $\y$, the factor $1/(1+\q)^2$ 
is cancelled by contributions of other genera, whereas the factor $1/(1-\q)^2$ remains. 
If it multiplies a dominant contribution around $m=0$, 
it generates a plateau because effectively it extends it to all $m>0$. 
Since the contribution discussed in the previous subsection starts rapidly decreasing at positive $m$, 
eventually the plateau starts dominating, precisely as we observe on the plots of $\log| \PT(d,m)|$.

\begin{figure}[h]
	\begin{center}
		\includegraphics[height=5.2cm]{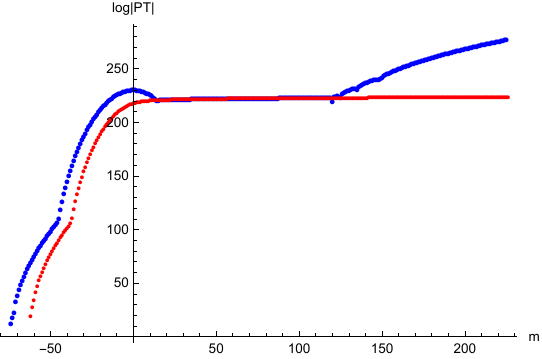}\hspace{2mm}
		\includegraphics[height=5.2cm]{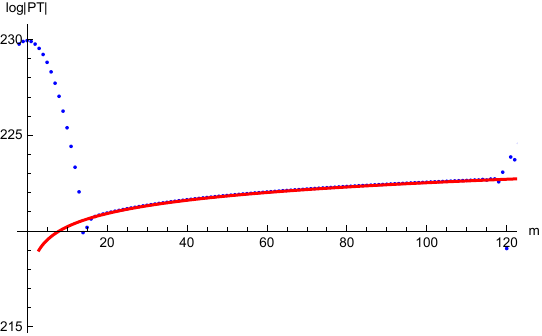}
	\end{center}
	\vspace{-0.7cm}
	\caption{The plots of $\log| \PT(d,m)|$ for $X_{4,2}$, $d=31$ (blue) and its approximations (red). 
		On the left, the approximation is obtained from MNOP formula where
		one takes into account only the contribution of $\GV^{(g)}_{29}$ invariants with $g\geq 2$ 
		with the plethystic exponential replaced by the usual one 
		and $\GV^{(0)}_{1}$ with the plethystic exponential replaced by its $k=2$ contribution. 
		On the right, it is the logarithm of \eqref{genplateau2}. 
		\label{fig-approxPT-X42-4}}
\end{figure}

As shown in Fig. \ref{fig-approxPT-X42-4}, to capture PT invariants at the plateau, 
it is sufficient to restrict to the $k=2$ contribution of the plethystic exponential considered above 
and in addition to pick up only the term with $d=1$. 
Then the dominant contribution, which is multiplied by this genus 0 contribution, 
should be taken at degree $d-2$. 
This approximation leads to the following generating function\footnote{We dropped also the contribution 
	from $g=1$ which is negligible compared to higher genera.}
\be 
Z^{(d)}_{\rm pl}(\q)=
-\hf\, \GV^{(0)}_1 \sum_{g\geq 2} \frac{(1+\q)^{2g-4}}{\q^{g-3}(1-\q)^2}\, \GV^{(g)}_{d-2}.
\ee 
Expanding in $\q$, one finds
\be
\PT(d,m)\approx 
 -\frac{1}{2}\, \GV^{(0)}_1\sum_{k=1-g_{d}}^{\min(m,g_{d})-1} (m-k) \Si{2}(d-2,k) ,
\label{genplateau}
\ee
where $g_{d}=g_{\rm max}(d-2)-1$ and $\Si{2}$ was defined in \eqref{PTbefore}.
For $m\geq g_d$, one can further simplify this result by exchanging the sums over $k$ in \eqref{genplateau} 
and the sum over $g$ in the definition of $\Si{2}$,
and noticing that the term linear in $k$ cancels due to symmetry reasons. In the remaining term
the sum over $k$ can be explicitly evaluated and gives 
\be
\PT(d,m)\approx 
 -\frac{m}{32}\GV^{(0)}_1\sum_{g=2}^{g_d+1} 2^{2g}\GV^{(g)}_{d-2}.
\label{genplateau2}
\ee
We see that PT invariants are not really constant at the plateau, 
but have a slow linear growth (which becomes logarithmic in our graphs as we plot $\log|\PT|$).
The right plot in Fig. \ref{fig-approxPT-X42-4} demonstrates that this approximation works
extremely well all the way through the plateau.

An interesting question is what are the macroscopic objects which provide the dominant contribution 
in this regime and have the entropy given by the logarithm of \eqref{genplateau2}?
However, this question goes beyond the scope of this paper.

\subsection{After the plateau}

As was already noticed in Fig. \ref{Fig-PTN}, PT invariants after 
the last critical point are perfectly captured by halo states, 
i.e. by the contribution of GV invariants of genus 0 only, 
and ignoring the plethystic nature of the exponential in \eqref{PTGVpleth},
i.e. keeping only the $k=1$ term in \eqref{defPE}. 
With these restrictions, the MNOP formula gives rise to the following generating function at degree $d$
\be 
\begin{split}
Z^{(d)}_0(\q)=&\, \sum_{\sum_{i=1}^n k_id_i=d \atop d_1<\cdots < d_n,\ k_i,d_i>0}
\(\prod_{i=1}^n \frac{(\GV^{(0)}_{d_i})^{k_i}}{k_i!}\) \(\frac{\q}{(1+\q)^2}\)^{\sum_{i=1}^n k_i}
\\
=&\, \sum_{m=d}^\infty \q^m \sum_{\sum_{i=1}^n k_id_i=d \atop d_1<\cdots < d_n}
\(\prod_{i=1}^n \frac{(\GV^{(0)}_{d_i})^{k_i}}{k_i!}\) C\(\sum_{i=1}^n k_i,m\)  ,
\end{split}
\label{genfunhalo}
\ee 
where
\be
C(d,m)= \frac{(-1)^{m-d} (m+d-1)! }{(2d-1)!(m-d)!}\, .
\ee

\begin{figure}[h]
	\begin{center}
		\includegraphics[height=5cm]{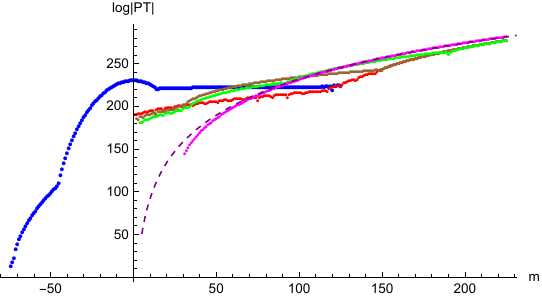} 
	\end{center}
	\vspace{-0.7cm}
	\caption{The plots of $\log| \PT(d,m)|$ for $X_{4,2}$, $d=31$ and its approximations:
		the actual PT invariants (blue), 
		the contribution of genus 0 GV invariants with the plethystic exponential replaced by the usual one (red),
		additional restriction to degrees $d\leq 20$ (brown), $d\leq 10$ (green)
		and $d=1$ (magenta). Besides, the purple dashed curve is the approximation \eqref{asymphalo}. 
		\label{fig-approxPT-X42}}
\end{figure}

Moreover, the asymptotics of the generating function is determined by the contribution of only one GV invariant, 
namely $\GV^{(0)}_1$! 
Indeed, due to Stirling's formula, at large $m$ the coefficient $C(d,m)$ scales as
\be 
C(d,m)\sim \frac{(-1)^{m-d}}{(2d-1)!}\, m^{2d-1}.
\ee 
Therefore, different terms in \eqref{genfunhalo} scale as $m^{2\sum_{i=1}^n k_i-1}$, 
so that the dominant contribution is the one that maximizes the sum of $k_i$'s. 
It is clear that it corresponds to $d_1=1$, $k_1=d$.
Picking up this contribution in \eqref{genfunhalo}, one finds that
the asymptotics of PT invariants at large $m$ is given by
\be
\PT(d,m) \sim (-1)^{m-d}\,\frac{(\GV^{(0)}_1)^d}{d!(2d-1)!}\, m^{2d-1}\, , \qquad m\gg 1.
\label{asymphalo}
\ee 
Thus, as was anticipated,  $\log|\PT|$ grows logarithmically with  coefficient  $2d-1$. 
We illustrate contributions to $\log|\PT|$ from different degrees in Fig. \ref{fig-approxPT-X42}.
Unfortunately, as in the case of the plateau, the macroscopic interpretation
of the objects generating the entropy \eqref{asymphalo} remains puzzling for us.

\section{Growth of DT invariants}
\label{sec-DT}

The Donaldson-Thomas invariants $\DT(d,m)$ count
bound states of a single D6-brane with $d$ D2-branes and $m$ D0-branes.
They satisfy the same Castelnuovo bound \eqref{mCast} as PT invariants.
Indeed, their partition functions are related by~\cite{Pandharipande:2007kc,toda2010curve,bridgeland2011hall}
\be
Z_{DT} (\y,\q)= M(-\q)^{\chi_X} \, Z_{PT} (\y,\q),
\label{eqn:DTPTrelation}
\ee
where $M(\q)=\prod_{k>0}(1-\q^k)^{-k}$ is the MacMahon function. 

\begin{figure}[h]
\begin{center}
\includegraphics[height=5.4cm]{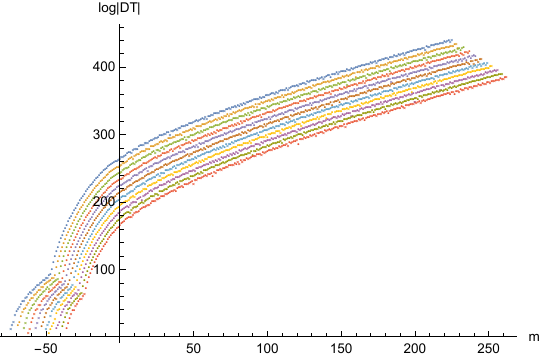}
\hspace{0.2mm}
\includegraphics[height=5.4cm]{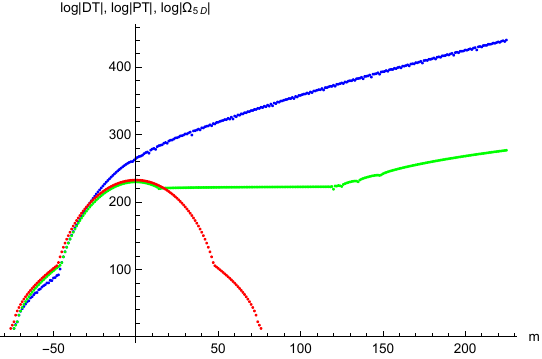}
\end{center}
\vspace{-0.7cm}
\caption{Left: $\log|\DT(d,m)|$ as a function of $m$ for 10 maximal available degrees for $X_{4,2}$. 
	Different degrees correspond to different colors. Positive invariants are shown by dots, 
	while negative ones by crosses. 
Right: $\log|\DT(d,m)|$ (blue), $\log|\PT(d,m)|$ (green) and $\log|\Omega_{5D}|$ (red) 
for the maximal available degree $d=31$.
\label{fig-DT}}
\end{figure}

We computed and plotted DT invariants for $X_{4,2}$ in Fig. \ref{fig-DT}
where we also contrasted them against PT invariants and the 5D index $\Omega_{5D}$. 
Here we summarize our observations:
\begin{itemize} 
\item 
The curves interpolating between the values at fixed degree $d$ 
appear to exhibit only
one phase transition corresponding to the usual kink at negative $m$.
However, inspecting the behavior of the sign, one finds that 
there is another transition around $m=0$.

\item 
In contrast to PT invariants, DT invariants do not exhibit any plateau.
Instead, at positive $m$ they immediately start growing even faster 
than PT invariants at large $m$.

\item
Another difference is that in both regions where DT invariants take positive and negative values 
(before the kink and at positive $m$), the sign is not strictly alternating
but exhibits a more complicated pattern.

\item
At negative $m$, PT invariants provide a better approximation for $\Omega_{5D}$
than DT invariants.

\end{itemize} 

In fact, the behavior of DT invariants at large $m$ is dominated 
by the MacMahon factor $M(-\q)^{\chi_X}$ in \eqref{eqn:DTPTrelation}.  
The mechanism here is very similar to the one producing the plateau for PT invariants: 
the MacMahon function multiplies the contribution
of positive genera determining the PT invariants around $m=0$ and extends them to all positive $m$.
Thus, an approximate formula for  $\DT(d,m)$ in this region can be obtained 
by multiplying the MacMahon function and the generating function \eqref{genleft}.
Then expanding $M(-\q)^{\chi}Z^{(d)}_{\rm \geq 1}(\q)$ in $\q$ (for $\chi=\chi_X$)
and denoting the Taylor coefficients of $M(\q)^{\chi}$ by $\omega_{\chi}(n)$, one obtains (cf. \eqref{genplateau})
\be 
\DT(d,m)\approx 
\sum_{k= 1-g_{\rm max}(d)}^{\min(m,g_{\rm max}(d)-1)} (-1)^{m-k}\omega_\chi(m-k)
\Si{1}(d,k)\,,
\label{genDTap}
\ee 
where it is important to take into account the sign $(-1)^{m-k}$
arising due to the minus in the argument of the MacMahon function.

To get a more explicit expression, we need the asymptotics of the coefficients $\omega_{\chi}(n)$.
If the power $\chi$ was positive, the asymptotics would be easily determined 
from Meinardus' theorem (see e.g. \cite[\S 6.2]{Benvenuti:2006qr}). 
For one-parameter CY threefolds, however, $\chi$ is negative and 
the derivation of the asymptotics is more delicate. We relegate it to Appendix \ref{ap-McMahon} 
and the final result can be found in \eqref{assMcMneg}. Up to constant and cosine factors, 
it is given by 
\be
\begin{split} 
\omega_{\chi}(n) \sim & 
\exp \left[ \frac32  \left( \frac{ |\chi| \zeta(3) n^2}{4} \right)^{1/3}   + \frac{|\chi|-24}{36} \log n \right].
\end{split}
\label{asomchi}
\ee
Note that the cosine factor is the one responsible for a complicated sign behavior of DT invariants at positive $m$.
Substituting \eqref{asomchi} into \eqref{genDTap}, we conclude that 
\be 
\log|\DT(d,m)|\sim \frac32  \left( \frac{ |\chi_X| \zeta(3) }{4} \right)^{1/3} m^{2/3},
\qquad
m\gg 1.
\label{asDT}
\ee 

\begin{figure}[h]
	\begin{center}
		\includegraphics[height=5.3cm]{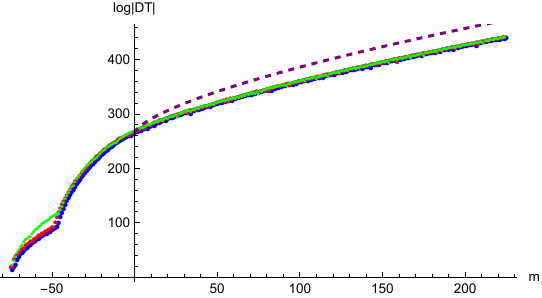}
	\end{center}
	\vspace{-0.7cm}
	\caption{The plots of $\log| \DT(d,m)|$ for $X_{4,2}$, $d=31$ (blue) and its approximations given by \eqref{genDTap}
		with either exact Fourier coefficients $\omega_\chi$ (red) or replaced by their asymptotics \eqref{assMcMneg} (green).
		Besides, the purple dashed curve is the approximation \eqref{asDT} multiplied by $\DT(d,0)$. 
		\label{fig-DTas}}
\end{figure}

While the discussion above assumes that the degree $d$ is fixed, a natural
question is to study the behavior of $\DT(d,m)$ as $m^2/d^3$ is held fixed.
Indeed, the authors of \cite{Denef:2007vg} argued that, in order 
for the OSV conjecture in its original form~\cite{Ooguri:2004zv} to have a chance to hold true, the logarithm of $|\DT(\lambda^2 d,\lambda^3 m)|$
should grow like $\lambda^2$, implying miraculous cancellations.  Unfortunately, despite our 
attempts, we have not been able to improve on the results presented in~\cite{Huang:2007sb}.  
The status of this important question therefore remains inconclusive.

\section{Discussion}
\label{sec-disc}

In this paper we analyzed enumerative invariants of one-parameter hypergeometric CY threefolds 
using both analytical and numerical methods.
Here we summarize our main results and observations.
\begin{enumerate}
\item 
We derived the asymptotic form \eqref{allgenusleading} of GW and GV invariants at large degree and fixed genus,
improving on earlier results by fixing the overall coefficient $a_g$.
This formula follows from the singular behavior of the free energy $\cF^{(g)}$ near the conifold point, 
and was extensively checked to high genus.
% A similar asymptotics can be derived for local CY threefolds \cite{akp-to-appear}, but it differs in the overall factor $a_g$, 
% which no longer depends on the conifold period $\cV$, 
% as a result of the D0-brane central charge being constant in such models.

\item 
The values of all invariants such as GV, PT, DT and $\Omega_{5D}$, at fixed but large degree $d$ can be interpolated by piece-wise smooth curves 
and exhibit several phase transitions.

\item 
One phase transition is common to all invariants. For the 5D index, it is interpreted as a transition between phases
where the leading contribution to the index is due to either black holes or black rings, as anticipated in \cite{Halder:2023kza}.
For the 4D indices (namely, PT and DT invariants), it is a transition between single-centered black holes and two-centered bound states.
We found that the kink takes place at the value \eqref{omkink}, 
correcting the value conjectured in \cite{Halder:2023kza}.

\item 
Perhaps unexpectedly, PT invariants exhibit two additional transitions at positive values of the D0-brane charge $m$.
Using the GV/PT relation \eqref{PTGVpleth}, we obtained approximate expressions for PT invariants in terms of GV invariants 
in all regimes, which can be found in \eqref{PTbefore}, \eqref{genplateau2} and \eqref{asymphalo}. DT invariants
appear to exhibit only one additional transition, with asymptotics at large D0-brane charge given in \eqref{asDT}, 
dominated by D6-D0 bound states.

\item 
We also observed that in the phase dominated by two-centered black holes, PT invariants 
are accurately captured by MSW invariants with unit D4-brane charge or, more precisely, by the combination \eqref{PT1}.
While this relation is exact very close to the Castelnuovo bound, it is an unexpected surprise that it continues to hold with high accuracy
throughout this phase.

\item 
We noticed a general relation between $\Omega_{5D}$ and PT invariants \eqref{ZPTOm5}, which is a direct consequence 
of the MNOP formula and can be seen as a manifestation of the 4D/5D lift \cite{Gaiotto:2005gf}.
Furthermore, in the regime where PT invariants are dominated by black holes (before the plateau),
this relation is approximated by the equality \eqref{relOmPT} identifying the two indices up to a shift in the D0-brane charge.

\item 
We have checked the behavior of the 5D index against supergravity predictions for the entropy of five-dimensional 
black holes and black rings. In the static case, we have found a perfect agreement with the classical 
entropy and the first higher derivative correction.
In the rotating case, our results strongly support the validity of the prediction \eqref{Sbhfull} from \cite{Cassani:2024tvk}, over earlier computations in the literature.
Unfortunately, our results for the coefficient of the logarithmic correction are inconclusive and suggest,
if anything, that this coefficient may vanish, rather than taking the value predicted by \cite{Sen:2012cj}.

\item 
In the black ring regime, the 5D index is in perfect 
agreement with the quantum corrected black ring entropy \eqref{Sbrc2}
with dipole charge $r$ set to 1. By the 4D/5D lift this entropy 
is the same as the entropy of D4-D2-D0 black hole with unit D4-brane charge
and therefore the agreement appears to be a consequence of the approximate relation \eqref{OmPT1}.
On the other hand, we find evidence that black rings with $r>1$ should not contribute to the index.

\item 
Returning to GV invariants, we proposed an approximate formula \eqref{GV-gauss-new-res}, 
which extends the fixed genus asymptotics \eqref{allgenusleading} to arbitrary genus up to $\gkink(d)$.
It is based on the observation that GV invariants at fixed degree are well approximated by a Gaussian.
The main parameter $\cA$ in the formula is fixed in terms of 
the asymptotics 
of the 5D index at zero angular momentum \eqref{invAOm}.

\item 
Finally, assuming that this approximate formula for GV invariants holds, we proved a conjecture by Mari\~no  
about the asymptotics of topological free energies at fixed but large degree (and complex string coupling), 
and pointed out an intriguing transient behavior at intermediate degree.  
\end{enumerate}

These results also raise various issues which we discuss next.
\begin{enumerate}

\item[a.] 
An important problem is to resolve the tension between
the supergravity prediction for the coefficient of the logarithmic correction to the entropy 
of static BMPV black holes and our numerical results. One possibility is that this tension 
disappears upon going to higher degree, which requires 
computing GV invariants to higher genus.
If the deviations persist, one would have to reconsider the supergravity analysis, 
perhaps taking into account specificities of theories with eight supercharges.

\item[b.]
Another important but subtle issue is the relation between the 5D index, computed in asymptotically flat spacetime 
and in a specific ensemble, and the actual microcanonical degrees
of freedom localized near the horizon, which can be meaningfully compared with the black hole entropy. To our knowledge, this issue has only been studied 
so far in string vacua with higher supersymmetry \cite{Banerjee:2009uk,Jatkar:2009yd}
and it deserves further attention. 

\item[c.] Alternatively, one may try to compute the 5D index macroscopically 
by identifying the complex saddle points which dominate the gravitational path integral, along the lines of 
\cite{Iliesiu:2022kny,Anupam:2023yns,Cassani:2024kjn,Boruch:2025qdq,Cassani:2025iix,Boruch:2025sie}.
In this respect, it is interesting to note that the authors of \cite{Boruch:2025sie} have found that black rings generally dominate over black holes in the canonical ensemble. It would be interesting to understand how this is consistent with the microcanonical picture which emerges from our work. 

\item[d.] 
Our results strongly indicate that black rings with dipole charge $r>1$ do not contribute to the 5D index,
but what prevents them from contributing remains unclear to us. 
It becomes even more puzzling taking into account the relation \eqref{relOmPT} to PT invariants, which in turn
can be expressed through MSW invariants via wall-crossing. 
The latter have exactly the same entropy \eqref{Sbrc2} as black rings and MSW invariants with D4-brane charge $r>1$
are known to contribute to PT invariants in certain 
regimes~\cite{Alexandrov:2023ltz}.

\item[e.] 
The existence of new phases where (the logarithm of) PT and DT invariants behave differently from 
the black hole entropy came as a surprise. It also raised the natural question: what is 
the macroscopic interpretation of these phases?
A hint to the answer can be gleaned from the approximate expressions for the invariants, 
which we found from the MNOP formula,
and, in particular, which GV invariants they involve.
For example, it is striking that, due to \eqref{asymphalo}, at large D0-brane charge the asymptotics of PT invariants 
is determined by a single GV invariant $\GV^{(0)}_1$!

\item[f.] 
There are several features of the curves interpolating invariants at fixed degree that call for an explanation.
For example, why are GV invariants at fixed degree
well approximated by a Gaussian? 
What is the mathematical origin of the slope discontinuity 
at $g=\gkink(d)$? What is the origin of the kink for PT invariants and DT invariants?

\item[g.] 
An open problem is to fix the prefactor $\cP(d,g)$
in the approximate formula \eqref{GV-gauss-new-res} for GV invariants. While one obtains satisfactory results by
replacing it by its large degree limit $\cP_0(d,g)$ \eqref{xi-P}, one may hope
to extend the regime of validity of the formula by including
further corrections.
Another interesting related problem is to find a similar approximation for GV invariants at $g>\gkink(d)$.

\item[h.] Despite our attempts, we have not been able to determine the growth of DT invariants $\DT(\lambda^2 d,\lambda^3 m)$ at large $\lambda$, a question which was identified
in \cite{Denef:2007vg} as key to the  validity of the OSV conjecture~\cite{Ooguri:2004zv},
and studied in \cite{Huang:2007sb} inconclusively. It remains an important question
for future studies.

\item[i.] 	
In this work we restricted ourselves to one-parameter hypergeometric models. 
Clearly, it would be interesting to extend this analysis to the one-parameter CY threefolds 
in the database \cite{cycluster}, for which the knowledge of higher genus GV invariants is unfortunately scarce. 
In particular, it would be interesting to extend the fixed genus asymptotic formula \eqref{allgenusleading}
and the approximation \eqref{GV-gauss-new-res}
to examples with hyperconifold singularities. 

\item[j.]
Eventually, one may hope that the qualitative features identified in our study, 
and in particular the unreasonable effectiveness of the simplest PT/MSW relation noted in \S\ref{subsec-effPT}, 
could help in fixing the holomorphic ambiguities in the direct integration method, and allow us to 
push it to higher genus than currently accessible.

\item[k.]
Finally, it is interesting to extend our analysis to the case of local non-compact CY threefolds, where 5D BPS states can be counted at fixed angular momenta $(j_L,j_R)$, without tracing over $j_R$. 
In \cite{akp-to-appear}  we shall establish an analogue of the asymptotic
formula \eqref{allgenusleading} for refined GV invariants at fixed degree, and analyze the
profile of the 5D index and refined DT/PT invariants. 
% Our preliminary analysis show that the 5D index still grows exponentially fast and exhibits a kink at finite $m$, despite the fact that no kink appears in the profile of GV invariants.

\end{enumerate}

\bigskip 

\noindent
{\bf Acknowledgments.}
We are grateful to Iosif Bena, Tom Bridgeland, Andrea Brini, Davide Cassani, Alejandra Castro,  Jie Gu, Sameer Murthy, Eric Pichon-Pharabod, Alejandro Ruip\'erez, Thorsten Schimannek, Ashoke Sen and Amitabh Virmani for
 valuable discussions and correspondence.
 The research of BP was  supported by the Agence Nationale de la Recherche under contract number ANR-21-CE31-0021. AK research is   
 supported by the Leverhulme Grant LIP LIP-2023-007 ``Quantum Geometry and Arithmetic.''    
 \textit{For the purpose of Open Access, a CC-BY public copyright license has been applied by the authors 
 	to the present document and will be applied to all subsequent versions up
 	to the Author Accepted Manuscript arising from this submission. }

\appendix

\section{Revisiting the Schwinger one-loop computation}
\label{sec_Schwinger}

The topological free energy $F^{(g)}$ on $X$ computes the coefficient of the term $\cR_-^2 F_-^{2g-2}$   
in the effective action of type IIA string compactified on a CY threefold $X$~\cite{Antoniadis:1993ze},
where $\cR_-$ and $F_-$ are the   anti-selfdual parts of the Riemann tensor and graviphoton fields strength, respectively.
 According to~\cite{Gopakumar:1998ii,Gopakumar:1998jq}, in the large volume limit the whole tower of couplings can be computed by a Schwinger-type one-loop integral describing contributions of BPS states propagating in a constant anti-selfdual graviphoton field-strength. This results in the formula\footnote{Since the original work of Gopakumar and Vafa, this computation has been revisited on several occasions, including \cite{Dedushenko:2014nya,Blumenhagen:2023tev,Hattab:2024ewk,Hattab:2024ssg}. 
 While some steps in the derivation remain unclear, we ignore them since our aim is mainly to fix the normalization of the conifold gap condition.}
\be
\cR_-^2 \sum_{g=0}^\infty (\I g_s F_-)^{2 g-2} \cF_{\rm inst}^{(g)}(t)=\cR^2_-\int_{\epsilon\rightarrow 0}^\infty \frac{\d s}{s}\,
\frac{  \Tr' 
\left[ (-1)^{2(j^z_L+j^z_R)} e^{2 \I s j_L^z F_- }  e^{- s Z/g_s} \right]
}{\(2\sin\frac{s F_- }{2}\)^2}\, ,
\ee      
where the trace runs over all light BPS multiplets\footnote{The contributions 
	of the descendants in a fixed multiplet, which are included in the trace in \eqref{Om5Dindex}, 
are expected to cancel against the insertions of the two anti-self dual gravitons.}, 
namely D2-D0 bound states with charge $d\geq 0$, $m\in \IZ$. Here $Z= 2 \pi   \left( t d + m \right)$ is the central charge and $j_L^z, j_R^z$ are the spins of the 5D BPS states whose reduction on the M-theory circle produces the D2-D0 bound states. Note that the anti-self-dual graviphoton background couples only to the left spin. Decomposing into powers of the representation $[1/2]\oplus [0] \oplus [0]$ as in \eqref{eqNjLjRGV}, 
we get
\be
\sum_{g=0}^\infty \hat \lambda^{2g-2} \cF^{(g)}_{\rm inst} (t)
= \displaystyle{\sum_{d= 0}^\infty \sum_{m\in\IZ} \sum_{g=0}^\infty \GV^{(g)}_d \int_{\epsilon\rightarrow 0}^\infty 
\frac{\d s}{s} \(2 \sin\tfrac{s}{2}\)^{2g-2} e^{2 \pi \I  \frac{s}{\hat \lambda} (t d +m)}}\, ,
\ee 
where we denoted  $\hat \lambda=\I g_s  F_-$ and rescaled $s\rightarrow s/F_-$. Poisson resumming over $m$ (or equivalently using $\sum_{m\in\IZ} e^{ 2\pi i  m \frac{s}{\lambda}}=\sum_{k\in\IZ}\delta (\frac{s}{\lambda}-k)$), one arrives
at~\cite{Gopakumar:1998ii,Gopakumar:1998jq}  
\be
\cF_{\rm inst}(\hat \lambda,t)=\displaystyle{\sum_{d=0}^\infty\sum_{g=0}^\infty \sum_{k=1}^\infty \frac{\GV^{(g)}_d}{k} \left(2 \sin\tfrac{k\hat \lambda }{2}\right)^{2g-2} e^{2 \pi \I  k t d}. }
\ee
Equating with the Gromov-Witten expansion~\eqref{defFg} and identifying $\hat \lambda=(2 \pi \I )^{3/2} \lambda$,
we recover the  multi covering formula  \eqref{multicover}.
Setting $d=0$, identifying $\GV_0^{(0)}=-\chi_X/2$ and using the identity 
\be
\frac{1}{4 x  (\sin(x/2))^2}
=\frac{1}{x^3}+\sum_{g=1} x^{2g-3}  \, \frac{(-1)^{g+1} B_{2g}}{2 g (2g-2)!}
\ee
together with the fact that 
the analytic continuation of the Zeta function yields $\zeta(1-n)=-B_{n}/n$, 
we also obtain the constant term in \eqref{Fgpol}. 

Let us now work away from the large volume limit, and consider the contribution of a single BPS state of central charge $t_c$. Expanding in powers
of  $\hat \lambda$ and integrating term by term, we find
\be 
\begin{split}
\int_{\epsilon\rightarrow 0}^\infty  \frac{\d s}{s}  \frac{e^{ 2\pi \I s t_c}}{4 \bigl(\sin(\hat \lambda s/2)\bigr)^2}= & -
\frac{1}{\lambda^2}\left( \frac{t_c^2}{2 (2 \pi \I)} \log(t_c)+ \ldots\right)- \left( \frac{1}{12} \log(t_c)+  \ldots \right) \\
& + \sum_{g>1}\lambda^{2g-2} 
\left(\frac{(-1)^{g-1}(2 \pi \I )^{g-1} B_{2g}}{ 2g(2g-2)}\, \frac{1}{ t_c^{2g-2}}  + {\cal O}(t^0_c)\right) ,
\end{split}
\label{eq:schwingerloop2} 
\ee         
where we made the same identification $\hat \lambda=(2 \pi \I )^{3/2} \lambda$ and,
as in \cite{Huang:2006hq,Huang:2010kf}, regularized the first two terms by isolating and dropping contributions proportional to $\epsilon^{-k}$ with $k=1,2$. 
In particular, identifying the BPS state with the one becoming massless at the conifold point, we find the behavior of the topological free energy in the conifold frame, 
\be
\label{Fgap2}
\cF_c^{(0)}(t_c ) \stackrel{t_c\to 0}{\sim}
\begin{cases}  
-  \frac{t_c^2 }{2 (2 \pi \I)} \log(t_c),
& g=0, 
\\
-\frac{1}{12} \log t_c, 
& g=1, 
\\
 \frac{(-1)^{g-1}( 2\pi \I)^{g-1} B_{2g}}{ 2g(2g-2)
}\, \frac{1}{t_c^{2g-2}} + \cO(t_c^0),
\quad 
&  g\geq 2.
\end{cases}
\ee
In the case of a hyperconifold, the massless state carries multiplicity $m_c$, so that the r.h.s. of \eqref{Fgap2}
should be multiplied by a factor of $m_c$, recovering the prescription in \eqref{Fgap}.

\section{Richardson transform and $E$-algorithm}
\label{sec_Richardson}

In this appendix, we recall some useful techniques for accelerating the convergence of 
slowly convergent series $\{s_d\}$, following \cite{Huang:2007sb,Caliceti:2007ra}. 
In practical applications, $s_d$ often arises as the partial sum $s_d=\sum_{k=1}^d r_k$ 
of some series $\sum_{k\geq 1} r_k$, but this need not be the case. 
For us $d$ will typically be the degree of GV invariants or the black hole charge, 
and $s_d$ is known only up to some maximal number $d_{\rm max}$. 
The challenge is then to guess  $\lim_{d\to \infty} s_d$, assuming it exists, from the
knowledge of a finite number of terms.

The simplest technique, known as the Richardson transform, assumes that the series 
$s_d$ converges to a finite value $s_\infty$, such that for $d\to\infty$ the error is suppressed by 
inverse powers of $d$, 
\be
\label{flimd}
s_d = s_\infty + \sum_{k\geq 1} \frac{a_k}{d^k}\, .
\ee
For any $N\geq 1$, the depth $N$ Richardson transform is a new series 
\be
\label{defRichardN} 
\(R_N [s]\)_d := \sum_{k=0}^N
(-1)^{k+N} \frac{(d+k)^N}{k! (N-k)!}\,
s_{d+k},
\ee
which converges to the same value $s_\infty$, but such that the error is now suppressed by $1/d^{N+1}$.
Indeed, it is easy to check that the linear combination in \eqref{defRichardN} annihilates all terms
with $k\leq N$ in \eqref{flimd}. Note that $R_0 [s]= s$, and that the composition $R_{N_1}[ R_{N_2} [s]]$ 
achieves the same acceleration as $R_{N_1+N_2}[s]$ (i.e. both eliminate all terms with $k\leq N_1+N_2$), 
although the two prescriptions yield different remainders.

Similarly, one may consider a series $s_d$, which converges to a finite value $s_\infty$, 
but with an error suppressed by inverse powers of $\log d$, 
\be
\label{flimdlog}
s_d = s_\infty + \sum_{k\geq 1} \frac{b_k}{(\log d)^k}\, .
\ee
The analogue of \eqref{defRichardN} is the depth $N$ logarithmic Richardson transform given by
\be 
\(L_N [s]\)_d 
:=\sum_{m=0}^N  \frac{(\log(d+m))^N}{\prod_{l\neq m}\log\left(\frac{d+m}{d+l}\right)}\,
s_{d+m}.
\label{logrichardson}
\ee
One may check that it
converges to the same value $s_\infty$, but such that the error is now suppressed by $1/(\log d)^{N+1}$. 

More generally, consider a series $s_d$, 
which converges to a finite value $s_\infty$, with an error given by 
\be
\label{flimdgen}
s_d = s_\infty + \sum_{k\geq 1} f_k(d),
\ee
where $f_k(d)$ are functions of $d$, ordered such that $f_{k+1}(d)/f_k(d)\to 0$ as $d\to \infty$.
For any $N$, the $E$-algorithm \cite{haavie1979generalized,brezinski1980general}, 
(see \cite[\S 2.2.7]{Caliceti:2007ra}) provides a linear combination of $s_{d+k}$ with $0\leq k\leq N$
which annihilates all $f_k(d)$ with $1\leq k\leq N$, such that the error is suppressed by $f_{N+1}(d)$. 
While the linear combination can be found in principle by solving a linear problem, 
it is more efficient to construct it through the recursion
\be
\begin{split}
\(E_{N}[s]\)_d = &\, \(E_{N-1}[s]\)_d
- \frac{ \(E_{N-1}[s]\)_{d+1} - \(E_{N-1}[s]\)_d}{f_{N-1,k}(d+1)-f_{N-1,N}(d)} 
\, f_{N-1,N}(d), 
\\
f_{N,k}(d) = &\, f_{N-1,k}(d) 
- \frac{ f_{N-1,k}(d+1)  - f_{N-1,k}(d)}{f_{N-1,N}(d+1)-f_{N-1,N}(d)}
\, f_{N-1,N}(d),
\qquad 
k\geq N+1,
\end{split}
\ee
initialized with
\be
\(E_0[s]\)_d =  s_d, 
\qquad 
f_{0,k}(d) = f_k(d).
\ee
This implements a sequence of projections, such that $f_{N,k}(d)=0$ for $k\leq N$ 
and $\(E_{N}[s]\)_d$ is the desired linear combination.

\section{Growth of topological free energies at fixed degree}
\label{ap-growthFg}

In this appendix, we investigate the convergence of the infinite sum over degrees 
in the topological free energy obtained by the sum over genera of $\cF^{(g)}$ \eqref{defFg}, 
assuming that this sum is performed first. Using \eqref{multicover}, the free energy can then be written as 
\be
\label{eqFGVFd}
\cF^{\rm GV}(t,\lambda)=
\sum_{g=0}^\infty \lambda^{2g-2}\cF^{(g)}_{\rm pol}(t) + 
\sum_{d\geq 1} F_d(\hat \lambda) q^d,
\ee 
where $\hat \lambda=(2 \pi \I )^{3/2} \lambda$ as in Appendix \ref{sec_Schwinger} and
\be 
F_d(\lambda)=\sum_{g=0}^{\gmax(d)} \sum_{k|d} \frac{\GV^{(g)}_{d/k}}{k}\(2\sin\frac{k\lambda}{2}\)^{2g-2}.
\label{defFd}
\ee 
Since the sum in \eqref{defFd} is obviously finite due to the Castelnuovo bound, 
the convergence of the sum over degrees in \eqref{eqFGVFd}, if it held, 
would give a non-perturbative definition of the topological free energy. 
Unfortunately, it was observed in \cite{Hatsuda:2013oxa,Marino:2015nla}
that for local $\IP^2$, the summand \eqref{defFd} grows super-exponentially, namely 
\be 
\log |F_d|\sim d^2,
\qquad 
d\gg 1,
\label{expectFd}
\ee 
as long as $\lambda$ has a non-vanishing (not too small) imaginary part.\footnote{For $\lambda$ real 
	and positive, $\log |F_d|$ appears to grow linearly, so the series would have finite radius of convergence, 
	but it would exhibit a dense set of poles on the $\lambda$ axis, which are problematic.}
It was argued in \cite{Marino:2024tbx} that this behavior is in fact generic, 
and is a direct consequence of the Castelnuovo bound. Here
we shall verify this prediction for compact hypergeometric CY threefolds, and  
show that in fact,
\be 
\log |F_d|\approx \hf\(\frac{d^2}{\kappa}+d\)|\Im \lambda| 
\label{leadCast}
\ee 
for large enough $d$, with interesting transient behavior at finite $d$.
As a result, the series \eqref{eqFGVFd} is asymptotic, with an ambiguity of order 
$e^{-2\pi^2\kappa (\Im t)^2/|\Im\lambda|}$, obtained by truncating the sum at the optimal value 
$d\sim 2\pi\kappa\, \frac{\Im t}{|\Im \lambda|}$, where the general term is smallest in absolute value.
This is similar to the  divergence  of the sum over D-instanton corrections 
to the hypermultiplet moduli space (which, in the case of compact CY threefolds, is of the form
$\sum_{Q} e^{Q^2-2|Q|/\lambda}$), 
signaling the existence of Neveu-Schwarz five-brane 
instanton corrections of order $e^{-1/\lambda^2}$~\cite{Pioline:2009ia}. 
In the present case, the ambiguity is recognized as a Euclidean D3-brane instanton, but note that the expansion is at large volume rather than 
small string coupling.

\medskip

To extract the large $d$ asymptotics of $F_d$, note that the main contribution to this limit
is expected to arise from large $g$. Therefore, we can approximate the sine function 
in \eqref{defFd} by the leading exponential and, as in \eqref{Om5D-intap-new}, 
the sum over $g$ by an integral, which gives
\be 
F_d(\lambda) \approx \sum_{k|d}\frac{1}{k}\int_{g=0}^{g_{\rm max}(d/k)}\de g\, 
\GV^{(g)}_{d/k}\, e^{(g-1)\(k|\Im \lambda|-\I s \phi \)} ,
\label{Fdint}
\ee
where $s=\sign(\Im \lambda)$ and $\phi=2\pi\mbox{Frac}\(\frac{k \Re \lambda}{2\pi}-\frac12\)$ 
is the fractional part belonging to $(-\pi,\pi]$.
We have chosen this definition so as to minimize the imaginary contribution to the exponential. 
Evaluating the integral by saddle point, one gets two competing contributions.

The first contribution comes from the upper limit of integration and is 
equal to\footnote{Note that the existence of this contribution is in stark 
	contrast with the absence of a similar contribution for the integral \eqref{Om5D-intap-new} 
	representing the 5D index (see footnote \ref{foot-Cast-cancel}). 
The difference is that no cancellations are expected to happen in \eqref{Fdint} near the Castelnuovo bound.} 
\be 
\GV^{(\gmax(d/k))}_{d/k}\, e^{(\gmax(d/k)-1)\(k|\Im \lambda|-\I s  \phi \)},
\label{contrCast}
\ee 
where $\gmax(d)$ is defined in \eqref{gCast}. Since this function is quadratic 
and GV invariants at the Castelnuovo bound are of order 1, 
it is clear that the contribution \eqref{contrCast} is maximized at $k=1$, leading
precisely to the behavior \eqref{leadCast}. This provides
a precise version of the argument presented in \cite{Marino:2024tbx}.

However, one should also analyze the competing contribution coming from the saddle point
of the integral \eqref{Fdint}, which we denote $g_*$. Assuming that\footnote{This assumption 
	can be justified as follows. Since for a given degree $g_*<\gmax$, 
the saddle point contribution can win against \eqref{contrCast} only if $\GV^{(g_*)}_{d/k}$ 
is sufficiently large. However, at least in the available range of degrees, 
one always has $\gtop<\gkink$, i.e. the maximum for a fixed $d$ is achieved before the kink. 
Thus, only the saddle points with $g_*<\gkink$ have a chance to govern the asymptotics at large $d$. 
Of course, for moderate $d$ this argument does not apply and a saddle with $g_*>\gkink$
can turn out to be dominating. \label{foot-saddle}} $g_*<\gkink(d)$, 
one can apply the approximate formula \eqref{GV-gauss-new-res}.
Then the saddle point is found to be at
\be  
g_*= \frac{\cA^{-1}}{\(\log\(\frac{\cV}{2\pi^2}\, d_k\log d_k\)\)^2}\, 
\(\log\(\frac{\cV}{2\pi^2}\, d_k\log d_k\)+\hf\,(k|\Im\lambda|-\I s \phi)\),
\label{gsp-Fd}
\ee 
where $d_k=d/k$,
and the corresponding contribution is given by
\be 
\frac{\sqrt{\pi}\cP(g_*,d_k) \cA^{-1/2}}{\(\log\(\frac{\cV}{2\pi^2}\, d_k\log d_k\)\)^2}\,
\exp\[2\pi \cV d_k+\cA^{-1}
\(\tfrac{\log\(\frac{\cV}{4\pi^2} \,d_k\log d_k\)+\hf\,(k|\Im\lambda|-s \I \phi)}
{\log\(\frac{\cV}{2\pi^2} \,d_k\log d_k\)}\)^2
-k|\Im\lambda|+ \I s \phi\],
\label{contr-saddle} 
\ee 
where $\cA^{-1}$ can be found either in \eqref{invAOm} or in \eqref{invA} and should be evaluated at $d_k$. 

\begin{figure}[h]
	\begin{center}
	\def\leng{3.4cm}
    \def\lengg{3.32cm}
		\includegraphics[height=\leng]{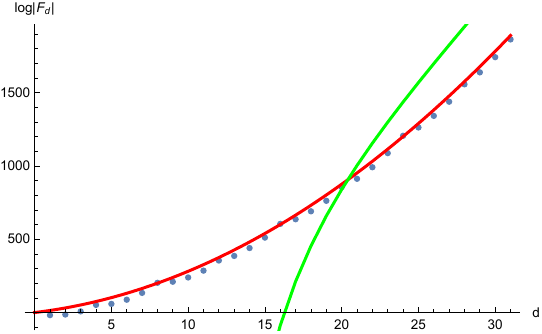} 
		\includegraphics[height=\leng]{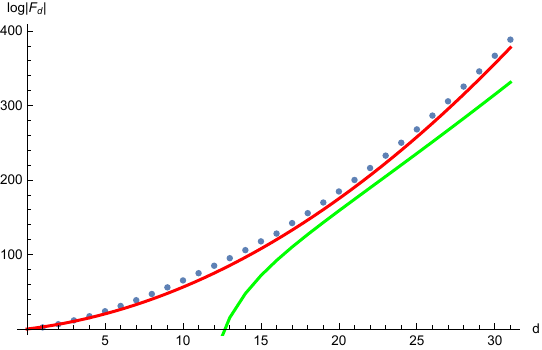} 
		\includegraphics[height=\leng]{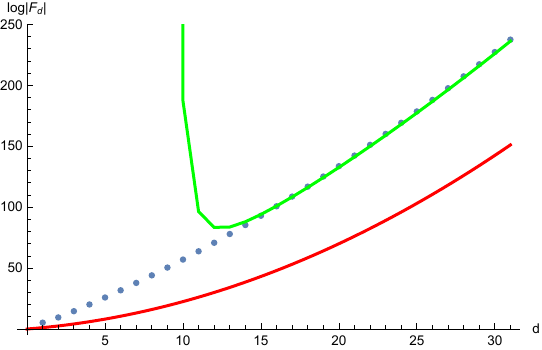}\\ 
		\;\includegraphics[height=\lengg]{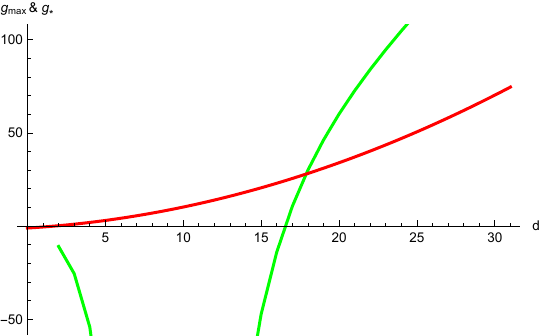} 
		\;\includegraphics[height=\lengg]{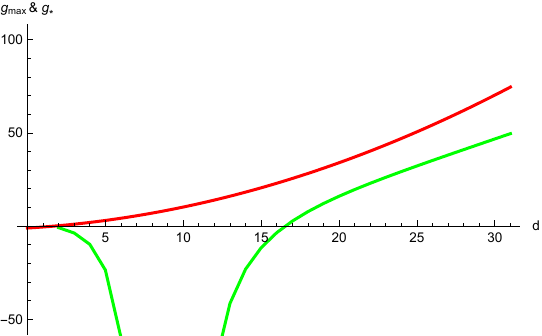} 
		\includegraphics[height=\lengg]{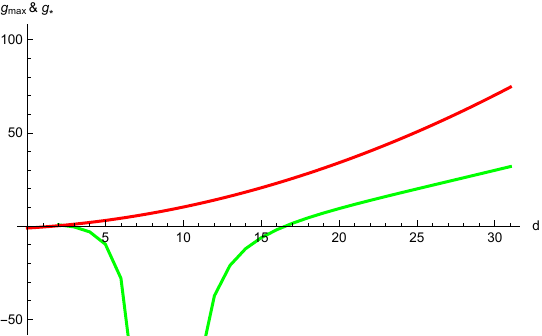} 
	\end{center}
\vspace{-0.7cm}
\caption{All graphs are made for $X_{4,2}$ and $\lambda=5+25\I$ (left), $5+5\I$ (middle) and $5+2\I$ (right). 
Top: $\log|F_d|$ (blue) and two contributions, \eqref{leadCast} (red) and \eqref{lead-saddle} (green).
Bottom: $\gmax(d)$ (red) and $g_*(d)$ (green).
Although on the left the green curve is dominating for $d>20$, 
it does not contribute since $d_*>\gmax$ for $d>18$. 
In contrast, on the right it provides the correct approximation as $d_*$ remains always 
less than $\gmax$. For other hypergeometric CY threefolds, similar graphs look exactly the same.\label{Fig-Fd}}
\end{figure}

Let us consider first $k=1$. In this case the saddle point contribution to $\log |F_d|$
is found to be
\be 
\frac{\(\log\(\frac{\cV}{4\pi^2} \,d\log d\)+\hf\,|\Im\lambda|\)^2-\frac14\, \phi^2}
{\cA\(\log\(\frac{\cV}{2\pi^2} \,d\log d\)\)^2}
+2\pi \cV d -|\Im\lambda|,
\label{lead-saddle}
\ee 
where we ignored the logarithm of the prefactor in \eqref{contr-saddle}
which scales as $\cO(\log d)$. Since $\cA\sim d^{-3/2}$, the contribution \eqref{lead-saddle}
behaves as $\sim d^{3/2}$ and therefore it is always subdominant compared to \eqref{leadCast}.
Of course, if $d$ is not sufficiently large, especially if $|\Im\lambda|$ is small, 
\eqref{lead-saddle} can easily become larger than \eqref{leadCast}.
In fact, for large $|\Im\lambda|$ it also increases faster than 
the boundary contribution \eqref{contrCast} because it is quadratic in this variable.
However, at this point one should remember that the saddle point must satisfy
the condition\footnote{In fact, to apply \eqref{gsp-Fd}, as indicated above, 
	we need the stronger condition $\Re g_*<\gkink(d)$.} $\Re g_*<\gmax(d)$, otherwise it does not contribute.
This severely restricts the possibilities for \eqref{lead-saddle} to be the dominating contribution.
In Fig. \ref{Fig-Fd}, we present several cases which demonstrate that typically 
\eqref{lead-saddle} and \eqref{leadCast} provide a very good approximation for $\log |F_d|$
at small and large values of $|\Im\lambda|$, respectively.

Finally, let us consider the saddle point contribution assuming that $k\sim d^\gamma$
and verify whether one can tune $\gamma$ so that the contribution \eqref{contr-saddle} 
becomes the dominant one. In this case the leading asymptotics is determined by the term quadratic
in $|\Im\lambda|$ and is given by 
\be 
\sim \frac{d^{(3+\gamma)/2}}{(\log d)^2}\, |\Im\lambda|^2.
\ee 
Thus, it seems that, taking $\gamma=1$, one can achieve the scaling which, up to the logarithm in the denominator, will be 
comparable to \eqref{leadCast}. However, the condition on the saddle point, $\Re g_*<\gmax(d_k)$, 
at large $d$ requires that
\be 
\frac{d^{(3-\gamma)/2}}{(\log d)^2}< d^{2-2\gamma}
\ \Rightarrow\ 
\gamma\leq 1/3.
\ee 
Hence, even with non-trivial $k$, the saddle point contribution cannot scale faster than $\frac{d^{5/3}}{(\log d)^2}$. 
This completes the proof of the asymptotics \eqref{leadCast}, showing that it is indeed 
controlled by the contribution from GV invariants near the Castelnuovo bound.

\section{Fourier coefficients of powers of the MacMahon function}
\label{ap-McMahon}

In this appendix, we derive the asymptotics of the Fourier coefficients of the $\chi$-th power of
the MacMahon function, where $\chi$ is an arbitrary integer,
\be
M(\q)^{\chi}= \prod_{k>0}(1-\q^k)^{-k \chi} :=\sum_{n\geq 0} \omega_{\chi}(n) \q^n.
\ee
For this, we need an asymptotic expansion of $f(t)=-\log M(e^{-t})$ as $t\to 0$. This can be
obtained from the Mellin transform \cite[(4.35)]{Pioline:2006ni}
\be
g(s) = \int_0^s \frac{\de t}{t^{1-s}}\, f(t) = - \zeta(s+1)\zeta(s-1) \Gamma(s)
\ee
by shifting the contour in the Mellin inversion formula
\be
f(t) = \frac{1}{2\pi\I} \int_{s_0+\I \IR} \de s \, t^{-s} g(s),
\ee
where $s_0>2$. The pole at $s=2$ produces
\be
\Res_{s=2} \left[ t^{-s} g(s) \right]  = -\frac{\zeta(3)}{t^2}\, ,
\ee
while the pole at $s=0$ gives
\be
\Res_{s=0} \left[ t^{-s} g(s)  \right] = -\frac{1}{12}\log t -\zeta'(-1).
\ee
Poles at negative integer $s$ give further analytic terms in $t$, hence one obtains
\be
f(t) = -\frac{\zeta(3)}{t^2} -\frac{1}{12}\log t -\zeta'(-1) +\frac{t^2}{2880}
+\frac{t^4}{725760}+ \cdots.
\ee

We can now compute the coefficients $\omega_\chi(n)$ by integrating along
a small circle around $\q=0$,
\be
\omega_\chi(n) = \frac{1}{2\pi\I} \oint \frac{\de \q}{\q^{-n-1}}M(\q)^{\chi} = -\frac{1}{2\pi\I} 
\int_{t=-\I\pi + \eps}^{t=\I\pi +\eps}
e^{-\chi f(t) +  n t}.
\ee
For $n\gg 1$, the integral is dominated by a saddle point at $\chi f'(t)=n$, i.e. 
$t^3=2\chi\zeta(3)/n$. For $\chi>0$, the relevant solution is $(2\chi\zeta(3)/n)^{1/3}$, leading to 
\be
\begin{split} 
\omega_\chi(n) \sim & 
\frac{2 \zeta(3)^{\frac{6+\chi}{36}}}
{\sqrt{12\pi \chi}}
\left(\frac{n}{2\chi}\right)^{-\frac{\chi+24}{36}}
\exp \left[ 3 \left( \frac{ \chi \zeta(3) n^2}{4} \right)^{1/3}  
+ \chi \zeta'(-1) 
\right] .
\end{split}
\label{assMcMpos}
\ee
This formula also easily follows from Meinardus' theorem 
(see e.g. \cite[\S 6.2]{Benvenuti:2006qr}), which applies for $\chi>0$ because all coefficients $\omega_\chi(n)$ are non-negative.
For $\chi<0$, there are instead two saddle points $(2|\chi|\zeta(3)/n)^{1/3} e^{\pm \I \pi/3}$,
leading to\footnote{This result differs from the one quoted in \cite[(6.2)]{Denef:2007vg}, and was obtained by   
Charles Cosnier-Horeau and the last-named author in May 2015.} 
\be
\begin{split} 
\omega_\chi(n) \sim & 
\frac{2 \zeta(3)^{\frac{6-|\chi|}{36}}}
{\sqrt{12\pi |\chi|}}
\left(\frac{n}{2|\chi|}\right)^{\frac{|\chi|-24}{36}}
\exp \left[ \frac32 \left( \frac{ |\chi| \zeta(3) n^2}{4} \right)^{1/3}  
- |\chi| \zeta'(-1) 
\right] \\
& \times 
\cos\left[ \frac{3\sqrt{3}}{2} \left( \frac{ |\chi| \zeta(3) n^2}{4} \right)^{1/3} +  \frac{\pi}{36} ( 6 - |\chi| ) 
\right].
\end{split}
\label{assMcMneg}
\ee

\begin{figure}[h]
\begin{center}
\includegraphics[height=5cm]{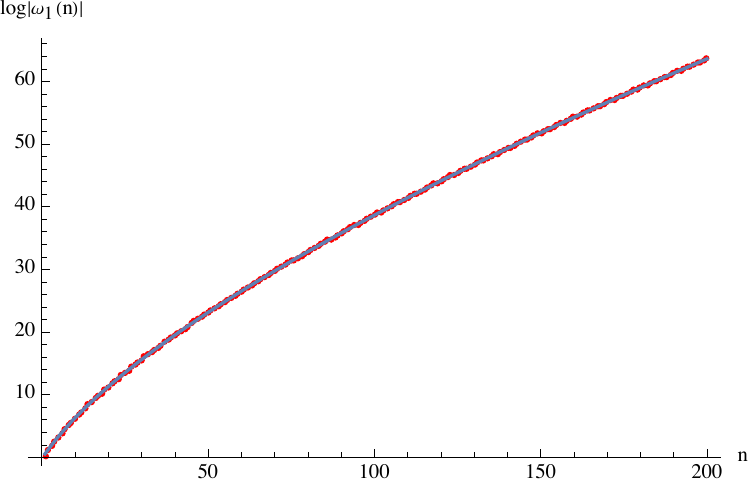} 
\includegraphics[height=5cm]{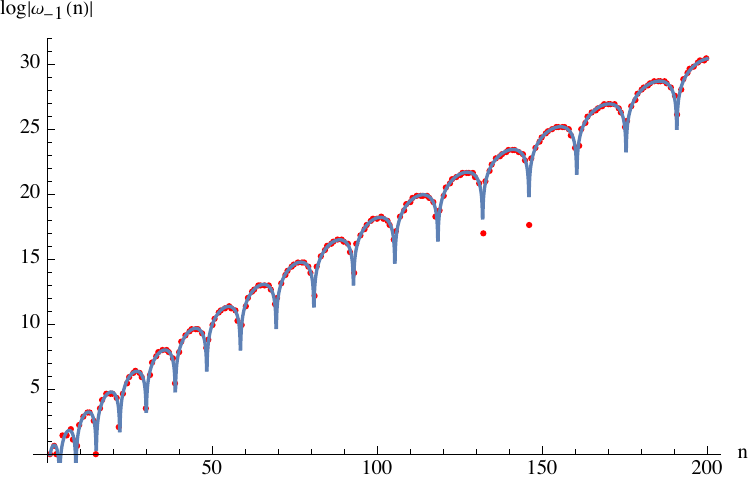} 
\end{center}
\vspace{-0.7cm}
\caption{Logarithm of the Fourier coefficients
of the $\chi$-th power of the MacMahon function (in red), 
compared to the asymptotic formula (in blue) \eqref{assMcMpos} for $\chi=1$ (left) 
or \eqref{assMcMneg} for $\chi=-1$ (right).
\label{fig_McM}}
\end{figure}

In Fig. \ref{fig_McM} we compare this asymptotic formula to the exact coefficients 
for $\chi=\pm 1$; a similar agreement holds for higher values of $|\chi|$. 
It would be interesting to obtain a convergent Rademacher-type expansion of 
$\omega_\chi(n)$ by applying the circle method, along the lines of \cite{Govindarajan:2013pza}.

%\bibliography{combined}
%\bibliographystyle{utphys}

\providecommand{\href}[2]{#2}\begingroup\raggedright\endgroup

\end{document}